# *Open Source Fundamental Industry Classification*

Zura Kakushadze[§†1] and Willie Yu[¶2]

[§] *Quantigic® Solutions LLC,[3] 1127 High Ridge Road, #135, Stamford, CT 06905*

[†] *Free University of Tbilisi, Business School & School of Physics*
*240, David Agmashenebeli Alley, Tbilisi, 0159, Georgia*

[¶] *Centre for Computational Biology, Duke-NUS Medical School*
*8 College Road, Singapore 169857*

Abstract

We provide complete source code for building a fundamental industry classification based on publically available and freely downloadable data. We compare various fundamental industry classifications by running a horserace of short-horizon trading signals (alphas) utilizing open source heterotic risk models (https://ssrn.com/abstract=2600798) built using such industry classifications. Our source code includes various stand-alone and portable modules, e.g., for downloading/parsing web data, etc.

## 1. Introduction

Fundamental industry classifications such as GICS, BICS, ICB, NAICS, SIC, TRBC, etc.[4] are widely used in a variety of fields, including economic applications,[5] general population and healthcare related studies,[6] and (quantitative) finance/trading (including risk modeling).[7] Industry classification (i.e., taxonomy) groups companies into baskets (e.g., industries) based on some kind of a similarity criterion or criteria, which differ from one classification to another. Such fundamental industry classifications generally are expected to be based on pertinent fundamental/economic data, such as companies' products and services, revenue sources, suppliers, competitors, partners, etc. They are essentially independent of the pricing data and, if well-built, tend to be rather stable out-of-sample as companies seldom jump industries.

Many industry classifications are developed commercially and acquiring such data is associated with nontrivial costs. Even government-developed classifications such as NAICS or even SIC (see below) are not exactly free. This is for two main reasons. First, simply specifying a hierarchical structure (e.g., a complete list of, say, sectors, industries and sub-industries as in BICS) is only the tip of the iceberg; many (qualified) man-hours are required to assign to the actual companies the nomenclature within such structure (i.e., which sector, industry and sub-industry each company belongs to, using the BICS terminology). Second, the resultant data is not necessarily provided (by government agencies) as a simple one-click/single-file download.

---

[4] GICS = Global Industry Classification Standard (by MSCI and Standard & Poor's); BICS = Bloomberg Industry Classification Systems; ICB = Industry Classification Benchmark (by London Stock Exchange FTSE); NAICS = North American Industry Classification System (by Mexico's Instituto Nacional de Estadística y Geografía, Statistics Canada aka Statistique Canada, and the United States Office of Management and Budget); SIC = Standard Industrial Classification (by the United States government agencies); TRBC = Thomson Reuters Business Classification.

[5] For economics, financial economics and accounting related literature, see, e.g., [Clarke, 1989], [Cotterman & Peracchi, 1992], [DellaVigna & Pollet, 2007], [Evangelista, 2000], [Guibert *et al*, 1971], [Heston & Rouwenhorst, 1994], [Hicks, 2011], [Hill, 1999], [Hrazdil & Scott, 2013], [Hrazdil *et al*, 2013, 2014], [Hrazdil & Zhang, 2012], [Kile & Phillips, 2009], [Kort, 2001], [Krishnan & Press, 2002], [Laestadius, 2005], [Ojala, 2005], [Pavitt, 1984], [Peneder, 2003], [Perry *et al*, 1985], [Pol *et al*, 2002], [Schröder & Yim, 2012], [Scislaw, 2015]. For a recent review, see, e.g., [Phillips & Ormsby, 2016]. For other applications and more generally related literature, see, e.g., [Arbuckle, 1997], [Boettcher, 1999], [Bowker & Star, 2000], [Guenther & Rosman, 1994], [Katzen, 1995], [Mross & McGuigan, 2016], [Murphy, 1998], [O'Connor, 2000], [Quint, 1996], [Sabroski, 2000], [Walker & Murphy, 2001].

[6] See, e.g., ['t Mannetje & Kromhout, 2003] and references therein.

[7] For related literature, see, e.g., [Alford, 1992], [Asness *et al*, 2014], [Asness & Stevens, 1995], [Bhojraj *et al*, 2003], [Carhart, 1997], [Cavaglia *et al*, 2000], [Chan *et al*, 2007], [Chou *et al*, 2012], [Chung *et al*, 2014], [Cizeau *et al*, 2001], [Elliott *et al*, 2005], [Fama & French, 1993, 1997], [Hong *et al*, 2007], [Horrell & Meraz, 2009], [Kahle & Walkling, 1996], [Kakushadze, 2015a, 2016], [King, 1966], [Lamponi, 2014], [Nadig & Crigger, 2011], [Vermorken, 2011], [Yang *et al*, 2006]. For applications to risk modeling within quantitative finance, see, e.g., [Grinold & Kahn, 2000], [Kakushadze, 2015b], [Kakushadze & Yu, 2016a]. For statistical/data mining related methods, see, e.g., [Bao *et al*, 2008], [Hunink *et al*, 2010], [Kakushadze & Yu, 2016b], [Mantegna, 1999], [Miccichè *et al*, 2005], [Yaros & Imieliński, 2015], [Lee *et al*, 2015].

In this paper, among other things, we fill a gap in this space. We provide open source code for freely downloading SIC data directly from the U.S. Securities and Exchange Commission (SEC) without any APIs, accounts, logins, passwords, etc. Furthermore, since this data is provided by the U.S. government, its downloads by the public are unrestricted in any fashion.

The data provided by the SEC, among other things, contains company names and SIC codes. SIC codes are 4-digit numeric identifiers corresponding to the SIC hierarchy (Division → Major Group → Industry Group → Industry). As we discuss below, there is an efficient and fast way for downloading all companies with SIC codes. However, the SEC does not maintain accurate ticker data.[8] So, the data downloaded from the SEC website must be matched to tickers. At the end, our code matches tickers to SIC codes.[9] There are various nuances in the underlying SEC data, such as peculiarities with the SIC codes used by the SEC, which we also discuss in detail.

SIC is not the best classification around. And, contrary to an apparent misconception, this is not because SIC is not "granular" (or detailed) enough. On the face of it, it is more granular than BICS. However, as mentioned above, the hierarchical structure itself is only the tip of the iceberg. What is arguably even more important is the assignment of companies into such hierarchy. In the case of SIC apparently this assignment does not necessarily always follow the aforesaid criteria based on companies' products and services, revenue sources, suppliers, competitors, partners, etc. Instead, it appears that at least in some cases such assignments are made on more superficial grounds (this could, e.g., be the companies own assessment, etc.).

However, SIC is not a "disaster" by any stretch. It is widely used by academics (see, e.g., [Fama and French, 1997]), and perhaps to a lesser extent by quant traders (who usually prefer more robust industry classifications such as GICS and BICS), but still used. For a (young, but not only) researcher who wishes to test, e.g., some trading ideas involving industry classification but does not wish to commit to costly commercial data subscriptions, SIC can be a good zeroth-order approximation, so long it is free and easily obtained.[10] Our code provides such a solution.

We quantify a comparison between different industry classifications by utilizing them for building short-horizon mean-reversion trading signals (alphas) via (open source) heterotic risk models [Kakushadze, 2015b]. We find that GICS slightly outperforms BICS, and SIC performs worse, including when restricted to the so-called Fama-French [1997] industry classification.

The remainder is organized as follows. Section 2 discusses data (downloads). Section 3 discusses backtests. Section 4 briefly concludes. Data and source code are in Appendices.

---

[8] There is a way to search for companies by tickers on the SEC website. However, many tickers are missing.

[9] We discuss how to source all listed and OTC (over-the-counter) U.S. tickers.

[10] SIC data is available through various data providers; however, most are not free, and not all have access to them.

## 2. Data (Downloads)

### *2.1. SIC Hierarchical Structure*

SIC structure has 4 levels: Division → Major Group → Industry Group → Industry. This structure can be download from the website https://www.osha.gov/pls/imis/sic_manual.html of the Occupational Safety and Health Administration (OSHA) of the U.S. Department of Labor using the R function `sec.osha()` in Appendix C.[11] This function outputs a single tab-delimited file `SIC.table.txt`, which contains the SIC hierarchy given in Appendix A. More precisely, in Appendix A, for reading convenience, the data is separated by `" > "`. Also, Appendix A contains lines (in bold italic font), which are not present in `SIC.table.txt` and pertain to additional SIC codes present in the SEC data, which we will discuss in detail below.

The 10 SIC Divisions are labeled by characters A through J. The Major Groups are labeled by 2-digit numeric codes XY, where both X and Y can take values 0 through 9. It is convenient to label Major Groups by 4-digit codes XY00. The Industry Groups are labeled by 3-digit numeric codes XYZ. Unlike X and Y, Z can only take values between 1 and 9. Again, it is convenient to label Industry Groups by 4-digit codes XYZ0. Thus, Industry Group XYZ0 belongs to Major Group XY00. Finally, Industries are labeled by 4-digit codes XYZW, where W also takes values between 1 and 9. Industry XYZW belongs to Industry Group XYZ0. This is the SIC hierarchical structure.

### *2.2. SEC Data Download*

The SEC website allows data to be searched for based on a company name, CIK number,[12] ticker (albeit, as mentioned above, ticker search is unreliable), etc.[13] Happily, we can also search by SIC codes. There are two approaches we can take here. A priori we do not know what SIC codes to expect in the data. So, we can scan all sic codes 0100 through 9999. Once we have all SIC codes used by the SEC, we can limit the downloads to these predefined SIC codes to reduce the download time. As a precautionary measure, periodically we may wish to run full scans to check whether new SIC codes crop up in the data. The `sec.all.sic()` R function in Appendix C downloads the data. For its first argument `run.all.sic = T` it downloads all sic codes 0100 through 9999, whereas for `run.all.sic = F` the download is limited to the SIC codes in a tab-delimited input file `SIC.Codes.txt`, which contains the SIC codes currently used by the SEC along with the corresponding Industry names.[14] Appendix B contains this file. Here some remarks are in order. Thus, some of the actual Industry names in the downloaded data (column 4 of the output file `SIC.Download.txt` – see below) are (immaterially) different or bad.[15] Therefore, the Industry names in `SIC.Codes.txt` (Appendix B) are based on the table provided at https://www.sec.gov/info/edgar/siccodes.htm combined[16] with those in `SIC.table.txt` (Appendix A) downloaded from OSHA (see above).

---

[11] The source code in Appendix C hereof is not written to be "fancy" or optimized for speed or in any other way. Its sole purpose is to illustrate the algorithms described and/or discussion in the main text in a simple-to-understand fashion. See Appendix D for some important legalese.

[12] According to the SEC, https://www.sec.gov/edgar/searchedgar/cik.htm: "The Central Index Key (CIK) is used on the SEC's computer systems to identify corporations and individual people who have filed disclosure with the SEC."

[13] There are some nuances. E.g., an automated search by name via scanning through all single and double ASCII and non-ASCII characters captures most filers/companies. However, due to a constraint on the number of pages that can be scanned and inefficiencies in the data ordering, this gets complicated relatively quickly. *Así es la vida…*

The `sec.all.sic()` function outputs a tab-delimited file `SIC.Download.txt`. The first column is the CIK number, the second column is the company name, the third column is the SIC code, the fourth column is the Industry name as it appears in the SEC data (which, as mentioned above, is not necessarily the same as the corresponding Industry name in the file `SIC.Codes.txt`), and the fifth column is the location (U.S. state, Canadian province, foreign county, etc.) code.[17] The page https://www.sec.gov/edgar/searchedgar/edgarstatecodes.htm contains most location codes. The data also contains legacy locations codes (to wit, E6, L4, I8, I9, E7, U2, L5 and LO ["L-O" as opposed to L0 = "L-zero"]). These old codes are all described at https://www.sec.gov/edgar/searchedgar/edgarstatecodes.htm. One additional code present in the data is X9. Only two companies have this code, and these appear to be German entities.[18]

As mentioned above, some entries in Appendix A (in bold italic font) are not included in the OSHA download `SIC.table.txt`. These correspond to the additional SIC codes present in the SEC data. Appendix A is obtained by amending `SIC.table.txt` with these additional codes. Most of them fit nicely into the SIC hierarchical structure. A couple of potential hiccups are: 1) SIC code = 6025, with only a single company, PNB BANCSHARES INC (CIK = 0001230585); and 2) SIC code = 9995.[19] Three SIC codes 0888, 8880 and 8888 do not fit in the 4-digit hierarchy described above, so we appended them at the end of Appendix A. All lines with non-OSHA SIC codes in Appendix A (i.e., those in bold italic font) end with *our* descriptor "(SEC)". Overall, the SEC data is reasonably "clean" barring the aforesaid manually-to-deal-with glitches.

---

[14] If a SIC code is of the form XYZ0, then the Industry name is the same as the corresponding Industry Group name. If a SIC code is of the form XY00, then the Industry name is the same as the corresponding Major Group name.

[15] E.g., some well-defined SIC codes in the SEC data download are erroneously marked as "unknown", etc.

[16] The table at https://www.sec.gov/info/edgar/siccodes.htm is missing the SIC codes 0888, 1044, 6025, 6120.

[17] E.g., for Microsoft Corporation we have the following data: CIK = 0000789019, company name = MICROSOFT CORP, SIC code = 7372, Industry name = SERVICES-PREPACKAGED SOFTWARE, location code = WA.

[18] To wit, CYBERMIND AG (CIK = 0001135128) and KPMG DEUTSCHE TREUHAND GESELLSCHAFT AG (CIK = 0001184474). These companies do not have SIC codes, but some companies with the legacy location codes do.

[19] Also, see https://www.sec.gov/fast-answers/answers-blankcheckhtm.html for "Blank Checks" (SIC code = 6770).

*2.3. Matching to Tickers*

In practical quantitative finance/trading applications assignment of SIC codes to company names is only partially useful as most (e.g., pricing) data is labeled by tickers. So, we need to match our data in the file `SIC.Download.txt` to tickers. This is done in the R function `sec.sic()` in Appendix C. Its sole argument is `incl.otc`, which is the second argument of the function `sec.all.sic()` (see above). For `incl.otc = F` only listed U.S. tickers are matched, and for `incl.otc = T` the over-the-counter (OTC) tickers are also included. The input files of `sec.sic()` are: i) `NQ_AMEX.csv`, `NQ_NYSE.csv`, `NQ_NASDAQ.csv` (these files can be downloaded daily from www.nasdaq.com via, e.g., the `wget` utility – see Appendix C.1); `NT_otherlisted.txt`, `NT_nasdaqlisted.txt` (these files can be downloaded daily from www.nasdaqtrader.com via, e.g., the `wget` utility – see Appendix C.1); and iii) if we wish to include OTCs, i.e., if `incl.otc = T`, then also `otctickers.csv` (this file can be manually downloaded daily from http://www.otcmarkets.com/reports/symbol_info.csv). Note that the source code is straightforward to modify to accommodate other sources of ticker lists.

The function `sec.sic()` matches tickers in the above files to SIC codes by matching the company names in the SEC data to those in the ticker lists. It goes without saying that there are cases where a match is not reasonably attainable. There are two main categories here. First, the SEC data simply may not have a SIC code assigned to a given company, or the company name used by the SEC may be different from that in the ticker lists, hence no match (e.g., ETRADE v. E*TRADE). Second, there may be more than one match (after stripping the company names of various extraneous attributes that muddy the waters, e.g., CORPORATION v. CORP). The statistics for the occurrences are given in Table 1, which also provides the number of total tickers, and how many of the tickers without the SIC code correspond to funds, trusts, ETFs and similar as-is difficult-to-classify vehicles, which the code attempts to detect using a heuristic.[20] Overall, the ticker matching (especially if we exclude funds, etc.) yields a reasonable coverage.

The function `sec.sic()` outputs 3 files. The file `TICKER.SIC.txt` contains the tickers for which SIC codes are matched: first column = ticker; second column = exchange (A = AMEX, N = NYSE, Q = NASDAQ, the others relate to OTCs); third column = SIC code; fourth column = Industry name (as it appears in the SEC data – see above); fifth column = market cap (for listed tickers only as provided in the aforesaid three www.nasdaq.com files). The file `NO.SIC.txt` contains the tickers for which SIC codes are not matched: first column = ticker; second column = exchange (see above); third column = market cap (see above); fourth column = TRUE (FALSE) if the ticker is (not) a fund, etc.; fifth column = TRUE (FALSE) if there is no match (multiple matches). The file `SIC.IND.CLASS.txt` contains a binary industry classification.

---

[20] Which may be improved by using the ETF field in `NT_otherlisted.txt` and `NT_nasdaqlisted.txt`.

*2.4. Some Stats*

The function `sec.all.sic()` at the end calls the function `sec.sic()`. Before doing so it outputs a log file, which records the start and end times for measuring the time required for downloading the SEC data. Such measurements are reported in Table 2. Downloading all SIC codes 0100-9999 takes about 30 minutes, while downloading the SIC codes restricted to the `SIC.Codes.txt` file takea about 5 minutes. Ticker matching is fast (on a quad core, 3.1 GHz CPU machine). Finally, let us mention some statistics for the binary industry classification file `SIC.IND.CLASS.txt`. On 04/14/2017, for 4734 tickers with SIC codes, there were 392 Industries with the following ticker counts: Min = 1, 1st Quartile = 2, Median = 5, Mean = 12.08, 3rd Quartile = 10, Max = 376. I.e., there are many small (with 1 or 2 tickers) Industries present.

## 3. Horserace (Backtests)

So, we can download the SEC data and build an industry classification based on SIC codes. How does this industry classification compare with others? A general comparison (granularity, etc.) was performed in [Hrazdil *et al*, 2014] for GICS, NAICS and SIC. A comparison of how well industry classifications explain the co-movement of stock returns at long horizons was done in [Chan *et al*, 2007], which found that GICS – not surprisingly – does better than the Fama-French [1997] industry classification with 48 "industries" (FF48), which is based on SIC. However, FF48 is not the same as SIC, it aggregates tickers with multiple SIC codes into these 48 "industries". I.e., FF48 is much less granular than GICS, BICS and SIC at their respective most granular levels.

Here we compare GICS, BICS, SIC, FF48 and FF49 (which is another variation of FF48 – see below) at short horizons. However, here we take the comparison to another level – instead of looking at the co-movement of stock returns, we look at the performance of short-horizon mean-reversion trading signals (alphas) computed using otherwise identical risk models except for the industry classifications these risk models use to define the industry risk factors. The risk models we use are the open source heterotic risk models of [Kakushadze, 2015b]. Heterotic risk models do not use any style factors, so our horserace compares the underlying industry classifications without muddling them up with other extraneous factors. To our knowledge, this is the first such comparison of industry classifications appearing in published literature. The remainder of this section closely follows most parts of Section 6 of [Kakushadze, 2015b].[21]

In brief, heterotic risk models for equities combine: i) granularity of an industry classification; ii) diagonality of the principal component factor covariance matrix for any sub-universe of stocks; and iii) dramatic reduction of the factor covariance matrix size in the Russian-doll risk model construction [Kakushadze, 2015c]. Thus, in the heterotic risk model construction, using (for the sake of definiteness) the BICS nomenclature, one breaks the universe of stocks into subsets based on BICS sub-industries, computes the first principal components of the return sample correlation matrices for these subsets, and uses these first principal components as weights for computing the corresponding factor returns. Typically, for short lookbacks, the factor return sample covariance matrix is singular. One then further breaks the universe of these factors into subsets based on BICS industries, computes the first principal components of the return sample correlations matrices for these subsets, and uses these first principal components as weights for computing the corresponding factor returns. This nested embedding (the Russian-doll construction) is repeated (e.g., by going from BICS industries to BICS sectors to the "broad market" as defining fewer and fewer risk factors, as needed) until the number of the resultant factor returns is small enough such that the corresponding factor covariance matrix is nonsingular and sufficiently stable. This proves a powerful approach for constructing out-of-sample stable short-lookback equity risk models.

---
[21] We "rehash" it here not to be repetitive but so that our presentation here is self-contained.

### *3.1. Notations*

Let $P_{is}$ be the time series of stock prices, where $i = 1, \ldots, N$ labels the stocks, and $s = 1, 2, \ldots$ labels the trading dates, with $s = 1$ corresponding to the most recent date in the time series. The superscripts $O$ and $C$ (unadjusted open and close prices) and $AO$ and $AC$ (open and close prices fully adjusted for splits and dividends) will distinguish the corresponding prices, so, e.g., $P_{is}^C$ is the unadjusted close price. $V_{is}$ is the unadjusted daily volume (in shares). Also, for each date $s$ we define the overnight return as the previous-close-to-open return:

$$E_{is} = \ln\left(\frac{P_{is}^{AO}}{P_{i,s+1}^{AC}}\right) \qquad (1)$$

This return will be used in the definition of the expected return in our mean-reversion alpha. We will also need the close-to-close return

$$R_{is} = \ln\left(\frac{P_{is}^{AC}}{P_{i,s+1}^{AC}}\right) \qquad (2)$$

An out-of-sample (see below) time series of these returns will be used in constructing the risk models. Note that all prices in the definitions of the returns $E_{is}$ and $R_{is}$ are fully adjusted.

We assume that: i) the portfolio is established at the open[22] with fills at the open prices $P_{is}^O$; ii) it is liquidated at the close on the same day – so this is a purely intraday alpha – with fills at the close prices $P_{is}^C$; and iii) there are no transaction costs or slippage – our aim here is not to build a realistic trading strategy, but to test *relative* performance of various industry classifications. The P&L for each stock

$$\Pi_{is} = H_{is} \left( \frac{P_{is}^C}{P_{is}^O} - 1 \right) \tag{3}$$

---

[22] This is a so-called "delay-0" alpha: the same price, $P_{is}^O$ (or adjusted $P_{is}^{AO}$), is used in computing the expected return (via $E_{is}$) and as the establishing fill price.

Here $H_{is}$ are the *dollar* holdings. The shares bought plus sold (establishing plus liquidating trades) for each stock on each day are computed via $Q_{is} = 2\,|H_{is}|\,/\,P_{is}^O$.

*3.2. Universe Selection*

For the sake of simplicity,[23] we select our universe based on the average daily dollar volume (ADDV) defined via (note that $A_{is}$ is out-of-sample for each date $s$):

$$A_{is} = \frac{1}{m} \sum_{r=1}^{m} V_{i,s+r}\, P_{i,s+r}^C \tag{4}$$

We take $m = 21$ (i.e., one month), and then take our universe to be the top 2000 tickers by ADDV. To ensure that we do not inadvertently introduce a universe selection bias, we rebalance monthly (every 21 trading days, to be precise). I.e., we break our 5-year backtest period (see below) into 21-day intervals, we compute the universe using ADDV (which, in turn, is computed based on the 21-day period immediately preceding such interval), and use this universe during the entire such interval. We do have the survivorship bias as we take the data for the universe of tickers as of 9/6/2014 that have historical pricing data sourced from http://finance.yahoo.com (accessed on 9/6/2014) for the period 8/1/2008 through 9/5/2014. We restrict this universe to include only U.S. listed common stocks and class shares (no OTCs, preferred shares, etc.) with GICS,[24] BICS and SIC industry assignments as of 9/8/2014.[25] However, as discussed in detail in Section 7 of [Kakushadze, 2015a], the survivorship bias is not a leading effect in such backtests.[26]

*3.3. Backtesting*

We run our simulations over a period of 5 years (more precisely, 1260 trading days going back from 9/5/2014, inclusive). The annualized return-on-capital (ROC) is computed as the average daily P&L divided by the intraday investment level $I$ (with no leverage) and multiplied by 252. The annualized Sharpe Ratio (SR) is computed as the daily Sharpe ratio multiplied by $\sqrt{252}$. Cents-per-share (CPS) is computed as the total P&L in cents (not dollars) divided by the total shares traded.

---

[23] In practice, the trading universe is selected based on market cap, liquidity (ADDV), price and other criteria.
[24] ZK would like to thank Jim Liew for sharing the GICS data on 9/8/2014.

[25] The choice of the backtesting window is based on what data was readily available.

[26] Furthermore, here we are after the *relative outperformance*, and it is reasonable to assume that, to the leading order, individual performances are affected by the survivorship bias approximately equally as the construction of all alphas and risk models (see below) is "statistical" and oblivious to the trading universe.

### 3.4. Optimized Alphas

The optimized alphas are based on the expected returns $E_{is}$ defined in Eq. (1) optimized via Sharpe ratio maximization using heterotic risk models [Kakushadze, 2015b] based on the industry classifications we are testing, which are GICS, BICS, SIC, FF48 (see [French, 2017a] for the definition thereof) and FF49 (see [French, 2017b] for the definition thereof). FF48 and FF49 have just one level by construction. GICS, BICS and SIC have multiple levels. We take these industry classifications at their respective most granular levels (which are called sub-industries for GICS and BICS, and industries for SIC – see above). We then run the R function `qrm.het()` in Appendix B of [Kakushadze, 2015b] with the following inputs: `ret` is the matrix of returns $R_{is}$ defined in Eq. (2); `ind` is the binary $N \times K$ industry classification matrix (each element equals 1 if the corresponding ticker belongs to the corresponding industry, and 0 otherwise – for SIC this matrix is in the file `SIC.IND.CLASS.txt` we discuss above); `mkt.fac = T` (this adds the "market" factor, so that effectively we have a 2-level industry classification and the risk factor covariance matrix is invertible – see [Kakushadze, 2015b] for details); and `rm.sing.tkr = F` (the default). As in [Kakushadze, 2015b], we compute the heterotic risk model covariance matrix $\Gamma_{ij}$ every 21 trading days (same as for the universe). For each date (we omit the index $s$) we maximize the Sharpe ratio subject to the dollar neutrality constraint and position bounds:[27]

$$S = \frac{\sum_{i=1}^{N} H_i E_i}{\sqrt{\sum_{i,j=1}^{N} \Gamma_{ij} H_i H_j}} \to \max \qquad (5)$$

$$\sum_{i=1}^{N} H_i = 0 \qquad (6)$$

$$|H_{is}| \leq 0.01 \, A_{is} \qquad (7)$$

$$\sum_{i=1}^{N} |H_{is}| = I \qquad (8)$$

---

[27] Which in this case are the same as trading bounds as the strategy is purely intraday.

Here $A_{is}$ is ADDV defined in Eq. (4). Eq. (6) is the dollar neutrality constraint. Eq. (7) imposes the aforesaid trading bounds. Eq. (8) ensures that the portfolio has the investment level $I$. In the presence of bounds computing $H_i$ requires an iterative procedure and we use the R code in Appendix C of [Kakushadze, 2015b] (which also contains detailed documentation).

*3.5. Simulation Results*

Table 3 (also see Figure 1) summarizes the simulation results.[28] GICS sub-industries slightly outperform BICS sub-industries, which outperform SIC industries. FF48 and FF49 underperform SIC industries – simply due to lesser granularity. For comparison purposes in Table 3 we also include simulation results for BICS where we replace BICS sub-industries by BICS industries and BICS sectors. Generally, reducing granularity leads to underperformance in such backtests.

## 4. Conclusions

To our knowledge, the industry classification comparison at short-horizons we present here – especially using actual trading signals (as opposed to co-movements of returns) – is the first of its kind in the published literature. GICS remains to be a safe choice. It is unfortunate that BICS is no longer supported by Bloomberg. Also, SIC is not a "disaster" by any stretch and – now that we have provided open source code for it – can be used widely in, e.g., preliminary research.

Finally, let us comment that (well-built) industry classifications are intrinsically stable (w.r.t. temporal changes). This is primarily due to the fact that companies seldom jump industries (let alone less granular structures such as sectors). After all, fundamental industry classifications generally are expected to be based on pertinent fundamental/economic data, such as companies' products and services, revenue sources, suppliers, competitors, partners, etc. This simplifies things, including as they relate to out-of-sample backtesting. Thus, if one obtains an industry classification for a universe of stocks today and uses this industry classification for backtesting for the same universe of stocks for the past year (that is, accounting for any ticker changes, etc., that may have transpired in the interim), the backtest can still be deemed valid as the likelihood of stocks jumping industries during this period is small. This is not to say that the industry classification provider may not have retroactively changed, say, sub-industry assignments for a few tickers. However, this kind of "in-sampleness" can be present in virtually any third-party data unless it is collected by the end-user on each and every historical day.

**Acknowledgments**

We would like to thank anonymous reviewers for valuable suggestions, which have improved the manuscript.

---

[28] The result for BICS sub-industries differs from that in [Kakushadze, 2015b] as the universes are different. In [Kakushadze, 2015b] the universe is based on BICS only. Here we require GICS, BICS and SIC assignments.

# Appendix A: SIC Hierarchical Structure

```
Code > Division > Major Group > Industry Group > Industry
0111 > Agriculture, Forestry, And Fishing > Agricultural Production Crops > Cash Grains > Wheat
0112 > Agriculture, Forestry, And Fishing > Agricultural Production Crops > Cash Grains > Rice
0115 > Agriculture, Forestry, And Fishing > Agricultural Production Crops > Cash Grains > Corn
0116 > Agriculture, Forestry, And Fishing > Agricultural Production Crops > Cash Grains > Soybeans
0119 > Agriculture, Forestry, And Fishing > Agricultural Production Crops > Cash Grains > Cash Grains, Not Elsewhere Classified
0131 > Agriculture, Forestry, And Fishing > Agricultural Production Crops > Field Crops, Except Cash Grains > Cotton
0132 > Agriculture, Forestry, And Fishing > Agricultural Production Crops > Field Crops, Except Cash Grains > Tobacco
0133 > Agriculture, Forestry, And Fishing > Agricultural Production Crops > Field Crops, Except Cash Grains > Sugarcane and Sugar Beets
0134 > Agriculture, Forestry, And Fishing > Agricultural Production Crops > Field Crops, Except Cash Grains > Irish Potatoes
0139 > Agriculture, Forestry, And Fishing > Agricultural Production Crops > Field Crops, Except Cash Grains > Field Crops, Except Cash Grains, Not Elsewhere Classified
0161 > Agriculture, Forestry, And Fishing > Agricultural Production Crops > Vegetables And Melons > Vegetables and Melons
0171 > Agriculture, Forestry, And Fishing > Agricultural Production Crops > Fruits And Tree Nuts > Berry Crops
0172 > Agriculture, Forestry, And Fishing > Agricultural Production Crops > Fruits And Tree Nuts > Grapes
0173 > Agriculture, Forestry, And Fishing > Agricultural Production Crops > Fruits And Tree Nuts > Tree Nuts
0174 > Agriculture, Forestry, And Fishing > Agricultural Production Crops > Fruits And Tree Nuts > Citrus Fruits
0175 > Agriculture, Forestry, And Fishing > Agricultural Production Crops > Fruits And Tree Nuts > Deciduous Tree Fruits
0179 > Agriculture, Forestry, And Fishing > Agricultural Production Crops > Fruits And Tree Nuts > Fruits and Tree Nuts, Not Elsewhere Classified
0181 > Agriculture, Forestry, And Fishing > Agricultural Production Crops > Horticultural Specialties > Ornamental Floriculture and Nursery Products
0182 > Agriculture, Forestry, And Fishing > Agricultural Production Crops > Horticultural Specialties > Food Crops Grown Under Cover
0191 > Agriculture, Forestry, And Fishing > Agricultural Production Crops > General Farms, Primarily Crop > General Farms, Primarily Crop
0211 > Agriculture, Forestry, And Fishing > Agriculture production livestock and animal specialties > Livestock, Except Dairy And Poultry > Beef Cattle Feedlots
0212 > Agriculture, Forestry, And Fishing > Agriculture production livestock and animal specialties > Livestock, Except Dairy And Poultry > Beef Cattle, Except Feedlots
0213 > Agriculture, Forestry, And Fishing > Agriculture production livestock and animal specialties > Livestock, Except Dairy And Poultry > Hogs
0214 > Agriculture, Forestry, And Fishing > Agriculture production livestock and animal specialties > Livestock, Except Dairy And Poultry > Sheep and Goats
0219 > Agriculture, Forestry, And Fishing > Agriculture production livestock and animal specialties > Livestock, Except Dairy And Poultry > General Livestock, Except Dairy and Poultry
0241 > Agriculture, Forestry, And Fishing > Agriculture production livestock and animal specialties > Dairy Farms > Dairy Farms
0251 > Agriculture, Forestry, And Fishing > Agriculture production livestock and animal specialties > Poultry And Eggs > Broiler, Fryer, and Roaster Chickens
0252 > Agriculture, Forestry, And Fishing > Agriculture production livestock and animal specialties > Poultry And Eggs > Chicken Eggs
0253 > Agriculture, Forestry, And Fishing > Agriculture production livestock and animal specialties > Poultry And Eggs > Turkeys and Turkey Eggs
0254 > Agriculture, Forestry, And Fishing > Agriculture production livestock and animal specialties > Poultry And Eggs > Poultry Hatcheries
0259 > Agriculture, Forestry, And Fishing > Agriculture production livestock and animal specialties > Poultry And Eggs > Poultry and Eggs, Not Elsewhere Classified
0271 > Agriculture, Forestry, And Fishing > Agriculture production livestock and animal specialties > Animal Specialties > Fur-Bearing Animals and Rabbits
```

```
0272 > Agriculture, Forestry, And Fishing > Agriculture production livestock and animal specialties > Animal Specialties > Horses and
Other Equines
0273 > Agriculture, Forestry, And Fishing > Agriculture production livestock and animal specialties > Animal Specialties > Animal
Aquaculture
0279 > Agriculture, Forestry, And Fishing > Agriculture production livestock and animal specialties > Animal Specialties > Animal
Specialties, Not Elsewhere Classified
0291 > Agriculture, Forestry, And Fishing > Agriculture production livestock and animal specialties > General Farms, Primarily
Livestock And Animal > General Farms, Primarily Livestock and Animal Specialties
0711 > Agriculture, Forestry, And Fishing > Agricultural Services > Soil Preparation Services > Soil Preparation Services
0721 > Agriculture, Forestry, And Fishing > Agricultural Services > Crop Services > Crop Planting, Cultivating, and Protecting
0722 > Agriculture, Forestry, And Fishing > Agricultural Services > Crop Services > Crop Harvesting, Primarily by Machine
0723 > Agriculture, Forestry, And Fishing > Agricultural Services > Crop Services > Crop Preparation Services for Market, Except
Cotton Ginning
0724 > Agriculture, Forestry, And Fishing > Agricultural Services > Crop Services > Cotton Ginning
0741 > Agriculture, Forestry, And Fishing > Agricultural Services > Veterinary Services > Veterinary Services for Livestock
0742 > Agriculture, Forestry, And Fishing > Agricultural Services > Veterinary Services > Veterinary Services for Animal Specialties
0751 > Agriculture, Forestry, And Fishing > Agricultural Services > Animal Services, Except Veterinary > Livestock Services, Except
Veterinary
0752 > Agriculture, Forestry, And Fishing > Agricultural Services > Animal Services, Except Veterinary > Animal Specialty Services,
Except Veterinary
0761 > Agriculture, Forestry, And Fishing > Agricultural Services > Farm Labor And Management Services > Farm Labor Contractors and
Crew Leaders
0762 > Agriculture, Forestry, And Fishing > Agricultural Services > Farm Labor And Management Services > Farm Management Services
0781 > Agriculture, Forestry, And Fishing > Agricultural Services > Landscape And Horticultural Services > Landscape Counseling and
Planning
0782 > Agriculture, Forestry, And Fishing > Agricultural Services > Landscape And Horticultural Services > Lawn and Garden Services
0783 > Agriculture, Forestry, And Fishing > Agricultural Services > Landscape And Horticultural Services > Ornamental Shrub and Tree
Services
0811 > Agriculture, Forestry, And Fishing > Forestry > Timber Tracts > Timber Tracts
0831 > Agriculture, Forestry, And Fishing > Forestry > Forest Nurseries And Gathering Of Forest > Forest Nurseries and Gathering of
Forest Products
0851 > Agriculture, Forestry, And Fishing > Forestry > Forestry Services > Forestry Services
0912 > Agriculture, Forestry, And Fishing > Fishing, hunting, and trapping > Commercial Fishing > Finfish
0913 > Agriculture, Forestry, And Fishing > Fishing, hunting, and trapping > Commercial Fishing > Shellfish
0919 > Agriculture, Forestry, And Fishing > Fishing, hunting, and trapping > Commercial Fishing > Miscellaneous Marine Products
0921 > Agriculture, Forestry, And Fishing > Fishing, hunting, and trapping > Fish Hatcheries And Preserves > Fish Hatcheries and
Preserves
0971 > Agriculture, Forestry, And Fishing > Fishing, hunting, and trapping > Hunting And Trapping, And Game Propagation > Hunting and
Trapping, and Game Propagation
1011 > Mining > Metal Mining > Iron Ores > Iron Ores
1021 > Mining > Metal Mining > Copper Ores > Copper Ores
1031 > Mining > Metal Mining > Lead And Zinc Ores > Lead and Zinc Ores
1041 > Mining > Metal Mining > Gold And Silver Ores > Gold Ores
1044 > Mining > Metal Mining > Gold And Silver Ores > Silver Ores
1061 > Mining > Metal Mining > Ferroalloy Ores, Except Vanadium > Ferroalloy Ores, Except Vanadium
1081 > Mining > Metal Mining > Metal Mining Services > Metal Mining Services
1094 > Mining > Metal Mining > Miscellaneous Metal Ores > Uranium-Radium-Vanadium Ores
1099 > Mining > Metal Mining > Miscellaneous Metal Ores > Miscellaneous Metal Ores, Not Elsewhere Classified
1221 > Mining > Coal Mining > Bituminous Coal And Lignite Mining > Bituminous Coal and Lignite Surface Mining
1222 > Mining > Coal Mining > Bituminous Coal And Lignite Mining > Bituminous Coal Underground Mining
1231 > Mining > Coal Mining > Anthracite Mining > Anthracite Mining
```

1241 > Mining > Coal Mining > Coal Mining Services > Coal Mining Services
1311 > Mining > Oil And Gas Extraction > Crude Petroleum And Natural Gas > Crude Petroleum and Natural Gas
1321 > Mining > Oil And Gas Extraction > Natural Gas Liquids > Natural Gas Liquids
1381 > Mining > Oil And Gas Extraction > Oil And Gas Field Services > Drilling Oil and Gas Wells
1382 > Mining > Oil And Gas Extraction > Oil And Gas Field Services > Oil and Gas Field Exploration Services
1389 > Mining > Oil And Gas Extraction > Oil And Gas Field Services > Oil and Gas Field Services, Not Elsewhere Classified
1411 > Mining > Mining And Quarrying Of Nonmetallic Minerals, Except Fuels > Dimension Stone > Dimension Stone
1422 > Mining > Mining And Quarrying Of Nonmetallic Minerals, Except Fuels > Crushed And Broken Stone, Including Riprap > Crushed and Broken Limestone
1423 > Mining > Mining And Quarrying Of Nonmetallic Minerals, Except Fuels > Crushed And Broken Stone, Including Riprap > Crushed and Broken Granite
1429 > Mining > Mining And Quarrying Of Nonmetallic Minerals, Except Fuels > Crushed And Broken Stone, Including Riprap > Crushed and Broken Stone, Not Elsewhere Classified
1442 > Mining > Mining And Quarrying Of Nonmetallic Minerals, Except Fuels > Sand And Gravel > Construction Sand and Gravel
1446 > Mining > Mining And Quarrying Of Nonmetallic Minerals, Except Fuels > Sand And Gravel > Industrial Sand
1455 > Mining > Mining And Quarrying Of Nonmetallic Minerals, Except Fuels > Clay, Ceramic, And Refractory Minerals > Kaolin and Ball Clay
1459 > Mining > Mining And Quarrying Of Nonmetallic Minerals, Except Fuels > Clay, Ceramic, And Refractory Minerals > Clay, Ceramic, and Refractory Minerals, Not Elsewhere Classified
1474 > Mining > Mining And Quarrying Of Nonmetallic Minerals, Except Fuels > Chemical And Fertilizer Mineral Mining > Potash, Soda, and Borate Minerals
1475 > Mining > Mining And Quarrying Of Nonmetallic Minerals, Except Fuels > Chemical And Fertilizer Mineral Mining > Phosphate Rock
1479 > Mining > Mining And Quarrying Of Nonmetallic Minerals, Except Fuels > Chemical And Fertilizer Mineral Mining > Chemical and Fertilizer Mineral Mining, Not Elsewhere Classified
1481 > Mining > Mining And Quarrying Of Nonmetallic Minerals, Except Fuels > Nonmetallic Minerals Services, Except Fuels > Nonmetallic Minerals Services, Except Fuels
1499 > Mining > Mining And Quarrying Of Nonmetallic Minerals, Except Fuels > Miscellaneous Nonmetallic Minerals, Except > Miscellaneous Nonmetallic Minerals, Except Fuels
1521 > Construction > Building Construction General Contractors And Operative Builders > General Building Contractors-residential > General Contractors-Single-Family Houses
1522 > Construction > Building Construction General Contractors And Operative Builders > General Building Contractors-residential > General Contractors-Residential Buildings, Other Than Single-Family
1531 > Construction > Building Construction General Contractors And Operative Builders > Operative Builders > Operative Builders
1541 > Construction > Building Construction General Contractors And Operative Builders > General Building Contractors-nonresidential > General Contractors-Industrial Buildings and Warehouses
1542 > Construction > Building Construction General Contractors And Operative Builders > General Building Contractors-nonresidential > General Contractors-Nonresidential Buildings, Other than Industrial Buildings and Warehouses
1611 > Construction > Heavy Construction Other Than Building Construction Contractors > Highway And Street Construction, Except > Highway and Street Construction, Except Elevated Highways
1622 > Construction > Heavy Construction Other Than Building Construction Contractors > Heavy Construction, Except Highway And Street > Bridge, Tunnel, and Elevated Highway Construction
1623 > Construction > Heavy Construction Other Than Building Construction Contractors > Heavy Construction, Except Highway And Street > Water, Sewer, Pipeline, and Communications and Power Line Construction
1629 > Construction > Heavy Construction Other Than Building Construction Contractors > Heavy Construction, Except Highway And Street > Heavy Construction, Not Elsewhere Classified
1711 > Construction > Construction Special Trade Contractors > Plumbing, Heating And Air-conditioning > Plumbing, Heating and Air-Conditioning
1721 > Construction > Construction Special Trade Contractors > Painting And Paper Hanging > Painting and Paper Hanging
1731 > Construction > Construction Special Trade Contractors > Electrical Work > Electrical Work
1741 > Construction > Construction Special Trade Contractors > Masonry, Stonework, Tile Setting, And Plastering > Masonry, Stone Setting, and Other Stone Work 1742 > Construction > Construction Special Trade Contractors > Masonry, Stonework, Tile Setting, And Plastering > Plastering, Drywall, Acoustical, and Insulation Work
1743 > Construction > Construction Special Trade Contractors > Masonry, Stonework, Tile Setting, And Plastering > Terrazzo, Tile, Marble, and Mosaic Work
1751 > Construction > Construction Special Trade Contractors > Carpentry And Floor Work > Carpentry Work
1752 > Construction > Construction Special Trade Contractors > Carpentry And Floor Work > Floor Laying and Other Floor Work, Not Elsewhere Classified
1761 > Construction > Construction Special Trade Contractors > Roofing, Siding, And Sheet Metal Work > Roofing, Siding, and Sheet Metal Work
1771 > Construction > Construction Special Trade Contractors > Concrete Work > Concrete Work
1781 > Construction > Construction Special Trade Contractors > Water Well Drilling > Water Well Drilling
1791 > Construction > Construction Special Trade Contractors > Miscellaneous Special Trade Contractors > Structural Steel Erection
1793 > Construction > Construction Special Trade Contractors > Miscellaneous Special Trade Contractors > Glass and Glazing Work
1794 > Construction > Construction Special Trade Contractors > Miscellaneous Special Trade Contractors > Excavation Work
1795 > Construction > Construction Special Trade Contractors > Miscellaneous Special Trade Contractors > Wrecking and Demolition Work
1796 > Construction > Construction Special Trade Contractors > Miscellaneous Special Trade Contractors > Installation or Erection of Building Equipment, Not Elsewhere
1799 > Construction > Construction Special Trade Contractors > Miscellaneous Special Trade Contractors > Special Trade Contractors, Not Elsewhere Classified
2011 > Manufacturing > Food And Kindred Products > Meat Products > Meat Packing Plants
2013 > Manufacturing > Food And Kindred Products > Meat Products > Sausages and Other Prepared Meat Products
2015 > Manufacturing > Food And Kindred Products > Meat Products > Poultry Slaughtering and Processing
2021 > Manufacturing > Food And Kindred Products > Dairy Products > Creamery Butter
2022 > Manufacturing > Food And Kindred Products > Dairy Products > Natural, Processed, and Imitation Cheese
2023 > Manufacturing > Food And Kindred Products > Dairy Products > Dry, Condensed, and Evaporated Dairy Products
2024 > Manufacturing > Food And Kindred Products > Dairy Products > Ice Cream and Frozen Desserts
2026 > Manufacturing > Food And Kindred Products > Dairy Products > Fluid Milk
2032 > Manufacturing > Food And Kindred Products > Canned, Frozen, And Preserved Fruits, Vegetables, and Food Specialties > Canned Specialties
2033 > Manufacturing > Food And Kindred Products > Canned, Frozen, And Preserved Fruits, Vegetables, and Food Specialties > Canned Fruits, Vegetables, Preserves, Jams, and Jellies
2034 > Manufacturing > Food And Kindred Products > Canned, Frozen, And Preserved Fruits, Vegetables, and Food Specialties > Dried and Dehydrated Fruits, Vegetables, and Soup Mixes
2035 > Manufacturing > Food And Kindred Products > Canned, Frozen, And Preserved Fruits, Vegetables, and Food Specialties > Pickled Fruits and Vegetables, Vegetable Sauces and Seasonings, and Salad Dressings
2037 > Manufacturing > Food And Kindred Products > Canned, Frozen, And Preserved Fruits, Vegetables, and Food Specialties > Frozen Fruits, Fruit Juices, and Vegetables
2038 > Manufacturing > Food And Kindred Products > Canned, Frozen, And Preserved Fruits, Vegetables, and Food Specialties > Frozen Specialties, Not Elsewhere Classified
2041 > Manufacturing > Food And Kindred Products > Grain Mill Products > Flour and Other Grain Mill Products
2043 > Manufacturing > Food And Kindred Products > Grain Mill Products > Cereal Breakfast Foods
2044 > Manufacturing > Food And Kindred Products > Grain Mill Products > Rice Milling
2045 > Manufacturing > Food And Kindred Products > Grain Mill Products > Prepared Flour Mixes and Doughs
2046 > Manufacturing > Food And Kindred Products > Grain Mill Products > Wet Corn Milling
2047 > Manufacturing > Food And Kindred Products > Grain Mill Products > Dog and Cat Food
2048 > Manufacturing > Food And Kindred Products > Grain Mill Products > Prepared Feed and Feed Ingredients for Animals and Fowls, Except Dogs and Cats
2051 > Manufacturing > Food And Kindred Products > Bakery Products > Bread and Other Bakery Products, Except Cookies and Crackers
2052 > Manufacturing > Food And Kindred Products > Bakery Products > Cookies and Crackers
2053 > Manufacturing > Food And Kindred Products > Bakery Products > Frozen Bakery Products, Except Bread
2061 > Manufacturing > Food And Kindred Products > Sugar And Confectionery Products > Cane Sugar, Except Refining

```
2062 > Manufacturing > Food And Kindred Products > Sugar And Confectionery Products > Cane Sugar Refining
2063 > Manufacturing > Food And Kindred Products > Sugar And Confectionery Products > Beet Sugar
2064 > Manufacturing > Food And Kindred Products > Sugar And Confectionery Products > Candy and Other Confectionery Products
2066 > Manufacturing > Food And Kindred Products > Sugar And Confectionery Products > Chocolate and Cocoa Products
2067 > Manufacturing > Food And Kindred Products > Sugar And Confectionery Products > Chewing Gum
2068 > Manufacturing > Food And Kindred Products > Sugar And Confectionery Products > Salted and Roasted Nuts and Seeds
2074 > Manufacturing > Food And Kindred Products > Fats And Oils > Cottonseed Oil Mills
2075 > Manufacturing > Food And Kindred Products > Fats And Oils > Soybean Oil Mills
2076 > Manufacturing > Food And Kindred Products > Fats And Oils > Vegetable Oil Mills, Except Corn, Cottonseed, and Soybean
2077 > Manufacturing > Food And Kindred Products > Fats And Oils > Animal and Marine Fats and Oils
2079 > Manufacturing > Food And Kindred Products > Fats And Oils > Shortening, Table Oils, Margarine, and Other Edible Fats and Oils, Not Elsewhere Classified
2082 > Manufacturing > Food And Kindred Products > Beverages > Malt Beverages
2083 > Manufacturing > Food And Kindred Products > Beverages > Malt
2084 > Manufacturing > Food And Kindred Products > Beverages > Wines, Brandy, and Brandy Spirits
2085 > Manufacturing > Food And Kindred Products > Beverages > Distilled and Blended Liquors
2086 > Manufacturing > Food And Kindred Products > Beverages > Bottled and Canned Soft Drinks and Carbonated Waters
2087 > Manufacturing > Food And Kindred Products > Beverages > Flavoring Extracts and Flavoring Syrups, Not Elsewhere Classified
2091 > Manufacturing > Food And Kindred Products > Miscellaneous Food Preparations And Kindred > Canned and Cured Fish and Seafoods
2092 > Manufacturing > Food And Kindred Products > Miscellaneous Food Preparations And Kindred > Prepared Fresh or Frozen Fish and Seafoods
2095 > Manufacturing > Food And Kindred Products > Miscellaneous Food Preparations And Kindred > Roasted Coffee
2096 > Manufacturing > Food And Kindred Products > Miscellaneous Food Preparations And Kindred > Potato Chips, Corn Chips, and Similar Snacks
2097 > Manufacturing > Food And Kindred Products > Miscellaneous Food Preparations And Kindred > Manufactured Ice
2098 > Manufacturing > Food And Kindred Products > Miscellaneous Food Preparations And Kindred > Macaroni, Spaghetti, Vermicelli, and Noodles
2099 > Manufacturing > Food And Kindred Products > Miscellaneous Food Preparations And Kindred > Food Preparations, Not Elsewhere Classified
2111 > Manufacturing > Tobacco Products > Cigarettes > Cigarettes
2121 > Manufacturing > Tobacco Products > Cigars > Cigars
2131 > Manufacturing > Tobacco Products > Chewing And Smoking Tobacco And Snuff > Chewing and Smoking Tobacco and Snuff
2141 > Manufacturing > Tobacco Products > Tobacco Stemming And Redrying > Tobacco Stemming and Redrying
2211 > Manufacturing > Textile Mill Products > Broadwoven Fabric Mills, Cotton > Broadwoven Fabric Mills, Cotton
2221 > Manufacturing > Textile Mill Products > Broadwoven Fabric Mills, Manmade Fiber And Silk > Broadwoven Fabric Mills, Manmade Fiber and Silk
2231 > Manufacturing > Textile Mill Products > Broadwoven Fabric Mills, Wool (including Dyeing and Finishing) > Broadwoven Fabric Mills, Wool (Including Dyeing and Finishing)
2241 > Manufacturing > Textile Mill Products > Narrow Fabric And Other Smallwares Mills, Cotton, Wool, Silk, and Manmade Fiber > Narrow Fabric and Other Smallware Mills: Cotton, Wool, Silk, and Manmade Fiber
2251 > Manufacturing > Textile Mill Products > Knitting Mills > Women's Full-Length and Knee-Length Hosiery, Except Socks
2252 > Manufacturing > Textile Mill Products > Knitting Mills > Hosiery, Not Elsewhere Classified
2253 > Manufacturing > Textile Mill Products > Knitting Mills > Knit Outerwear Mills
2254 > Manufacturing > Textile Mill Products > Knitting Mills > Knit Underwear and Nightwear Mills
2257 > Manufacturing > Textile Mill Products > Knitting Mills > Weft Knit Fabric Mills
2258 > Manufacturing > Textile Mill Products > Knitting Mills > Lace and Warp Knit Fabric Mills
2259 > Manufacturing > Textile Mill Products > Knitting Mills > Knitting Mills, Not Elsewhere Classified
2261 > Manufacturing > Textile Mill Products > Dyeing And Finishing Textiles, Except Wool Fabrics > Finishers of Broadwoven Fabrics of Cotton
2262 > Manufacturing > Textile Mill Products > Dyeing And Finishing Textiles, Except Wool Fabrics > Finishers of Broadwoven Fabrics of Manmade Fiber and Silk
```

2269 > Manufacturing > Textile Mill Products > Dyeing And Finishing Textiles, Except Wool Fabrics > Finishers of Textiles, Not elsewhere Classified
2273 > Manufacturing > Textile Mill Products > Carpets And Rugs > Carpets and Rugs
2281 > Manufacturing > Textile Mill Products > Yarn And Thread Mills > Yarn Spinning Mills
2282 > Manufacturing > Textile Mill Products > Yarn And Thread Mills > Yarn Texturizing, Throwing, Twisting, and Winding Mills
2284 > Manufacturing > Textile Mill Products > Yarn And Thread Mills > Thread Mills
2295 > Manufacturing > Textile Mill Products > Miscellaneous Textile Goods > Coated Fabrics, Not Rubberized
2296 > Manufacturing > Textile Mill Products > Miscellaneous Textile Goods > Tire Cord and Fabrics
2297 > Manufacturing > Textile Mill Products > Miscellaneous Textile Goods > Non-woven Fabrics
2298 > Manufacturing > Textile Mill Products > Miscellaneous Textile Goods > Cordage and Twine
2299 > Manufacturing > Textile Mill Products > Miscellaneous Textile Goods > Textile goods, Not Elsewhere Classified
2311 > Manufacturing > Apparel And Other Finished Products Made From Fabrics And Similar Materials > Men's And Boys' Suits, Coats, And Overcoats > Men's and Boys' Suits, Coats, and Overcoats
2321 > Manufacturing > Apparel And Other Finished Products Made From Fabrics And Similar Materials > Men's And Boys' Furnishings, Work Clothing, And Allied Garments > Men's and Boys' Shirts, Except Work Shirts
2322 > Manufacturing > Apparel And Other Finished Products Made From Fabrics And Similar Materials > Men's And Boys' Furnishings, Work Clothing, And Allied Garments > Men's and Boys' Underwear and Nightwear
2323 > Manufacturing > Apparel And Other Finished Products Made From Fabrics And Similar Materials > Men's And Boys' Furnishings, Work Clothing, And Allied Garments > Men's and Boys' Neckwear
2325 > Manufacturing > Apparel And Other Finished Products Made From Fabrics And Similar Materials > Men's And Boys' Furnishings, Work Clothing, And Allied Garments > Men's and Boys' Separate Trousers and Slacks
2326 > Manufacturing > Apparel And Other Finished Products Made From Fabrics And Similar Materials > Men's And Boys' Furnishings, Work Clothing, And Allied Garments > Men's and Boys' Work Clothing
2329 > Manufacturing > Apparel And Other Finished Products Made From Fabrics And Similar Materials > Men's And Boys' Furnishings, Work Clothing, And Allied Garments > Men's and Boys' Clothing, Not Elsewhere Classified
2331 > Manufacturing > Apparel And Other Finished Products Made From Fabrics And Similar Materials > Women's, Misses', And Juniors' Outerwear > Women's, Misses', and Juniors' Blouses and Shirts
2335 > Manufacturing > Apparel And Other Finished Products Made From Fabrics And Similar Materials > Women's, Misses', And Juniors' Outerwear > Women's, Misses', and Juniors' Dresses
2337 > Manufacturing > Apparel And Other Finished Products Made From Fabrics And Similar Materials > Women's, Misses', And Juniors' Outerwear > Women's, Misses', and Juniors' Suits, Skirts, and Coats
2339 > Manufacturing > Apparel And Other Finished Products Made From Fabrics And Similar Materials > Women's, Misses', And Juniors' Outerwear > Women's, Misses', and Juniors' Outerwear, Not Elsewhere Classified
2341 > Manufacturing > Apparel And Other Finished Products Made From Fabrics And Similar Materials > Women's, Misses', Children's, And Infants' > Women's, Misses', Children's, and Infants' Underwear and Nightwear
2342 > Manufacturing > Apparel And Other Finished Products Made From Fabrics And Similar Materials > Women's, Misses', Children's, And Infants' > Brassieres, Girdles, and Allied Garments
2353 > Manufacturing > Apparel And Other Finished Products Made From Fabrics And Similar Materials > Hats, Caps, And Millinery > Hats, Caps, and Millinery
2361 > Manufacturing > Apparel And Other Finished Products Made From Fabrics And Similar Materials > Girls', Children's, And Infants' Outerwear > Girls', Children's, and Infants' Dresses, Blouses, and Shirts
2369 > Manufacturing > Apparel And Other Finished Products Made From Fabrics And Similar Materials > Girls', Children's, And Infants' Outerwear > Girls', Children's, and Infants' Outerwear, Not Elsewhere Classified
2371 > Manufacturing > Apparel And Other Finished Products Made From Fabrics And Similar Materials > Fur Goods > Fur Goods
2381 > Manufacturing > Apparel And Other Finished Products Made From Fabrics And Similar Materials > Miscellaneous Apparel And Accessories > Dress and Work Gloves, Except Knit and All-Leather
2384 > Manufacturing > Apparel And Other Finished Products Made From Fabrics And Similar Materials > Miscellaneous Apparel And Accessories > Robes and Dressing Gowns
2385 > Manufacturing > Apparel And Other Finished Products Made From Fabrics And Similar Materials > Miscellaneous Apparel And Accessories > Waterproof Outerwear 2386 > Manufacturing > Apparel And Other Finished Products Made From Fabrics And Similar Materials > Miscellaneous Apparel And Accessories > Leather and Sheep-Lined Clothing
2387 > Manufacturing > Apparel And Other Finished Products Made From Fabrics And Similar Materials > Miscellaneous Apparel And Accessories > Apparel belts
2389 > Manufacturing > Apparel And Other Finished Products Made From Fabrics And Similar Materials > Miscellaneous Apparel And Accessories > Apparel and Accessories, Not Elsewhere Classified
2391 > Manufacturing > Apparel And Other Finished Products Made From Fabrics And Similar Materials > Miscellaneous Fabricated Textile Products > Curtains and Draperies
2392 > Manufacturing > Apparel And Other Finished Products Made From Fabrics And Similar Materials > Miscellaneous Fabricated Textile Products > House furnishing, Except Curtains and Draperies
2393 > Manufacturing > Apparel And Other Finished Products Made From Fabrics And Similar Materials > Miscellaneous Fabricated Textile Products > Textile Bags
2394 > Manufacturing > Apparel And Other Finished Products Made From Fabrics And Similar Materials > Miscellaneous Fabricated Textile Products > Canvas and Related Products
2395 > Manufacturing > Apparel And Other Finished Products Made From Fabrics And Similar Materials > Miscellaneous Fabricated Textile Products > Pleating, Decorative and Novelty Stitching, and Tucking for the Trade
2396 > Manufacturing > Apparel And Other Finished Products Made From Fabrics And Similar Materials > Miscellaneous Fabricated Textile Products > Automotive Trimmings, Apparel Findings, and Related Products
2397 > Manufacturing > Apparel And Other Finished Products Made From Fabrics And Similar Materials > Miscellaneous Fabricated Textile Products > Schiffli Machine Embroideries
2399 > Manufacturing > Apparel And Other Finished Products Made From Fabrics And Similar Materials > Miscellaneous Fabricated Textile Products > Fabricated Textile Products, Not Elsewhere Classified
2411 > Manufacturing > Lumber And Wood Products, Except Furniture > Logging > Logging
2421 > Manufacturing > Lumber And Wood Products, Except Furniture > Sawmills And Planing Mills > Sawmills and Planing Mills, General
2426 > Manufacturing > Lumber And Wood Products, Except Furniture > Sawmills And Planing Mills > Hardwood Dimension and Flooring Mills
2429 > Manufacturing > Lumber And Wood Products, Except Furniture > Sawmills And Planing Mills > Special Product Sawmills, Not Elsewhere Classified
2431 > Manufacturing > Lumber And Wood Products, Except Furniture > Millwork, Veneer, Plywood, And Structural Wood > Millwork
2434 > Manufacturing > Lumber And Wood Products, Except Furniture > Millwork, Veneer, Plywood, And Structural Wood > Wood Kitchen Cabinets
2435 > Manufacturing > Lumber And Wood Products, Except Furniture > Millwork, Veneer, Plywood, And Structural Wood > Hardwood Veneer and Plywood
2436 > Manufacturing > Lumber And Wood Products, Except Furniture > Millwork, Veneer, Plywood, And Structural Wood > Softwood Veneer and Plywood
2439 > Manufacturing > Lumber And Wood Products, Except Furniture > Millwork, Veneer, Plywood, And Structural Wood > Structural Wood Members, Not Elsewhere Classified
2441 > Manufacturing > Lumber And Wood Products, Except Furniture > Wood Containers > Nailed and Lock Corner Wood Boxes and Shook
2448 > Manufacturing > Lumber And Wood Products, Except Furniture > Wood Containers > Wood Pallets and Skids
2449 > Manufacturing > Lumber And Wood Products, Except Furniture > Wood Containers > Wood Containers, Not Elsewhere Classified
2451 > Manufacturing > Lumber And Wood Products, Except Furniture > Wood Buildings And Mobile Homes > Mobile Homes
2452 > Manufacturing > Lumber And Wood Products, Except Furniture > Wood Buildings And Mobile Homes > Prefabricated Wood Buildings and Components
2491 > Manufacturing > Lumber And Wood Products, Except Furniture > Miscellaneous Wood Products > Wood Preserving
2493 > Manufacturing > Lumber And Wood Products, Except Furniture > Miscellaneous Wood Products > Reconstituted Wood Products
2499 > Manufacturing > Lumber And Wood Products, Except Furniture > Miscellaneous Wood Products > Wood Products, Not Elsewhere Classified
2511 > Manufacturing > Furniture And Fixtures > Household Furniture > Wood Household Furniture, Except Upholstered
2512 > Manufacturing > Furniture And Fixtures > Household Furniture > Wood Household Furniture, Upholstered
2514 > Manufacturing > Furniture And Fixtures > Household Furniture > Metal Household Furniture
2515 > Manufacturing > Furniture And Fixtures > Household Furniture > Mattresses, Foundations, and Convertible Beds
2517 > Manufacturing > Furniture And Fixtures > Household Furniture > Wood Television, Radio, Phonograph, and Sewing Machine Cabinets

2519 > Manufacturing > Furniture And Fixtures > Household Furniture > Household Furniture, Not Elsewhere Classified
2521 > Manufacturing > Furniture And Fixtures > Office Furniture > Wood Office Furniture
2522 > Manufacturing > Furniture And Fixtures > Office Furniture > Office Furniture, Except Wood
2531 > Manufacturing > Furniture And Fixtures > Public Building And Related Furniture > Public Building and Related Furniture
2541 > Manufacturing > Furniture And Fixtures > Partitions, Shelving, Lockers, And Office And > Wood Office and Store Fixtures, Partitions, Shelving, and Lockers
2542 > Manufacturing > Furniture And Fixtures > Partitions, Shelving, Lockers, And Office And > Office and Store Fixtures, Partitions, Shelving, and Lockers, Except Wood
2591 > Manufacturing > Furniture And Fixtures > Miscellaneous Furniture And Fixtures > Drapery Hardware and Window Blinds and Shades
2599 > Manufacturing > Furniture And Fixtures > Miscellaneous Furniture And Fixtures > Furniture and Fixtures, Not Elsewhere Classified
2611 > Manufacturing > Paper And Allied Products > Pulp Mills > Pulp Mills
2621 > Manufacturing > Paper And Allied Products > Paper Mills > Paper Mills
2631 > Manufacturing > Paper And Allied Products > Paperboard Mills > Paperboard Mills
2652 > Manufacturing > Paper And Allied Products > Paperboard Containers And Boxes > Setup Paperboard Boxes
2653 > Manufacturing > Paper And Allied Products > Paperboard Containers And Boxes > Corrugated and Solid Fiber Boxes
2655 > Manufacturing > Paper And Allied Products > Paperboard Containers And Boxes > Fiber Cans, Tubes, Drums, and Similar Products
2656 > Manufacturing > Paper And Allied Products > Paperboard Containers And Boxes > Sanitary Food Containers, Except Folding
2657 > Manufacturing > Paper And Allied Products > Paperboard Containers And Boxes > Folding Paperboard Boxes, Including Sanitary
2671 > Manufacturing > Paper And Allied Products > Converted Paper And Paperboard Products, Except > Packaging Paper and Plastics Film, Coated and Laminated
2672 > Manufacturing > Paper And Allied Products > Converted Paper And Paperboard Products, Except > Coated and Laminated Paper, Not Elsewhere Classified
2673 > Manufacturing > Paper And Allied Products > Converted Paper And Paperboard Products, Except > Plastics, Foil, and Coated Paper Bags
2674 > Manufacturing > Paper And Allied Products > Converted Paper And Paperboard Products, Except > Uncoated Paper and Multiwall Bags
2675 > Manufacturing > Paper And Allied Products > Converted Paper And Paperboard Products, Except > Die-Cut Paper and Paperboard and Cardboard
2676 > Manufacturing > Paper And Allied Products > Converted Paper And Paperboard Products, Except > Sanitary Paper Products
2677 > Manufacturing > Paper And Allied Products > Converted Paper And Paperboard Products, Except > Envelopes
2678 > Manufacturing > Paper And Allied Products > Converted Paper And Paperboard Products, Except > Stationery, Tablets, and Related Products
2679 > Manufacturing > Paper And Allied Products > Converted Paper And Paperboard Products, Except > Converted Paper and Paperboard Products, Not Elsewhere Classified
2711 > Manufacturing > Printing, Publishing, And Allied Industries > Newspapers: Publishing, Or Publishing And Printing > Newspapers: Publishing, or Publishing and Printing
2721 > Manufacturing > Printing, Publishing, And Allied Industries > Periodicals: Publishing, Or Publishing And Printing > Periodicals: Publishing, or Publishing and Printing
2731 > Manufacturing > Printing, Publishing, And Allied Industries > Books > Books: Publishing, or Publishing and Printing
2732 > Manufacturing > Printing, Publishing, And Allied Industries > Books > Book Printing
2741 > Manufacturing > Printing, Publishing, And Allied Industries > Miscellaneous Publishing > Miscellaneous Publishing
2752 > Manufacturing > Printing, Publishing, And Allied Industries > Commercial Printing > Commercial Printing, Lithographic
2754 > Manufacturing > Printing, Publishing, And Allied Industries > Commercial Printing > Commercial Printing, Gravure
2759 > Manufacturing > Printing, Publishing, And Allied Industries > Commercial Printing > Commercial Printing, Not Elsewhere Classified
2761 > Manufacturing > Printing, Publishing, And Allied Industries > Manifold Business Forms > Manifold Business Forms
2771 > Manufacturing > Printing, Publishing, And Allied Industries > Greeting Cards > Greeting Cards
2782 > Manufacturing > Printing, Publishing, And Allied Industries > Blankbooks, Looseleaf Binders, And Bookbinding > Blankbooks, Looseleaf Binders and Devices
2789 > Manufacturing > Printing, Publishing, And Allied Industries > Blankbooks, Looseleaf Binders, And Bookbinding > Bookbinding and Related Work

```
2791 > Manufacturing > Printing, Publishing, And Allied Industries > Service Industries For The Printing Trade > Typesetting
2796 > Manufacturing > Printing, Publishing, And Allied Industries > Service Industries For The Printing Trade > Platemaking and
Related Services
2812 > Manufacturing > Chemicals And Allied Products > Industrial Inorganic Chemicals > Alkalies and Chlorine
2813 > Manufacturing > Chemicals And Allied Products > Industrial Inorganic Chemicals > Industrial Gases
2816 > Manufacturing > Chemicals And Allied Products > Industrial Inorganic Chemicals > Inorganic Pigments
2819 > Manufacturing > Chemicals And Allied Products > Industrial Inorganic Chemicals > Industrial Inorganic Chemicals, Not Elsewhere
Classified
2821 > Manufacturing > Chemicals And Allied Products > Plastics Materials And Synthetic Resins, Synthetic > Plastics Materials,
Synthetic Resins, and Nonvulcanizable Elastomers
2822 > Manufacturing > Chemicals And Allied Products > Plastics Materials And Synthetic Resins, Synthetic > Synthetic Rubber
(Vulcanizable Elastomers)
2823 > Manufacturing > Chemicals And Allied Products > Plastics Materials And Synthetic Resins, Synthetic > Cellulosic Manmade Fibers
2824 > Manufacturing > Chemicals And Allied Products > Plastics Materials And Synthetic Resins, Synthetic > Manmade Organic Fibers,
Except Cellulosic
2833 > Manufacturing > Chemicals And Allied Products > Drugs > Medicinal Chemicals and Botanical Products
2834 > Manufacturing > Chemicals And Allied Products > Drugs > Pharmaceutical Preparations
2835 > Manufacturing > Chemicals And Allied Products > Drugs > In Vitro and In Vivo Diagnostic Substances
2836 > Manufacturing > Chemicals And Allied Products > Drugs > Biological Products, Except Diagnostic Substances
2841 > Manufacturing > Chemicals And Allied Products > Soap, Detergents, And Cleaning Preparations; Perfumes, Cosmetics, and Other
Toilet Preparations > Soap and Other Detergents, Except Specialty Cleaners
2842 > Manufacturing > Chemicals And Allied Products > Soap, Detergents, And Cleaning Preparations; Perfumes, Cosmetics, and Other
Toilet Preparations > Specialty Cleaning, Polishing, and Sanitation Preparations
2843 > Manufacturing > Chemicals And Allied Products > Soap, Detergents, And Cleaning Preparations; Perfumes, Cosmetics, and Other
Toilet Preparations > Surface Active Agents, Finishing Agents, Sulfonated Oils, and Assistants
2844 > Manufacturing > Chemicals And Allied Products > Soap, Detergents, And Cleaning Preparations; Perfumes, Cosmetics, and Other
Toilet Preparations > Perfumes, Cosmetics, and Other Toilet Preparations
2851 > Manufacturing > Chemicals And Allied Products > Paints, Varnishes, Lacquers, Enamels, And Allied > Paints, Varnishes, Lacquers,
Enamels, and Allied Products
2861 > Manufacturing > Chemicals And Allied Products > Industrial Organic Chemicals > Gum and Wood Chemicals
2865 > Manufacturing > Chemicals And Allied Products > Industrial Organic Chemicals > Cyclic Organic Crudes and Intermediates, and
Organic Dyes and Pigments
2869 > Manufacturing > Chemicals And Allied Products > Industrial Organic Chemicals > Industrial Organic Chemicals, Not Elsewhere
Classified
2873 > Manufacturing > Chemicals And Allied Products > Agricultural Chemicals > Nitrogenous Fertilizers
2874 > Manufacturing > Chemicals And Allied Products > Agricultural Chemicals > Phosphatic Fertilizers
2875 > Manufacturing > Chemicals And Allied Products > Agricultural Chemicals > Fertilizers, Mixing Only
2879 > Manufacturing > Chemicals And Allied Products > Agricultural Chemicals > Pesticides and Agricultural Chemicals, Not Elsewhere
Classified
2891 > Manufacturing > Chemicals And Allied Products > Miscellaneous Chemical Products > Adhesives and Sealants
2892 > Manufacturing > Chemicals And Allied Products > Miscellaneous Chemical Products > Explosives
2893 > Manufacturing > Chemicals And Allied Products > Miscellaneous Chemical Products > Printing Ink
2895 > Manufacturing > Chemicals And Allied Products > Miscellaneous Chemical Products > Carbon Black
2899 > Manufacturing > Chemicals And Allied Products > Miscellaneous Chemical Products > Chemicals and Chemical Preparations, Not
Elsewhere Classified
2911 > Manufacturing > Petroleum Refining And Related Industries > Petroleum Refining > Petroleum Refining
2951 > Manufacturing > Petroleum Refining And Related Industries > Asphalt Paving And Roofing Materials > Asphalt Paving Mixtures and
Blocks
2952 > Manufacturing > Petroleum Refining And Related Industries > Asphalt Paving And Roofing Materials > Asphalt Felts and Coatings
2992 > Manufacturing > Petroleum Refining And Related Industries > Miscellaneous Products Of Petroleum And Coal > Lubricating Oils and
Greases
```

2999 > Manufacturing > Petroleum Refining And Related Industries > Miscellaneous Products Of Petroleum And Coal > Products of Petroleum and Coal, Not Elsewhere Classified
3011 > Manufacturing > Rubber And Miscellaneous Plastics Products > Tires And Inner Tubes > Tires and Inner Tubes
3021 > Manufacturing > Rubber And Miscellaneous Plastics Products > Rubber And Plastics Footwear > Rubber and Plastics Footwear
3052 > Manufacturing > Rubber And Miscellaneous Plastics Products > Gaskets, Packing, And Sealing Devices And Rubber > Rubber and Plastics Hose and Belting
3053 > Manufacturing > Rubber And Miscellaneous Plastics Products > Gaskets, Packing, And Sealing Devices And Rubber > Gaskets, Packing, and Sealing Devices
3061 > Manufacturing > Rubber And Miscellaneous Plastics Products > Fabricated Rubber Products, Not Elsewhere > Molded, Extruded, and Lathe-Cut Mechanical Rubber Goods
3069 > Manufacturing > Rubber And Miscellaneous Plastics Products > Fabricated Rubber Products, Not Elsewhere > Fabricated Rubber Products, Not Elsewhere Classified
3081 > Manufacturing > Rubber And Miscellaneous Plastics Products > Miscellaneous Plastics Products > Unsupported Plastics Film and Sheet
3082 > Manufacturing > Rubber And Miscellaneous Plastics Products > Miscellaneous Plastics Products > Unsupported Plastics Profile Shapes
3083 > Manufacturing > Rubber And Miscellaneous Plastics Products > Miscellaneous Plastics Products > Laminated Plastics Plate, Sheet, and Profile Shapes
3084 > Manufacturing > Rubber And Miscellaneous Plastics Products > Miscellaneous Plastics Products > Plastics Pipe
3085 > Manufacturing > Rubber And Miscellaneous Plastics Products > Miscellaneous Plastics Products > Plastics Bottles
3086 > Manufacturing > Rubber And Miscellaneous Plastics Products > Miscellaneous Plastics Products > Plastics Foam Products
3087 > Manufacturing > Rubber And Miscellaneous Plastics Products > Miscellaneous Plastics Products > Custom Compounding of Purchased Plastics Resins
3088 > Manufacturing > Rubber And Miscellaneous Plastics Products > Miscellaneous Plastics Products > Plastics Plumbing Fixtures
3089 > Manufacturing > Rubber And Miscellaneous Plastics Products > Miscellaneous Plastics Products > Plastics Products, Not Elsewhere Classified
3111 > Manufacturing > Leather And Leather Products > Leather Tanning And Finishing > Leather Tanning and Finishing
3131 > Manufacturing > Leather And Leather Products > Boot And Shoe Cut Stock And Findings > Boot and Shoe Cut Stock and Findings
3142 > Manufacturing > Leather And Leather Products > Footwear, Except Rubber > House Slippers
3143 > Manufacturing > Leather And Leather Products > Footwear, Except Rubber > Men's Footwear, Except Athletic
3144 > Manufacturing > Leather And Leather Products > Footwear, Except Rubber > Women's Footwear, Except Athletic
3149 > Manufacturing > Leather And Leather Products > Footwear, Except Rubber > Footwear, Except Rubber, Not Elsewhere Classified
3151 > Manufacturing > Leather And Leather Products > Leather Gloves And Mittens > Leather Gloves and Mittens
3161 > Manufacturing > Leather And Leather Products > Luggage > Luggage
3171 > Manufacturing > Leather And Leather Products > Handbags And Other Personal Leather Goods > Women's Handbags and Purses
3172 > Manufacturing > Leather And Leather Products > Handbags And Other Personal Leather Goods > Personal Leather Goods, Except Women's Handbags and Purses
3199 > Manufacturing > Leather And Leather Products > Leather Goods, Not Elsewhere Classified > Leather Goods, Not Elsewhere Classified
3211 > Manufacturing > Stone, Clay, Glass, And Concrete Products > Flat Glass > Flat Glass
3221 > Manufacturing > Stone, Clay, Glass, And Concrete Products > Glass And Glassware, Pressed Or Blown > Glass Containers
3229 > Manufacturing > Stone, Clay, Glass, And Concrete Products > Glass And Glassware, Pressed Or Blown > Pressed and Blown Glass and Glassware, Not Elsewhere Classified
3231 > Manufacturing > Stone, Clay, Glass, And Concrete Products > Glass Products, Made Of Purchased Glass > Glass Products, Made of Purchased Glass
3241 > Manufacturing > Stone, Clay, Glass, And Concrete Products > Cement, Hydraulic > Cement, Hydraulic
3251 > Manufacturing > Stone, Clay, Glass, And Concrete Products > Structural Clay Products > Brick and Structural Clay Tile
3253 > Manufacturing > Stone, Clay, Glass, And Concrete Products > Structural Clay Products > Ceramic Wall and Floor Tile
3255 > Manufacturing > Stone, Clay, Glass, And Concrete Products > Structural Clay Products > Clay Refractories
3259 > Manufacturing > Stone, Clay, Glass, And Concrete Products > Structural Clay Products > Structural Clay Products, Not Elsewhere Classified 3261 > Manufacturing > Stone, Clay, Glass, And Concrete Products > Pottery And Related Products > Vitreous China Plumbing Fixtures and China and Earthenware Fittings and Bathroom Accessories
3262 > Manufacturing > Stone, Clay, Glass, And Concrete Products > Pottery And Related Products > Vitreous China Table and Kitchen Articles
3263 > Manufacturing > Stone, Clay, Glass, And Concrete Products > Pottery And Related Products > Fine Earthenware (Whiteware) Table and Kitchen Articles
3264 > Manufacturing > Stone, Clay, Glass, And Concrete Products > Pottery And Related Products > Porcelain Electrical Supplies
3269 > Manufacturing > Stone, Clay, Glass, And Concrete Products > Pottery And Related Products > Pottery Products, Not Elsewhere Classified
3271 > Manufacturing > Stone, Clay, Glass, And Concrete Products > Concrete, Gypsum, And Plaster Products > Concrete Block and Brick
3272 > Manufacturing > Stone, Clay, Glass, And Concrete Products > Concrete, Gypsum, And Plaster Products > Concrete Products, Except Block and Brick
3273 > Manufacturing > Stone, Clay, Glass, And Concrete Products > Concrete, Gypsum, And Plaster Products > Ready-Mixed Concrete
3274 > Manufacturing > Stone, Clay, Glass, And Concrete Products > Concrete, Gypsum, And Plaster Products > Lime
3275 > Manufacturing > Stone, Clay, Glass, And Concrete Products > Concrete, Gypsum, And Plaster Products > Gypsum Products
3281 > Manufacturing > Stone, Clay, Glass, And Concrete Products > Cut Stone And Stone Products > Cut Stone and Stone Products
3291 > Manufacturing > Stone, Clay, Glass, And Concrete Products > Abrasive, Asbestos, And Miscellaneous > Abrasive Products
3292 > Manufacturing > Stone, Clay, Glass, And Concrete Products > Abrasive, Asbestos, And Miscellaneous > Asbestos Products
3295 > Manufacturing > Stone, Clay, Glass, And Concrete Products > Abrasive, Asbestos, And Miscellaneous > Minerals and Earths, Ground or Otherwise Treated
3296 > Manufacturing > Stone, Clay, Glass, And Concrete Products > Abrasive, Asbestos, And Miscellaneous > Mineral Wool
3297 > Manufacturing > Stone, Clay, Glass, And Concrete Products > Abrasive, Asbestos, And Miscellaneous > Nonclay Refractories
3299 > Manufacturing > Stone, Clay, Glass, And Concrete Products > Abrasive, Asbestos, And Miscellaneous > Nonmetallic Mineral Products, Not Elsewhere Classified
3312 > Manufacturing > Primary Metal Industries > Steel Works, Blast Furnaces, And Rolling And Finishing Mills > Steel Works, Blast Furnaces (Including Coke Ovens), and Rolling Mills
3313 > Manufacturing > Primary Metal Industries > Steel Works, Blast Furnaces, And Rolling And Finishing Mills > Electrometallurgical Products, Except Steel
3315 > Manufacturing > Primary Metal Industries > Steel Works, Blast Furnaces, And Rolling And Finishing Mills > Steel Wiredrawing and Steel Nails and Spikes
3316 > Manufacturing > Primary Metal Industries > Steel Works, Blast Furnaces, And Rolling And Finishing Mills > Cold-Rolled Steel Sheet, Strip, and Bars
3317 > Manufacturing > Primary Metal Industries > Steel Works, Blast Furnaces, And Rolling And Finishing Mills > Steel Pipe and Tubes
3321 > Manufacturing > Primary Metal Industries > Iron And Steel Foundries > Gray and Ductile Iron Foundries
3322 > Manufacturing > Primary Metal Industries > Iron And Steel Foundries > Malleable Iron Foundries
3324 > Manufacturing > Primary Metal Industries > Iron And Steel Foundries > Steel Investment Foundries
3325 > Manufacturing > Primary Metal Industries > Iron And Steel Foundries > Steel Foundries, Not Elsewhere Classified
3331 > Manufacturing > Primary Metal Industries > Primary Smelting And Refining Of Nonferrous > Primary Smelting and Refining of Copper
3334 > Manufacturing > Primary Metal Industries > Primary Smelting And Refining Of Nonferrous > Primary Production of Aluminum
3339 > Manufacturing > Primary Metal Industries > Primary Smelting And Refining Of Nonferrous > Primary Smelting and Refining of Nonferrous Metals, Except Copper and Aluminum
3341 > Manufacturing > Primary Metal Industries > Secondary Smelting And Refining Of Nonferrous > Secondary Smelting and Refining of Nonferrous Metals
3351 > Manufacturing > Primary Metal Industries > Rolling, Drawing, And Extruding Of Nonferrous > Rolling, Drawing, and Extruding Of Copper
3353 > Manufacturing > Primary Metal Industries > Rolling, Drawing, And Extruding Of Nonferrous > Aluminum Sheet, Plate, and Foil
3354 > Manufacturing > Primary Metal Industries > Rolling, Drawing, And Extruding Of Nonferrous > Aluminum Extruded Products
3355 > Manufacturing > Primary Metal Industries > Rolling, Drawing, And Extruding Of Nonferrous > Aluminum Rolling and Drawing, Not Elsewhere Classified 3356 > Manufacturing > Primary Metal Industries > Rolling, Drawing, And Extruding Of Nonferrous > Rolling, Drawing, and Extruding of Nonferrous Metals, Except Copper and Aluminum
3357 > Manufacturing > Primary Metal Industries > Rolling, Drawing, And Extruding Of Nonferrous > Drawing and Insulating of Nonferrous Wire
3363 > Manufacturing > Primary Metal Industries > Nonferrous Foundries (castings) > Aluminum Die-Castings
3364 > Manufacturing > Primary Metal Industries > Nonferrous Foundries (castings) > Nonferrous Die-Castings, Except Aluminum
3365 > Manufacturing > Primary Metal Industries > Nonferrous Foundries (castings) > Aluminum Foundries
3366 > Manufacturing > Primary Metal Industries > Nonferrous Foundries (castings) > Copper Foundries
3369 > Manufacturing > Primary Metal Industries > Nonferrous Foundries (castings) > Nonferrous Foundries, Except Aluminum and Copper
3398 > Manufacturing > Primary Metal Industries > Miscellaneous Primary Metal Products > Metal Heat Treating
3399 > Manufacturing > Primary Metal Industries > Miscellaneous Primary Metal Products > Primary Metal Products, Not Elsewhere Classified
3411 > Manufacturing > Fabricated Metal Products, Except Machinery And Transportation Equipment > Metal Cans And Shipping Containers > Metal Cans
3412 > Manufacturing > Fabricated Metal Products, Except Machinery And Transportation Equipment > Metal Cans And Shipping Containers > Metal Shipping Barrels, Drums, Kegs, and Pails
3421 > Manufacturing > Fabricated Metal Products, Except Machinery And Transportation Equipment > Cutlery, Handtools, And General Hardware > Cutlery
3423 > Manufacturing > Fabricated Metal Products, Except Machinery And Transportation Equipment > Cutlery, Handtools, And General Hardware > Hand and Edge Tools, Except Machine Tools and Handsaws
3425 > Manufacturing > Fabricated Metal Products, Except Machinery And Transportation Equipment > Cutlery, Handtools, And General Hardware > Saw Blades and Handsaws
3429 > Manufacturing > Fabricated Metal Products, Except Machinery And Transportation Equipment > Cutlery, Handtools, And General Hardware > Hardware, Not Elsewhere Classified
3431 > Manufacturing > Fabricated Metal Products, Except Machinery And Transportation Equipment > Heating Equipment, Except Electric And Warm Air; > Enameled Iron and Metal Sanitary Ware
3432 > Manufacturing > Fabricated Metal Products, Except Machinery And Transportation Equipment > Heating Equipment, Except Electric And Warm Air; > Plumbing Fixture Fittings and Trim
3433 > Manufacturing > Fabricated Metal Products, Except Machinery And Transportation Equipment > Heating Equipment, Except Electric And Warm Air; > Heating Equipment, Except Electric and Warm Air Furnaces
3441 > Manufacturing > Fabricated Metal Products, Except Machinery And Transportation Equipment > Fabricated Structural Metal Products > Fabricated Structural Metal
3442 > Manufacturing > Fabricated Metal Products, Except Machinery And Transportation Equipment > Fabricated Structural Metal Products > Metal Doors, Sash, Frames, Molding, and Trim Manufacturing
3443 > Manufacturing > Fabricated Metal Products, Except Machinery And Transportation Equipment > Fabricated Structural Metal Products > Fabricated Plate Work (Boiler Shops)
3444 > Manufacturing > Fabricated Metal Products, Except Machinery And Transportation Equipment > Fabricated Structural Metal Products > Sheet Metal Work
3446 > Manufacturing > Fabricated Metal Products, Except Machinery And Transportation Equipment > Fabricated Structural Metal Products > Architectural and Ornamental Metal Work
3448 > Manufacturing > Fabricated Metal Products, Except Machinery And Transportation Equipment > Fabricated Structural Metal Products > Prefabricated Metal Buildings and Components
3449 > Manufacturing > Fabricated Metal Products, Except Machinery And Transportation Equipment > Fabricated Structural Metal Products > Miscellaneous Structural Metal Work
3451 > Manufacturing > Fabricated Metal Products, Except Machinery And Transportation Equipment > Screw Machine Products, And Bolts, Nuts, Screws, Rivets, And Washers > Screw Machine Products
3452 > Manufacturing > Fabricated Metal Products, Except Machinery And Transportation Equipment > Screw Machine Products, And Bolts, Nuts, Screws, Rivets, And Washers > Bolts, Nuts, Screws, Rivets, and Washers
3462 > Manufacturing > Fabricated Metal Products, Except Machinery And Transportation Equipment > Metal Forgings And Stampings > Iron and Steel Forgings 3463 > Manufacturing > Fabricated Metal Products, Except Machinery And Transportation Equipment > Metal Forgings And Stampings > Nonferrous Forgings
3465 > Manufacturing > Fabricated Metal Products, Except Machinery And Transportation Equipment > Metal Forgings And Stampings > Automotive Stampings
3466 > Manufacturing > Fabricated Metal Products, Except Machinery And Transportation Equipment > Metal Forgings And Stampings > Crowns and Closures
3469 > Manufacturing > Fabricated Metal Products, Except Machinery And Transportation Equipment > Metal Forgings And Stampings > Metal Stampings, Not Elsewhere Classified
3471 > Manufacturing > Fabricated Metal Products, Except Machinery And Transportation Equipment > Coating, Engraving, And Allied Services > Electroplating, Plating, Polishing, Anodizing, and Coloring
3479 > Manufacturing > Fabricated Metal Products, Except Machinery And Transportation Equipment > Coating, Engraving, And Allied Services > Coating, Engraving, and Allied Services, Not Elsewhere Classified
3482 > Manufacturing > Fabricated Metal Products, Except Machinery And Transportation Equipment > Ordnance And Accessories, Except Vehicles And Guided Missiles > Small Arms Ammunition
3483 > Manufacturing > Fabricated Metal Products, Except Machinery And Transportation Equipment > Ordnance And Accessories, Except Vehicles And Guided Missiles > Ammunition, Except for Small Arms
3484 > Manufacturing > Fabricated Metal Products, Except Machinery And Transportation Equipment > Ordnance And Accessories, Except Vehicles And Guided Missiles > Small Arms
3489 > Manufacturing > Fabricated Metal Products, Except Machinery And Transportation Equipment > Ordnance And Accessories, Except Vehicles And Guided Missiles > Ordnance and Accessories, Not Elsewhere Classified
3491 > Manufacturing > Fabricated Metal Products, Except Machinery And Transportation Equipment > Miscellaneous Fabricated Metal Products > Industrial Valves
3492 > Manufacturing > Fabricated Metal Products, Except Machinery And Transportation Equipment > Miscellaneous Fabricated Metal Products > Fluid Power Valves and Hose Fittings
3493 > Manufacturing > Fabricated Metal Products, Except Machinery And Transportation Equipment > Miscellaneous Fabricated Metal Products > Steel Springs, Except Wire
3494 > Manufacturing > Fabricated Metal Products, Except Machinery And Transportation Equipment > Miscellaneous Fabricated Metal Products > Valves and Pipe Fittings, Not Elsewhere Classified
3495 > Manufacturing > Fabricated Metal Products, Except Machinery And Transportation Equipment > Miscellaneous Fabricated Metal Products > Wire Springs
3496 > Manufacturing > Fabricated Metal Products, Except Machinery And Transportation Equipment > Miscellaneous Fabricated Metal Products > Miscellaneous Fabricated Wire Products
3497 > Manufacturing > Fabricated Metal Products, Except Machinery And Transportation Equipment > Miscellaneous Fabricated Metal Products > Metal Foil and Leaf
3498 > Manufacturing > Fabricated Metal Products, Except Machinery And Transportation Equipment > Miscellaneous Fabricated Metal Products > Fabricated Pipe and Pipe Fittings
3499 > Manufacturing > Fabricated Metal Products, Except Machinery And Transportation Equipment > Miscellaneous Fabricated Metal Products > Fabricated Metal Products, Not Elsewhere Classified
3511 > Manufacturing > Industrial And Commercial Machinery And Computer Equipment > Engines And Turbines > Steam, Gas, and Hydraulic Turbines, and Turbine Generator Set Units
3519 > Manufacturing > Industrial And Commercial Machinery And Computer Equipment > Engines And Turbines > Internal Combustion Engines, Not Elsewhere Classified
3523 > Manufacturing > Industrial And Commercial Machinery And Computer Equipment > Farm And Garden Machinery And Equipment > Farm Machinery and Equipment
3524 > Manufacturing > Industrial And Commercial Machinery And Computer Equipment > Farm And Garden Machinery And Equipment > Lawn and Garden Tractors and Home Lawn and Garden Equipment
3531 > Manufacturing > Industrial And Commercial Machinery And Computer Equipment > Construction, Mining, And Materials Handling > Construction Machinery and Equipment
3532 > Manufacturing > Industrial And Commercial Machinery And Computer Equipment > Construction, Mining, And Materials Handling > Mining Machinery and Equipment, Except Oil and Gas Field Machinery and Equipment 3533 > Manufacturing > Industrial And Commercial Machinery And Computer Equipment > Construction, Mining, And Materials Handling > Oil and Gas Field Machinery and Equipment
3534 > Manufacturing > Industrial And Commercial Machinery And Computer Equipment > Construction, Mining, And Materials Handling > Elevators and Moving Stairways
3535 > Manufacturing > Industrial And Commercial Machinery And Computer Equipment > Construction, Mining, And Materials Handling > Conveyors and Conveying Equipment
3536 > Manufacturing > Industrial And Commercial Machinery And Computer Equipment > Construction, Mining, And Materials Handling > Overhead Traveling Cranes, Hoists, and Monorail Systems
3537 > Manufacturing > Industrial And Commercial Machinery And Computer Equipment > Construction, Mining, And Materials Handling > Industrial Trucks, Tractors, Trailers, and Stackers
3541 > Manufacturing > Industrial And Commercial Machinery And Computer Equipment > Metalworking Machinery And Equipment > Machine Tools, Metal Cutting Types
3542 > Manufacturing > Industrial And Commercial Machinery And Computer Equipment > Metalworking Machinery And Equipment > Machine Tools, Metal Forming Types
3543 > Manufacturing > Industrial And Commercial Machinery And Computer Equipment > Metalworking Machinery And Equipment > Industrial Patterns
3544 > Manufacturing > Industrial And Commercial Machinery And Computer Equipment > Metalworking Machinery And Equipment > Special Dies and Tools, Die Sets, Jigs and Fixtures, and Industrial Molds
3545 > Manufacturing > Industrial And Commercial Machinery And Computer Equipment > Metalworking Machinery And Equipment > Cutting Tools, Machine Tool Accessories, and Machinists' Precision Measuring Devices
3546 > Manufacturing > Industrial And Commercial Machinery And Computer Equipment > Metalworking Machinery And Equipment > Power-Driven Handtools
3547 > Manufacturing > Industrial And Commercial Machinery And Computer Equipment > Metalworking Machinery And Equipment > Rolling Mill Machinery and Equipment
3548 > Manufacturing > Industrial And Commercial Machinery And Computer Equipment > Metalworking Machinery And Equipment > Electric and Gas Welding and Soldering Equipment
3549 > Manufacturing > Industrial And Commercial Machinery And Computer Equipment > Metalworking Machinery And Equipment > Metalworking Machinery, Not Elsewhere Classified
3552 > Manufacturing > Industrial And Commercial Machinery And Computer Equipment > Special Industry Machinery, Except Metalworking > Textile Machinery
3553 > Manufacturing > Industrial And Commercial Machinery And Computer Equipment > Special Industry Machinery, Except Metalworking > Woodworking Machinery
3554 > Manufacturing > Industrial And Commercial Machinery And Computer Equipment > Special Industry Machinery, Except Metalworking > Paper Industries Machinery
3555 > Manufacturing > Industrial And Commercial Machinery And Computer Equipment > Special Industry Machinery, Except Metalworking > Printing Trades Machinery and Equipment
3556 > Manufacturing > Industrial And Commercial Machinery And Computer Equipment > Special Industry Machinery, Except Metalworking > Food Products Machinery
3559 > Manufacturing > Industrial And Commercial Machinery And Computer Equipment > Special Industry Machinery, Except Metalworking > Special Industry Machinery, Not Elsewhere Classified
3561 > Manufacturing > Industrial And Commercial Machinery And Computer Equipment > General Industrial Machinery And Equipment > Pumps and Pumping Equipment
3562 > Manufacturing > Industrial And Commercial Machinery And Computer Equipment > General Industrial Machinery And Equipment > Ball and Roller Bearings
3563 > Manufacturing > Industrial And Commercial Machinery And Computer Equipment > General Industrial Machinery And Equipment > Air and Gas Compressors
3564 > Manufacturing > Industrial And Commercial Machinery And Computer Equipment > General Industrial Machinery And Equipment > Industrial and Commercial Fans and Blowers and Air Purification Equipment
3565 > Manufacturing > Industrial And Commercial Machinery And Computer Equipment > General Industrial Machinery And Equipment > Packaging Machinery 3566 > Manufacturing > Industrial And Commercial Machinery And Computer Equipment > General Industrial Machinery And Equipment > Speed Changers, Industrial High-Speed Drives, and Gears
3567 > Manufacturing > Industrial And Commercial Machinery And Computer Equipment > General Industrial Machinery And Equipment > Industrial Process Furnaces and Ovens
3568 > Manufacturing > Industrial And Commercial Machinery And Computer Equipment > General Industrial Machinery And Equipment > Mechanical Power Transmission Equipment, Not Elsewhere Classified
3569 > Manufacturing > Industrial And Commercial Machinery And Computer Equipment > General Industrial Machinery And Equipment > General Industrial Machinery and Equipment, Not Elsewhere
3571 > Manufacturing > Industrial And Commercial Machinery And Computer Equipment > Computer And Office Equipment > Electronic Computers
3572 > Manufacturing > Industrial And Commercial Machinery And Computer Equipment > Computer And Office Equipment > Computer Storage Devices
3575 > Manufacturing > Industrial And Commercial Machinery And Computer Equipment > Computer And Office Equipment > Computer Terminals
*3576 > Manufacturing > Industrial And Commercial Machinery And Computer Equipment > Computer And Office Equipment > Computer Communications Equipment (SEC)*
3577 > Manufacturing > Industrial And Commercial Machinery And Computer Equipment > Computer And Office Equipment > Computer Peripheral Equipment, Not Elsewhere Classified
3578 > Manufacturing > Industrial And Commercial Machinery And Computer Equipment > Computer And Office Equipment > Calculating and Accounting Machines, Except Electronic Computers
3579 > Manufacturing > Industrial And Commercial Machinery And Computer Equipment > Computer And Office Equipment > Office Machines, Not Elsewhere Classified
3581 > Manufacturing > Industrial And Commercial Machinery And Computer Equipment > Refrigeration And Service Industry Machinery > Automatic Vending Machines
3582 > Manufacturing > Industrial And Commercial Machinery And Computer Equipment > Refrigeration And Service Industry Machinery > Commercial Laundry, Drycleaning, and Pressing Machines
3585 > Manufacturing > Industrial And Commercial Machinery And Computer Equipment > Refrigeration And Service Industry Machinery > Air-Conditioning and Warm Air Heating Equipment and Commercial and Industrial Refrigeration Equipment
3586 > Manufacturing > Industrial And Commercial Machinery And Computer Equipment > Refrigeration And Service Industry Machinery > Measuring and Dispensing Pumps
3589 > Manufacturing > Industrial And Commercial Machinery And Computer Equipment > Refrigeration And Service Industry Machinery > Service Industry Machinery, Not Elsewhere Classified
3592 > Manufacturing > Industrial And Commercial Machinery And Computer Equipment > Miscellaneous Industrial And Commercial > Carburetors, Pistons, Piston Rings, and Valves
3593 > Manufacturing > Industrial And Commercial Machinery And Computer Equipment > Miscellaneous Industrial And Commercial > Fluid Power Cylinders and Actuators
3594 > Manufacturing > Industrial And Commercial Machinery And Computer Equipment > Miscellaneous Industrial And Commercial > Fluid Power Pumps and Motors
3596 > Manufacturing > Industrial And Commercial Machinery And Computer Equipment > Miscellaneous Industrial And Commercial > Scales and Balances, Except Laboratory
3599 > Manufacturing > Industrial And Commercial Machinery And Computer Equipment > Miscellaneous Industrial And Commercial > Industrial and Commercial Machinery and Equipment, Not Elsewhere Classified
3612 > Manufacturing > Electronic And Other Electrical Equipment And Components, Except Computer Equipment > Electric Transmission And Distribution Equipment > Power, Distribution, and Specialty Transformers
3613 > Manufacturing > Electronic And Other Electrical Equipment And Components, Except Computer Equipment > Electric Transmission And Distribution Equipment > Switchgear and Switchboard Apparatus
3621 > Manufacturing > Electronic And Other Electrical Equipment And Components, Except Computer Equipment > Electrical Industrial Apparatus > Motors and Generators
3624 > Manufacturing > Electronic And Other Electrical Equipment And Components, Except Computer Equipment > Electrical Industrial Apparatus > Carbon and Graphite Products
3625 > Manufacturing > Electronic And Other Electrical Equipment And Components, Except Computer Equipment > Electrical Industrial Apparatus > Relays and Industrial Controls 3629 > Manufacturing > Electronic And Other Electrical Equipment And Components, Except Computer Equipment > Electrical Industrial Apparatus > Electrical Industrial Apparatus, Not Elsewhere Classified
3631 > Manufacturing > Electronic And Other Electrical Equipment And Components, Except Computer Equipment > Household Appliances > Household Cooking Equipment
3632 > Manufacturing > Electronic And Other Electrical Equipment And Components, Except Computer Equipment > Household Appliances > Household Refrigerators and HOme and Farm Freezers
3633 > Manufacturing > Electronic And Other Electrical Equipment And Components, Except Computer Equipment > Household Appliances > Household Laundry Equipment
3634 > Manufacturing > Electronic And Other Electrical Equipment And Components, Except Computer Equipment > Household Appliances > Electric Housewares and Fans
3635 > Manufacturing > Electronic And Other Electrical Equipment And Components, Except Computer Equipment > Household Appliances > Household Vacuum Cleaners
3639 > Manufacturing > Electronic And Other Electrical Equipment And Components, Except Computer Equipment > Household Appliances > Household Appliances, Not Elsewhere Classified
3641 > Manufacturing > Electronic And Other Electrical Equipment And Components, Except Computer Equipment > Electric Lighting And Wiring Equipment > Electric Lamp Bulbs and Tubes
3643 > Manufacturing > Electronic And Other Electrical Equipment And Components, Except Computer Equipment > Electric Lighting And Wiring Equipment > Current-Carrying Wiring Devices
3644 > Manufacturing > Electronic And Other Electrical Equipment And Components, Except Computer Equipment > Electric Lighting And Wiring Equipment > Noncurrent-Carrying Wiring Devices
3645 > Manufacturing > Electronic And Other Electrical Equipment And Components, Except Computer Equipment > Electric Lighting And Wiring Equipment > Residential Electric Lighting Fixtures
3646 > Manufacturing > Electronic And Other Electrical Equipment And Components, Except Computer Equipment > Electric Lighting And Wiring Equipment > Commercial, Industrial, and Institutional Electric Lighting Fixtures
3647 > Manufacturing > Electronic And Other Electrical Equipment And Components, Except Computer Equipment > Electric Lighting And Wiring Equipment > Vehicular Lighting Equipment
3648 > Manufacturing > Electronic And Other Electrical Equipment And Components, Except Computer Equipment > Electric Lighting And Wiring Equipment > Lighting Equipment, Not Elsewhere Classified
3651 > Manufacturing > Electronic And Other Electrical Equipment And Components, Except Computer Equipment > Household Audio And Video Equipment, And Audio > Household Audio and Video Equipment
3652 > Manufacturing > Electronic And Other Electrical Equipment And Components, Except Computer Equipment > Household Audio And Video Equipment, And Audio > Phonograph Records and Prerecorded Audio Tapes and Disks
3661 > Manufacturing > Electronic And Other Electrical Equipment And Components, Except Computer Equipment > Communications Equipment > Telephone and Telegraph Apparatus
3663 > Manufacturing > Electronic And Other Electrical Equipment And Components, Except Computer Equipment > Communications Equipment > Radio and Television Broadcasting and Communications Equipment
3669 > Manufacturing > Electronic And Other Electrical Equipment And Components, Except Computer Equipment > Communications Equipment > Communications Equipment, Not Elsewhere Classified
3671 > Manufacturing > Electronic And Other Electrical Equipment And Components, Except Computer Equipment > Electronic Components And Accessories > Electron Tubes
3672 > Manufacturing > Electronic And Other Electrical Equipment And Components, Except Computer Equipment > Electronic Components And Accessories > Printed Circuit Boards
3674 > Manufacturing > Electronic And Other Electrical Equipment And Components, Except Computer Equipment > Electronic Components And Accessories > Semiconductors and Related Devices
3675 > Manufacturing > Electronic And Other Electrical Equipment And Components, Except Computer Equipment > Electronic Components And Accessories > Electronic Capacitors
3676 > Manufacturing > Electronic And Other Electrical Equipment And Components, Except Computer Equipment > Electronic Components And Accessories > Electronic Resistors
3677 > Manufacturing > Electronic And Other Electrical Equipment And Components, Except Computer Equipment > Electronic Components And Accessories > Electronic Coils, Transformers, and Other Inductors 3678 > Manufacturing > Electronic And Other Electrical Equipment And Components, Except Computer Equipment > Electronic Components And Accessories > Electronic Connectors
3679 > Manufacturing > Electronic And Other Electrical Equipment And Components, Except Computer Equipment > Electronic Components And Accessories > Electronic Components, Not Elsewhere Classified
3691 > Manufacturing > Electronic And Other Electrical Equipment And Components, Except Computer Equipment > Miscellaneous Electrical Machinery, Equipment, and Supplies > Storage Batteries
3692 > Manufacturing > Electronic And Other Electrical Equipment And Components, Except Computer Equipment > Miscellaneous Electrical Machinery, Equipment, and Supplies > Primary Batteries, Dry and Wet
3694 > Manufacturing > Electronic And Other Electrical Equipment And Components, Except Computer Equipment > Miscellaneous Electrical Machinery, Equipment, and Supplies > Electrical Equipment for Internal Combustion Engines
3695 > Manufacturing > Electronic And Other Electrical Equipment And Components, Except Computer Equipment > Miscellaneous Electrical Machinery, Equipment, and Supplies > Magnetic And Optical Recording Media
3699 > Manufacturing > Electronic And Other Electrical Equipment And Components, Except Computer Equipment > Miscellaneous Electrical Machinery, Equipment, and Supplies > Electrical Machinery, Equipment, and Supplies, Not Elsewhere
3711 > Manufacturing > Transportation Equipment > Motor Vehicles And Motor Vehicle Equipment > Motor Vehicles and Passenger Car Bodies
3713 > Manufacturing > Transportation Equipment > Motor Vehicles And Motor Vehicle Equipment > Truck and Bus Bodies
3714 > Manufacturing > Transportation Equipment > Motor Vehicles And Motor Vehicle Equipment > Motor Vehicle Parts and Accessories
3715 > Manufacturing > Transportation Equipment > Motor Vehicles And Motor Vehicle Equipment > Truck Trailers
3716 > Manufacturing > Transportation Equipment > Motor Vehicles And Motor Vehicle Equipment > Motor Homes
3721 > Manufacturing > Transportation Equipment > Aircraft And Parts > Aircraft
3724 > Manufacturing > Transportation Equipment > Aircraft And Parts > Aircraft Engines and Engine Parts
3728 > Manufacturing > Transportation Equipment > Aircraft And Parts > Aircraft Parts and Auxiliary Equipment, Not Elsewhere Classified
3731 > Manufacturing > Transportation Equipment > Ship And Boat Building And Repairing > Ship Building and Repairing
3732 > Manufacturing > Transportation Equipment > Ship And Boat Building And Repairing > Boat Building and Repairing
3743 > Manufacturing > Transportation Equipment > Railroad Equipment > Railroad Equipment
3751 > Manufacturing > Transportation Equipment > Motorcycles, Bicycles, And Parts > Motorcycles, Bicycles, and Parts
3761 > Manufacturing > Transportation Equipment > Guided Missiles And Space Vehicles And Parts > Guided Missiles and Space Vehicles
3764 > Manufacturing > Transportation Equipment > Guided Missiles And Space Vehicles And Parts > Guided Missile and Space Vehicle Propulsion Units and Propulsion Unit Parts
3769 > Manufacturing > Transportation Equipment > Guided Missiles And Space Vehicles And Parts > Guided Missile Space Vehicle Parts and Auxiliary Equipment, Not Elsewhere Classified
3792 > Manufacturing > Transportation Equipment > Miscellaneous Transportation Equipment > Travel Trailers and Campers
3795 > Manufacturing > Transportation Equipment > Miscellaneous Transportation Equipment > Tanks and Tank Components
3799 > Manufacturing > Transportation Equipment > Miscellaneous Transportation Equipment > Transportation Equipment, Not Elsewhere Classified
3812 > Manufacturing > Measuring, Analyzing, And Controlling Instruments; Photographic, Medical And Optical Goods; Watches And Clocks > Search, Detection, Navigation, Guidance, Aeronautical, and Nautical Systems, Instruments, and Equipment > Search, Detection, Navigation, Guidance, Aeronautical, and Nautical Systems and Instruments
3821 > Manufacturing > Measuring, Analyzing, And Controlling Instruments; Photographic, Medical And Optical Goods; Watches And Clocks > Laboratory Apparatus And Analytical, Optical, Measuring, and Controlling Instruments > Laboratory Apparatus and Furniture
3822 > Manufacturing > Measuring, Analyzing, And Controlling Instruments; Photographic, Medical And Optical Goods; Watches And Clocks > Laboratory Apparatus And Analytical, Optical, Measuring, and Controlling Instruments > Automatic Controls for Regulating Residential and Commercial Environments and Appliances
3823 > Manufacturing > Measuring, Analyzing, And Controlling Instruments; Photographic, Medical And Optical Goods; Watches And Clocks > Laboratory Apparatus And Analytical, Optical, Measuring, and Controlling Instruments > Industrial Instruments for Measurement, Display, and Control of Process Variables; and Related Products
3824 > Manufacturing > Measuring, Analyzing, And Controlling Instruments; Photographic, Medical And Optical Goods; Watches And Clocks > Laboratory Apparatus And Analytical, Optical, Measuring, and Controlling Instruments > Totalizing Fluid Meters and Counting Devices 3825 > Manufacturing > Measuring, Analyzing, And Controlling Instruments; Photographic, Medical And Optical Goods; Watches And Clocks > Laboratory Apparatus And Analytical, Optical, Measuring, and Controlling Instruments > Instruments for Measuring and Testing of Electricity and Electrical Signals
3826 > Manufacturing > Measuring, Analyzing, And Controlling Instruments; Photographic, Medical And Optical Goods; Watches And Clocks > Laboratory Apparatus And Analytical, Optical, Measuring, and Controlling Instruments > Laboratory Analytical Instruments
3827 > Manufacturing > Measuring, Analyzing, And Controlling Instruments; Photographic, Medical And Optical Goods; Watches And Clocks > Laboratory Apparatus And Analytical, Optical, Measuring, and Controlling Instruments > Optical Instruments and Lenses
3829 > Manufacturing > Measuring, Analyzing, And Controlling Instruments; Photographic, Medical And Optical Goods; Watches And Clocks > Laboratory Apparatus And Analytical, Optical, Measuring, and Controlling Instruments > Measuring and Controlling Devices, Not Elsewhere Classified
3841 > Manufacturing > Measuring, Analyzing, And Controlling Instruments; Photographic, Medical And Optical Goods; Watches And Clocks > Surgical, Medical, And Dental Instruments And Supplies > Surgical and Medical Instruments and Apparatus
3842 > Manufacturing > Measuring, Analyzing, And Controlling Instruments; Photographic, Medical And Optical Goods; Watches And Clocks > Surgical, Medical, And Dental Instruments And Supplies > Orthopedic, Prosthetic, and Surgical Appliances and Supplies
3843 > Manufacturing > Measuring, Analyzing, And Controlling Instruments; Photographic, Medical And Optical Goods; Watches And Clocks > Surgical, Medical, And Dental Instruments And Supplies > Dental Equipment and Supplies
3844 > Manufacturing > Measuring, Analyzing, And Controlling Instruments; Photographic, Medical And Optical Goods; Watches And Clocks > Surgical, Medical, And Dental Instruments And Supplies > X-Ray Apparatus and Tubes and Related Irradiation Apparatus
3845 > Manufacturing > Measuring, Analyzing, And Controlling Instruments; Photographic, Medical And Optical Goods; Watches And Clocks > Surgical, Medical, And Dental Instruments And Supplies > Electromedical and Electrotherapeutic Apparatus
3851 > Manufacturing > Measuring, Analyzing, And Controlling Instruments; Photographic, Medical And Optical Goods; Watches And Clocks > Ophthalmic Goods > Ophthalmic Goods
3861 > Manufacturing > Measuring, Analyzing, And Controlling Instruments; Photographic, Medical And Optical Goods; Watches And Clocks > Photographic Equipment And Supplies > Photographic Equipment and Supplies
3873 > Manufacturing > Measuring, Analyzing, And Controlling Instruments; Photographic, Medical And Optical Goods; Watches And Clocks > Watches, Clocks, Clockwork Operated Devices, and Parts > Watches, Clocks, Clockwork Operated Devices, and Parts
3911 > Manufacturing > Miscellaneous Manufacturing Industries > Jewelry, Silverware, And Plated Ware > Jewelry, Precious Metal
3914 > Manufacturing > Miscellaneous Manufacturing Industries > Jewelry, Silverware, And Plated Ware > Silverware, Plated Ware, and Stainless Steel Ware
3915 > Manufacturing > Miscellaneous Manufacturing Industries > Jewelry, Silverware, And Plated Ware > Jewelers' Findings and Materials, and Lapidary Work
3931 > Manufacturing > Miscellaneous Manufacturing Industries > Musical Instruments > Musical Instruments
3942 > Manufacturing > Miscellaneous Manufacturing Industries > Dolls, Toys, Games And Sporting And Athletic > Dolls and Stuffed Toys
3944 > Manufacturing > Miscellaneous Manufacturing Industries > Dolls, Toys, Games And Sporting And Athletic > Games, Toys, and Children's Vehicles, Except Dolls and Bicycles
3949 > Manufacturing > Miscellaneous Manufacturing Industries > Dolls, Toys, Games And Sporting And Athletic > Sporting and Athletic Goods, Not Elsewhere Classified
3951 > Manufacturing > Miscellaneous Manufacturing Industries > Pens, Pencils, And Other Artists Materials > Pens, Mechanical Pencils, and Parts
3952 > Manufacturing > Miscellaneous Manufacturing Industries > Pens, Pencils, And Other Artists Materials > Lead Pencils, Crayons, and Artists' Materials
3953 > Manufacturing > Miscellaneous Manufacturing Industries > Pens, Pencils, And Other Artists Materials > Marking Devices
3955 > Manufacturing > Miscellaneous Manufacturing Industries > Pens, Pencils, And Other Artists Materials > Carbon Paper and Inked Ribbons
3961 > Manufacturing > Miscellaneous Manufacturing Industries > Costume Jewelry, Costume Novelties, Buttons, And Miscellaneous Notions, Ecept Precious Metal > Costume Jewelry and Costume Novelties, Except Precious Metal
3965 > Manufacturing > Miscellaneous Manufacturing Industries > Costume Jewelry, Costume Novelties, Buttons, And Miscellaneous Notions, Ecept Precious Metal > Fasteners, Buttons, Needles, and Pins
3991 > Manufacturing > Miscellaneous Manufacturing Industries > Miscellaneous Manufacturing Industries > Brooms and Brushes
3993 > Manufacturing > Miscellaneous Manufacturing Industries > Miscellaneous Manufacturing Industries > Signs and Advertising Specialties

3995 > Manufacturing > Miscellaneous Manufacturing Industries > Miscellaneous Manufacturing Industries > Burial Caskets
3996 > Manufacturing > Miscellaneous Manufacturing Industries > Miscellaneous Manufacturing Industries > Linoleum, Asphalted-Felt-Base, and Other Hard Surface Floor Coverings, Not Elsewhere Classified
3999 > Manufacturing > Miscellaneous Manufacturing Industries > Miscellaneous Manufacturing Industries > Manufacturing Industries, Not Elsewhere Classified
4011 > Transportation, Communications, Electric, Gas, And Sanitary Services > Railroad Transportation > Railroads > Railroads, Line-Haul Operating
4013 > Transportation, Communications, Electric, Gas, And Sanitary Services > Railroad Transportation > Railroads > Railroad Switching and Terminal Establishments
4111 > Transportation, Communications, Electric, Gas, And Sanitary Services > Local And Suburban Transit And Interurban Highway Passenger Transportation > Local And Suburban Passenger Transportation > Local and Suburban Transit
4119 > Transportation, Communications, Electric, Gas, And Sanitary Services > Local And Suburban Transit And Interurban Highway Passenger Transportation > Local And Suburban Passenger Transportation > Local Passenger Transportation, Not Elsewhere Classified
4121 > Transportation, Communications, Electric, Gas, And Sanitary Services > Local And Suburban Transit And Interurban Highway Passenger Transportation > Taxicabs > Taxicabs
4131 > Transportation, Communications, Electric, Gas, And Sanitary Services > Local And Suburban Transit And Interurban Highway Passenger Transportation > Intercity And Rural Bus Transportation > Intercity and Rural Bus Transportation
4141 > Transportation, Communications, Electric, Gas, And Sanitary Services > Local And Suburban Transit And Interurban Highway Passenger Transportation > Bus Charter Service > Local Bus Charter Service
4142 > Transportation, Communications, Electric, Gas, And Sanitary Services > Local And Suburban Transit And Interurban Highway Passenger Transportation > Bus Charter Service > Bus Charter Service, Except Local
4151 > Transportation, Communications, Electric, Gas, And Sanitary Services > Local And Suburban Transit And Interurban Highway Passenger Transportation > School Buses > School Buses
4173 > Transportation, Communications, Electric, Gas, And Sanitary Services > Local And Suburban Transit And Interurban Highway Passenger Transportation > Terminal And Service Facilities For Motor Vehicle > Terminal and Service Facilities for Motor Vehicle Passenger Transportation
4212 > Transportation, Communications, Electric, Gas, And Sanitary Services > Motor Freight Transportation And Warehousing > Trucking And Courier Services, Except Air > Local Trucking Without Storage
4213 > Transportation, Communications, Electric, Gas, And Sanitary Services > Motor Freight Transportation And Warehousing > Trucking And Courier Services, Except Air > Trucking, Except Local
4214 > Transportation, Communications, Electric, Gas, And Sanitary Services > Motor Freight Transportation And Warehousing > Trucking And Courier Services, Except Air > Local Trucking With Storage
4215 > Transportation, Communications, Electric, Gas, And Sanitary Services > Motor Freight Transportation And Warehousing > Trucking And Courier Services, Except Air > Courier Services, Except by Air
4221 > Transportation, Communications, Electric, Gas, And Sanitary Services > Motor Freight Transportation And Warehousing > Public Warehousing And Storage > Farm Product Warehousing and Storage
4222 > Transportation, Communications, Electric, Gas, And Sanitary Services > Motor Freight Transportation And Warehousing > Public Warehousing And Storage > Refrigerated Warehousing and Storage
4225 > Transportation, Communications, Electric, Gas, And Sanitary Services > Motor Freight Transportation And Warehousing > Public Warehousing And Storage > General Warehousing and Storage
4226 > Transportation, Communications, Electric, Gas, And Sanitary Services > Motor Freight Transportation And Warehousing > Public Warehousing And Storage > Special Warehousing and Storage, Not Elsewhere Classified
4231 > Transportation, Communications, Electric, Gas, And Sanitary Services > Motor Freight Transportation And Warehousing > Terminal And Joint Terminal Maintenance > Terminal and Joint Terminal Maintenance Facilities for Motor Freight Transportation
4311 > Transportation, Communications, Electric, Gas, And Sanitary Services > United States Postal Service > United States Postal Service > United States Postal Service
4412 > Transportation, Communications, Electric, Gas, And Sanitary Services > Water Transportation > Deep Sea Foreign Transportation Of Freight > Deep Sea Foreign Transportation of Freight
4424 > Transportation, Communications, Electric, Gas, And Sanitary Services > Water Transportation > Deep Sea Domestic Transportation Of Freight > Deep Sea Domestic Transportation of Freight 4432 > Transportation, Communications, Electric, Gas, And Sanitary Services > Water Transportation > Freight Transportation On The Great Lakes-st. > Freight Transportation on the Great Lakes-St. Lawrence Seaway
4449 > Transportation, Communications, Electric, Gas, And Sanitary Services > Water Transportation > Water Transportation Of Freight, Not Elsewhere > Water Transportation of Freight, Not Elsewhere Classified
4481 > Transportation, Communications, Electric, Gas, And Sanitary Services > Water Transportation > Water Transportation Of Passengers > Deep Sea Transportation of Passengers, Except by Ferry
4482 > Transportation, Communications, Electric, Gas, And Sanitary Services > Water Transportation > Water Transportation Of Passengers > Ferries
4489 > Transportation, Communications, Electric, Gas, And Sanitary Services > Water Transportation > Water Transportation Of Passengers > Water Transportation of Passengers, Not Elsewhere Classified
4491 > Transportation, Communications, Electric, Gas, And Sanitary Services > Water Transportation > Services Incidental To Water Transportation > Marine Cargo Handling
4492 > Transportation, Communications, Electric, Gas, And Sanitary Services > Water Transportation > Services Incidental To Water Transportation > Towing and Tugboat Services
4493 > Transportation, Communications, Electric, Gas, And Sanitary Services > Water Transportation > Services Incidental To Water Transportation > Marinas
4499 > Transportation, Communications, Electric, Gas, And Sanitary Services > Water Transportation > Services Incidental To Water Transportation > Water Transportation Services, Not Elsewhere Classified
4512 > Transportation, Communications, Electric, Gas, And Sanitary Services > Transportation By Air > Air Transportation, Scheduled, And Air Courier > Air Transportation, Scheduled
4513 > Transportation, Communications, Electric, Gas, And Sanitary Services > Transportation By Air > Air Transportation, Scheduled, And Air Courier > Air Courier Services
4522 > Transportation, Communications, Electric, Gas, And Sanitary Services > Transportation By Air > Air Transportation, Nonscheduled > Air Transportation, Nonscheduled
4581 > Transportation, Communications, Electric, Gas, And Sanitary Services > Transportation By Air > Airports, Flying Fields, And Airport Terminal > Airports, Flying Fields, and Airport Terminal Services
4612 > Transportation, Communications, Electric, Gas, And Sanitary Services > Pipelines, Except Natural Gas > Pipelines, Except Natural Gas > Crude Petroleum Pipelines
4613 > Transportation, Communications, Electric, Gas, And Sanitary Services > Pipelines, Except Natural Gas > Pipelines, Except Natural Gas > Refined Petroleum Pipelines
4619 > Transportation, Communications, Electric, Gas, And Sanitary Services > Pipelines, Except Natural Gas > Pipelines, Except Natural Gas > Pipelines, Not Elsewhere Classified
4724 > Transportation, Communications, Electric, Gas, And Sanitary Services > Transportation Services > Arrangement Of Passenger Transportation > Travel Agencies
4725 > Transportation, Communications, Electric, Gas, And Sanitary Services > Transportation Services > Arrangement Of Passenger Transportation > Tour Operators
4729 > Transportation, Communications, Electric, Gas, And Sanitary Services > Transportation Services > Arrangement Of Passenger Transportation > Arrangement of Passenger Transportation, Not Elsewhere Classified
4731 > Transportation, Communications, Electric, Gas, And Sanitary Services > Transportation Services > Arrangement Of Transportation Of Freight And Cargo > Arrangement of Transportation of Freight and Cargo
4741 > Transportation, Communications, Electric, Gas, And Sanitary Services > Transportation Services > Rental Of Railroad Cars > Rental of Railroad Cars
4783 > Transportation, Communications, Electric, Gas, And Sanitary Services > Transportation Services > Miscellaneous Services Incidental To Transportation > Packing and Crating
4785 > Transportation, Communications, Electric, Gas, And Sanitary Services > Transportation Services > Miscellaneous Services Incidental To Transportation > Fixed Facilities and Inspection and Weighing Services for Motor Vehicle Transportation
4789 > Transportation, Communications, Electric, Gas, And Sanitary Services > Transportation Services > Miscellaneous Services Incidental To Transportation > Transportation Services, Not Elsewhere Classified
4812 > Transportation, Communications, Electric, Gas, And Sanitary Services > Communications > Telephone Communications > Radiotelephone Communications 4813 > Transportation, Communications, Electric, Gas, And Sanitary Services > Communications > Telephone Communications > Telephone Communications, Except Radiotelephone
4822 > Transportation, Communications, Electric, Gas, And Sanitary Services > Communications > Telegraph And Other Message Communications > Telegraph and Other Message Communications
4832 > Transportation, Communications, Electric, Gas, And Sanitary Services > Communications > Radio And Television Broadcasting Stations > Radio Broadcasting Stations
4833 > Transportation, Communications, Electric, Gas, And Sanitary Services > Communications > Radio And Television Broadcasting Stations > Television Broadcasting Stations
4841 > Transportation, Communications, Electric, Gas, And Sanitary Services > Communications > Cable And Other Pay Television Services > Cable and Other Pay Television Services
4899 > Transportation, Communications, Electric, Gas, And Sanitary Services > Communications > Communications Services, Not Elsewhere > Communications Services, Not Elsewhere Classified
4911 > Transportation, Communications, Electric, Gas, And Sanitary Services > Electric, Gas, And Sanitary Services > Electric Services > Electric Services
4922 > Transportation, Communications, Electric, Gas, And Sanitary Services > Electric, Gas, And Sanitary Services > Gas Production And Distribution > Natural Gas Transmission
4923 > Transportation, Communications, Electric, Gas, And Sanitary Services > Electric, Gas, And Sanitary Services > Gas Production And Distribution > Natural Gas Transmission and Distribution
4924 > Transportation, Communications, Electric, Gas, And Sanitary Services > Electric, Gas, And Sanitary Services > Gas Production And Distribution > Natural Gas Distribution
4925 > Transportation, Communications, Electric, Gas, And Sanitary Services > Electric, Gas, And Sanitary Services > Gas Production And Distribution > Mixed, Manufactured, or Liquefied Petroleum Gas Production and/or
4931 > Transportation, Communications, Electric, Gas, And Sanitary Services > Electric, Gas, And Sanitary Services > Combination Electric And Gas, And Other Utility > Electric and Other Services Combined
4932 > Transportation, Communications, Electric, Gas, And Sanitary Services > Electric, Gas, And Sanitary Services > Combination Electric And Gas, And Other Utility > Gas and Other Services Combined
4939 > Transportation, Communications, Electric, Gas, And Sanitary Services > Electric, Gas, And Sanitary Services > Combination Electric And Gas, And Other Utility > Combination Utilities, Not Elsewhere Classified
4941 > Transportation, Communications, Electric, Gas, And Sanitary Services > Electric, Gas, And Sanitary Services > Water Supply > Water Supply
4952 > Transportation, Communications, Electric, Gas, And Sanitary Services > Electric, Gas, And Sanitary Services > Sanitary Services > Sewerage Systems
4953 > Transportation, Communications, Electric, Gas, And Sanitary Services > Electric, Gas, And Sanitary Services > Sanitary Services > Refuse Systems
***4955 > Transportation, Communications, Electric, Gas, And Sanitary Services > Electric, Gas, And Sanitary Services > Sanitary Services > Hazardous Waste Management (SEC)***
4959 > Transportation, Communications, Electric, Gas, And Sanitary Services > Electric, Gas, And Sanitary Services > Sanitary Services > Sanitary Services, Not Elsewhere Classified
4961 > Transportation, Communications, Electric, Gas, And Sanitary Services > Electric, Gas, And Sanitary Services > Steam And Air-conditioning Supply > Steam and Air-Conditioning Supply
4971 > Transportation, Communications, Electric, Gas, And Sanitary Services > Electric, Gas, And Sanitary Services > Irrigation Systems > Irrigation Systems
***4991 > Transportation, Communications, Electric, Gas, And Sanitary Services > Electric, Gas, And Sanitary Services > Cogeneration Services and Small Power Producers > Cogeneration Services and Small Power Producers (SEC)***
5012 > Wholesale Trade > Wholesale Trade-durable Goods > Motor Vehicles And Motor Vehicle Parts And Supplies > Automobiles and Other Motor Vehicles
5013 > Wholesale Trade > Wholesale Trade-durable Goods > Motor Vehicles And Motor Vehicle Parts And Supplies > Motor Vehicle Supplies and New Parts
5014 > Wholesale Trade > Wholesale Trade-durable Goods > Motor Vehicles And Motor Vehicle Parts And Supplies > Tires and Tubes
5015 > Wholesale Trade > Wholesale Trade-durable Goods > Motor Vehicles And Motor Vehicle Parts And Supplies > Motor Vehicle Parts, Used

5021 > Wholesale Trade > Wholesale Trade-durable Goods > Furniture And Home Furnishings > Furniture
5023 > Wholesale Trade > Wholesale Trade-durable Goods > Furniture And Home Furnishings > Home furnishings
5031 > Wholesale Trade > Wholesale Trade-durable Goods > Lumber And Other Construction Materials > Lumber, Plywood, Millwork, and Wood Panels
5032 > Wholesale Trade > Wholesale Trade-durable Goods > Lumber And Other Construction Materials > Brick, Stone, and Related Construction Materials
5033 > Wholesale Trade > Wholesale Trade-durable Goods > Lumber And Other Construction Materials > Roofing, Siding, and Insulation Materials
5039 > Wholesale Trade > Wholesale Trade-durable Goods > Lumber And Other Construction Materials > Construction Materials, Not Elsewhere Classified
5043 > Wholesale Trade > Wholesale Trade-durable Goods > Professional And Commercial Equipment And Supplies > Photographic Equipment and Supplies
5044 > Wholesale Trade > Wholesale Trade-durable Goods > Professional And Commercial Equipment And Supplies > Office Equipment
5045 > Wholesale Trade > Wholesale Trade-durable Goods > Professional And Commercial Equipment And Supplies > Computers and Computer Peripheral Equipment and Software
5046 > Wholesale Trade > Wholesale Trade-durable Goods > Professional And Commercial Equipment And Supplies > Commercial Equipment, Not Elsewhere Classified
5047 > Wholesale Trade > Wholesale Trade-durable Goods > Professional And Commercial Equipment And Supplies > Medical, Dental, and Hospital Equipment and Supplies
5048 > Wholesale Trade > Wholesale Trade-durable Goods > Professional And Commercial Equipment And Supplies > Ophthalmic Goods
5049 > Wholesale Trade > Wholesale Trade-durable Goods > Professional And Commercial Equipment And Supplies > Professional Equipment and Supplies, Not Elsewhere Classified
5051 > Wholesale Trade > Wholesale Trade-durable Goods > Metals And Minerals, Except Petroleum > Metals Service Centers and Offices
5052 > Wholesale Trade > Wholesale Trade-durable Goods > Metals And Minerals, Except Petroleum > Coal and Other Minerals and Ores
5063 > Wholesale Trade > Wholesale Trade-durable Goods > Electrical Goods > Electrical Apparatus and Equipment Wiring Supplies, and Construction Materials
5064 > Wholesale Trade > Wholesale Trade-durable Goods > Electrical Goods > Electrical Appliances, Television and Radio Sets
5065 > Wholesale Trade > Wholesale Trade-durable Goods > Electrical Goods > Electronic Parts and Equipment, Not Elsewhere Classified
5072 > Wholesale Trade > Wholesale Trade-durable Goods > Hardware, And Plumbing And Heating Equipment > Hardware
5074 > Wholesale Trade > Wholesale Trade-durable Goods > Hardware, And Plumbing And Heating Equipment > Plumbing and Heating Equipment and Supplies (Hydronics)
5075 > Wholesale Trade > Wholesale Trade-durable Goods > Hardware, And Plumbing And Heating Equipment > Warm Air Heating and Air-Conditioning Equipment and Supplies
5078 > Wholesale Trade > Wholesale Trade-durable Goods > Hardware, And Plumbing And Heating Equipment > Refrigeration Equipment and Supplies
5082 > Wholesale Trade > Wholesale Trade-durable Goods > Machinery, Equipment, And Supplies > Construction and Mining (Except Petroleum) Machinery and Equipment
5083 > Wholesale Trade > Wholesale Trade-durable Goods > Machinery, Equipment, And Supplies > Farm and Garden Machinery and Equipment
5084 > Wholesale Trade > Wholesale Trade-durable Goods > Machinery, Equipment, And Supplies > Industrial Machinery and Equipment
5085 > Wholesale Trade > Wholesale Trade-durable Goods > Machinery, Equipment, And Supplies > Industrial Supplies
5087 > Wholesale Trade > Wholesale Trade-durable Goods > Machinery, Equipment, And Supplies > Service Establishment Equipment and Supplies
5088 > Wholesale Trade > Wholesale Trade-durable Goods > Machinery, Equipment, And Supplies > Transportation Equipment and Supplies, Except Motor Vehicles
5091 > Wholesale Trade > Wholesale Trade-durable Goods > Miscellaneous Durable Goods > Sporting and Recreational Goods and Supplies
5092 > Wholesale Trade > Wholesale Trade-durable Goods > Miscellaneous Durable Goods > Toys and Hobby Goods and Supplies
5093 > Wholesale Trade > Wholesale Trade-durable Goods > Miscellaneous Durable Goods > Scrap and Waste Materials
5094 > Wholesale Trade > Wholesale Trade-durable Goods > Miscellaneous Durable Goods > Jewelry, Watches, Precious Stones, and Precious Metals
5099 > Wholesale Trade > Wholesale Trade-durable Goods > Miscellaneous Durable Goods > Durable Goods, Not Elsewhere Classified
5111 > Wholesale Trade > Wholesale Trade-non-durable Goods > Paper And Paper Products > Printing and Writing Paper

5112 > Wholesale Trade > Wholesale Trade-non-durable Goods > Paper And Paper Products > Stationery and Office Supplies
5113 > Wholesale Trade > Wholesale Trade-non-durable Goods > Paper And Paper Products > Industrial and Personal Service Paper
5122 > Wholesale Trade > Wholesale Trade-non-durable Goods > Drugs, Drug Proprietaries, And Druggists' Sundries > Drugs, Drug Proprietaries, and Druggists' Sundries
5131 > Wholesale Trade > Wholesale Trade-non-durable Goods > Apparel, Piece Goods, And Notions > Piece Goods, Notions, and Other Dry Good
5136 > Wholesale Trade > Wholesale Trade-non-durable Goods > Apparel, Piece Goods, And Notions > Men's and Boy's Clothing and Furnishings
5137 > Wholesale Trade > Wholesale Trade-non-durable Goods > Apparel, Piece Goods, And Notions > Women's, Children's, and Infants' Clothing and Accessories
5139 > Wholesale Trade > Wholesale Trade-non-durable Goods > Apparel, Piece Goods, And Notions > Footwear
5141 > Wholesale Trade > Wholesale Trade-non-durable Goods > Groceries And Related Products > Groceries, General Line
5142 > Wholesale Trade > Wholesale Trade-non-durable Goods > Groceries And Related Products > Packaged Frozen Foods
5143 > Wholesale Trade > Wholesale Trade-non-durable Goods > Groceries And Related Products > Dairy Products, Except Dried or Canned
5144 > Wholesale Trade > Wholesale Trade-non-durable Goods > Groceries And Related Products > Poultry and Poultry Products
5145 > Wholesale Trade > Wholesale Trade-non-durable Goods > Groceries And Related Products > Confectionery
5146 > Wholesale Trade > Wholesale Trade-non-durable Goods > Groceries And Related Products > Fish and Seafoods
5147 > Wholesale Trade > Wholesale Trade-non-durable Goods > Groceries And Related Products > Meats and Meat Products
5148 > Wholesale Trade > Wholesale Trade-non-durable Goods > Groceries And Related Products > Fresh Fruits and Vegetables
5149 > Wholesale Trade > Wholesale Trade-non-durable Goods > Groceries And Related Products > Groceries and Related Products, Not Elsewhere Classified
5153 > Wholesale Trade > Wholesale Trade-non-durable Goods > Farm-product Raw Materials > Grain and Field Beans
5154 > Wholesale Trade > Wholesale Trade-non-durable Goods > Farm-product Raw Materials > Livestock
5159 > Wholesale Trade > Wholesale Trade-non-durable Goods > Farm-product Raw Materials > Farm-Product Raw Materials, Not Elsewhere Classified
5162 > Wholesale Trade > Wholesale Trade-non-durable Goods > Chemicals And Allied Products > Plastics Materials and Basic Forms and Shapes
5169 > Wholesale Trade > Wholesale Trade-non-durable Goods > Chemicals And Allied Products > Chemicals and Allied Products, Not Elsewhere Classified
5171 > Wholesale Trade > Wholesale Trade-non-durable Goods > Petroleum And Petroleum Products > Petroleum Bulk stations and Terminals
5172 > Wholesale Trade > Wholesale Trade-non-durable Goods > Petroleum And Petroleum Products > Petroleum and Petroleum Products Wholesalers, Except Bulk Stations and Terminals
5181 > Wholesale Trade > Wholesale Trade-non-durable Goods > Beer, Wine, And Distilled Alcoholic Beverages > Beer and Ale
5182 > Wholesale Trade > Wholesale Trade-non-durable Goods > Beer, Wine, And Distilled Alcoholic Beverages > Wine and Distilled Alcoholic Beverages
5191 > Wholesale Trade > Wholesale Trade-non-durable Goods > Miscellaneous Non-durable Goods > Farm Supplies
5192 > Wholesale Trade > Wholesale Trade-non-durable Goods > Miscellaneous Non-durable Goods > Books, Periodicals, and Newspapers
5193 > Wholesale Trade > Wholesale Trade-non-durable Goods > Miscellaneous Non-durable Goods > Flowers, Nursery Stock, and Florists' Supplies
5194 > Wholesale Trade > Wholesale Trade-non-durable Goods > Miscellaneous Non-durable Goods > Tobacco and Tobacco Products
5198 > Wholesale Trade > Wholesale Trade-non-durable Goods > Miscellaneous Non-durable Goods > Paints, Varnishes, and Supplies
5199 > Wholesale Trade > Wholesale Trade-non-durable Goods > Miscellaneous Non-durable Goods > Nondurable Goods, Not Elsewhere Classified
5211 > Retail Trade > Building Materials, Hardware, Garden Supply, And Mobile Home Dealers > Lumber And Other Building Materials Dealers > Lumber and Other Building Materials Dealers
5231 > Retail Trade > Building Materials, Hardware, Garden Supply, And Mobile Home Dealers > Paint, Glass And Wallpaper Stores > Paint, Glass, and Wallpaper Stores
5251 > Retail Trade > Building Materials, Hardware, Garden Supply, And Mobile Home Dealers > Hardware Stores > Hardware Stores
5261 > Retail Trade > Building Materials, Hardware, Garden Supply, And Mobile Home Dealers > Retail Nurseries, Lawn And Garden Supply Stores > Retail Nurseries, Lawn and Garden Supply Stores
5271 > Retail Trade > Building Materials, Hardware, Garden Supply, And Mobile Home Dealers > Mobile Home Dealers > Mobile Home Dealers

```
5311 > Retail Trade > General Merchandise Stores > Department Stores > Department Stores
5331 > Retail Trade > General Merchandise Stores > Variety Stores > Variety Stores
5399 > Retail Trade > General Merchandise Stores > Miscellaneous General Merchandise Stores > Miscellaneous General Merchandise Stores
5411 > Retail Trade > Food Stores > Grocery Stores > Grocery Stores
```
*5412 > Retail Trade > Food Stores > Grocery Stores > Convenience Stores (SEC)*
```
5421 > Retail Trade > Food Stores > Meat And Fish (seafood) Markets, Including > Meat and Fish (Seafood) Markets, Including Freezer Provisioners
5431 > Retail Trade > Food Stores > Fruit And Vegetable Markets > Fruit and Vegetable Markets
5441 > Retail Trade > Food Stores > Candy, Nut, And Confectionery Stores > Candy, Nut, and Confectionery Stores
5451 > Retail Trade > Food Stores > Dairy Products Stores > Dairy Products Stores
5461 > Retail Trade > Food Stores > Retail Bakeries > Retail Bakeries
5499 > Retail Trade > Food Stores > Miscellaneous Food Stores > Miscellaneous Food Stores
5511 > Retail Trade > Automotive Dealers And Gasoline Service Stations > Motor Vehicle Dealers (new And Used) > Motor Vehicle Dealers (New and Used)
5521 > Retail Trade > Automotive Dealers And Gasoline Service Stations > Motor Vehicle Dealers (used Only) > Motor Vehicle Dealers (Used Only)
5531 > Retail Trade > Automotive Dealers And Gasoline Service Stations > Auto And Home Supply Stores > Auto and Home Supply Stores
5541 > Retail Trade > Automotive Dealers And Gasoline Service Stations > Gasoline Service Stations > Gasoline Service Stations
5551 > Retail Trade > Automotive Dealers And Gasoline Service Stations > Boat Dealers > Boat Dealers
5561 > Retail Trade > Automotive Dealers And Gasoline Service Stations > Recreational Vehicle Dealers > Recreational Vehicle Dealers
5571 > Retail Trade > Automotive Dealers And Gasoline Service Stations > Motorcycle Dealers > Motorcycle Dealers
5599 > Retail Trade > Automotive Dealers And Gasoline Service Stations > Automotive Dealers, Not Elsewhere Classified > Automotive Dealers, Not Elsewhere Classified
5611 > Retail Trade > Apparel And Accessory Stores > Men's And Boys' Clothing And Accessory Stores > Men's and Boys' Clothing and Accessory Stores
5621 > Retail Trade > Apparel And Accessory Stores > Women's Clothing Stores > Women's Clothing Stores
5632 > Retail Trade > Apparel And Accessory Stores > Women's Accessory And Specialty Stores > Women's Accessory and Specialty Stores
5641 > Retail Trade > Apparel And Accessory Stores > Children's And Infants' Wear Stores > Children's and Infants' Wear Stores
5651 > Retail Trade > Apparel And Accessory Stores > Family Clothing Stores > Family Clothing Stores
5661 > Retail Trade > Apparel And Accessory Stores > Shoe Stores > Shoe Stores
5699 > Retail Trade > Apparel And Accessory Stores > Miscellaneous Apparel And Accessory Stores > Miscellaneous Apparel and Accessory Stores
5712 > Retail Trade > Home Furniture, Furnishings, And Equipment Stores > Home Furniture And Furnishings Stores > Furniture Stores
5713 > Retail Trade > Home Furniture, Furnishings, And Equipment Stores > Home Furniture And Furnishings Stores > Floor Covering Stores
5714 > Retail Trade > Home Furniture, Furnishings, And Equipment Stores > Home Furniture And Furnishings Stores > Drapery, Curtain, and Upholstery Stores
5719 > Retail Trade > Home Furniture, Furnishings, And Equipment Stores > Home Furniture And Furnishings Stores > Miscellaneous home furnishings Stores
5722 > Retail Trade > Home Furniture, Furnishings, And Equipment Stores > Household Appliance Stores > Household Appliance Stores
5731 > Retail Trade > Home Furniture, Furnishings, And Equipment Stores > Radio, Television, Consumer Electronics, And Music Stores > Radio, Television, and Consumer Electronics Stores
5734 > Retail Trade > Home Furniture, Furnishings, And Equipment Stores > Radio, Television, Consumer Electronics, And Music Stores > Computer and Computer Software Stores
5735 > Retail Trade > Home Furniture, Furnishings, And Equipment Stores > Radio, Television, Consumer Electronics, And Music Stores > Record and Prerecorded Tape Stores
5736 > Retail Trade > Home Furniture, Furnishings, And Equipment Stores > Radio, Television, Consumer Electronics, And Music Stores > Musical Instrument Stores
5812 > Retail Trade > Eating And Drinking Places > Eating And Drinking Places > Eating Places
5813 > Retail Trade > Eating And Drinking Places > Eating And Drinking Places > Drinking Places (alcoholic Beverages)
5912 > Retail Trade > Miscellaneous Retail > Drug Stores And Proprietary Stores > Drug Stores and Proprietary Stores
```

```
5921 > Retail Trade > Miscellaneous Retail > Liquor Stores > Liquor Stores
5932 > Retail Trade > Miscellaneous Retail > Used Merchandise Stores > Used Merchandise Stores
5941 > Retail Trade > Miscellaneous Retail > Miscellaneous Shopping Goods Stores > Sporting Goods Stores and Bicycle Shops
5942 > Retail Trade > Miscellaneous Retail > Miscellaneous Shopping Goods Stores > Book Stores
5943 > Retail Trade > Miscellaneous Retail > Miscellaneous Shopping Goods Stores > Stationery Stores
5944 > Retail Trade > Miscellaneous Retail > Miscellaneous Shopping Goods Stores > Jewelry Stores
5945 > Retail Trade > Miscellaneous Retail > Miscellaneous Shopping Goods Stores > Hobby, Toy, and Game Shops
5946 > Retail Trade > Miscellaneous Retail > Miscellaneous Shopping Goods Stores > Camera and Photographic Supply Stores
5947 > Retail Trade > Miscellaneous Retail > Miscellaneous Shopping Goods Stores > Gift, Novelty, and Souvenir Shops
5948 > Retail Trade > Miscellaneous Retail > Miscellaneous Shopping Goods Stores > Luggage and Leather Goods Stores
5949 > Retail Trade > Miscellaneous Retail > Miscellaneous Shopping Goods Stores > Sewing, Needlework, and Piece Goods Stores
5961 > Retail Trade > Miscellaneous Retail > Nonstore Retailers > Catalog and Mail-Order Houses
5962 > Retail Trade > Miscellaneous Retail > Nonstore Retailers > Automatic Merchandising Machine Operators
5963 > Retail Trade > Miscellaneous Retail > Nonstore Retailers > Direct Selling Establishments
5983 > Retail Trade > Miscellaneous Retail > Fuel Dealers > Fuel Oil Dealers
5984 > Retail Trade > Miscellaneous Retail > Fuel Dealers > Liquefied Petroleum Gas (Bottled Gas) Dealers
5989 > Retail Trade > Miscellaneous Retail > Fuel Dealers > Fuel Dealers, Not Elsewhere Classified
5992 > Retail Trade > Miscellaneous Retail > Retail Stores, Not Elsewhere Classified > Florists
5993 > Retail Trade > Miscellaneous Retail > Retail Stores, Not Elsewhere Classified > Tobacco Stores and Stands
5994 > Retail Trade > Miscellaneous Retail > Retail Stores, Not Elsewhere Classified > News Dealers and Newsstands
5995 > Retail Trade > Miscellaneous Retail > Retail Stores, Not Elsewhere Classified > Optical Goods Stores
5999 > Retail Trade > Miscellaneous Retail > Retail Stores, Not Elsewhere Classified > Miscellaneous Retail Stores, Not Elsewhere Classified
6011 > Finance, Insurance, And Real Estate > Depository Institutions > Central Reserve Depository Institutions > Federal Reserve Banks
6019 > Finance, Insurance, And Real Estate > Depository Institutions > Central Reserve Depository Institutions > Central Reserve Depository Institutions, Not Elsewhere Classified
6021 > Finance, Insurance, And Real Estate > Depository Institutions > Commercial Banks > National Commercial Banks
6022 > Finance, Insurance, And Real Estate > Depository Institutions > Commercial Banks > State Commercial Banks
***6025 > Finance, Insurance, And Real Estate > Depository Institutions > Commercial Banks > National Banks, Federal Reserve System (SEC)***
6029 > Finance, Insurance, And Real Estate > Depository Institutions > Commercial Banks > Commercial Banks, Not Elsewhere Classified
6035 > Finance, Insurance, And Real Estate > Depository Institutions > Savings Institutions > Savings Institutions, Federally Chartered
6036 > Finance, Insurance, And Real Estate > Depository Institutions > Savings Institutions > Savings Institutions, Not Federally Chartered
6061 > Finance, Insurance, And Real Estate > Depository Institutions > Credit Unions > Credit Unions, Federally Chartered
6062 > Finance, Insurance, And Real Estate > Depository Institutions > Credit Unions > Credit Unions, Not Federally Chartered
6081 > Finance, Insurance, And Real Estate > Depository Institutions > Foreign Banking And Branches And Agencies Of > Branches and Agencies of Foreign Banks
6082 > Finance, Insurance, And Real Estate > Depository Institutions > Foreign Banking And Branches And Agencies Of > Foreign Trade and International Banking Institutions
6091 > Finance, Insurance, And Real Estate > Depository Institutions > Functions Related To Depository Banking > Non-deposit Trust Facilities,
6099 > Finance, Insurance, And Real Estate > Depository Institutions > Functions Related To Depository Banking > Functions Related to Depository Banking, Not Elsewhere Classified
6111 > Finance, Insurance, And Real Estate > Non-depository Credit Institutions > Federal And Federally-sponsored Credit Agencies > Federal and Federally-Sponsored Credit Agencies
***6120 > Finance, Insurance, And Real Estate > Non-depository Credit Institutions > Savings and Loan Associations (SEC)***
6141 > Finance, Insurance, And Real Estate > Non-depository Credit Institutions > Personal Credit Institutions > Personal Credit Institutions
6153 > Finance, Insurance, And Real Estate > Non-depository Credit Institutions > Business Credit Institutions > Short-Term Business Credit Institutions, Except Agricultural
```

6159 > Finance, Insurance, And Real Estate > Non-depository Credit Institutions > Business Credit Institutions > Miscellaneous business Credit Institutions
6162 > Finance, Insurance, And Real Estate > Non-depository Credit Institutions > Mortgage Bankers And Brokers > Mortgage Bankers and Loan Correspondents
6163 > Finance, Insurance, And Real Estate > Non-depository Credit Institutions > Mortgage Bankers And Brokers > Loan Brokers
*6172 > Finance, Insurance, And Real Estate > Non-depository Credit Institutions > Finance Lessors > Finance Lessors (SEC)*
*6189 > Finance, Insurance, And Real Estate > Non-depository Credit Institutions > Asset-Backed Securities > Asset-Backed Securities (SEC)*
*6199 > Finance, Insurance, And Real Estate > Non-depository Credit Institutions > Finance Services > Finance Services (SEC)*
6211 > Finance, Insurance, And Real Estate > Security And Commodity Brokers, Dealers, Exchanges, And Services > Security Brokers, Dealers, And Flotation > Security Brokers, Dealers, and Flotation Companies
6221 > Finance, Insurance, And Real Estate > Security And Commodity Brokers, Dealers, Exchanges, And Services > Commodity Contracts Brokers And Dealers > Commodity Contracts Brokers and Dealers
6231 > Finance, Insurance, And Real Estate > Security And Commodity Brokers, Dealers, Exchanges, And Services > Security And Commodity Exchanges > Security and Commodity Exchanges
6282 > Finance, Insurance, And Real Estate > Security And Commodity Brokers, Dealers, Exchanges, And Services > Services Allied With The Exchange Of Securities > Investment Advice
6289 > Finance, Insurance, And Real Estate > Security And Commodity Brokers, Dealers, Exchanges, And Services > Services Allied With The Exchange Of Securities > Services Allied With the Exchange of Securities or Commodities, Not Elsewhere Classified
6311 > Finance, Insurance, And Real Estate > Insurance Carriers > Life Insurance > Life Insurance
6321 > Finance, Insurance, And Real Estate > Insurance Carriers > Accident And Health Insurance And Medical > Accident and Health Insurance
6324 > Finance, Insurance, And Real Estate > Insurance Carriers > Accident And Health Insurance And Medical > Hospital and Medical Service Plans
6331 > Finance, Insurance, And Real Estate > Insurance Carriers > Fire, Marine, And Casualty Insurance > Fire, Marine, and Casualty Insurance
6351 > Finance, Insurance, And Real Estate > Insurance Carriers > Surety Insurance > Surety Insurance
6361 > Finance, Insurance, And Real Estate > Insurance Carriers > Title Insurance > Title Insurance
6371 > Finance, Insurance, And Real Estate > Insurance Carriers > Pension, Health, And Welfare Funds > Pension, Health, and Welfare Funds
6399 > Finance, Insurance, And Real Estate > Insurance Carriers > Insurance Carriers, Not Elsewhere Classified > Insurance Carriers, Not Elsewhere Classified
6411 > Finance, Insurance, And Real Estate > Insurance Agents, Brokers, And Service > Insurance Agents, Brokers, And Service > Insurance Agents, Brokers, and Service
6512 > Finance, Insurance, And Real Estate > Real Estate > Real Estate Operators (except Developers) And Lessors > Operators of Nonresidential Buildings
6513 > Finance, Insurance, And Real Estate > Real Estate > Real Estate Operators (except Developers) And Lessors > Operators or Apartment Buildings
6514 > Finance, Insurance, And Real Estate > Real Estate > Real Estate Operators (except Developers) And Lessors > Operators of Dwellings Other Than Apartment Buildings
6515 > Finance, Insurance, And Real Estate > Real Estate > Real Estate Operators (except Developers) And Lessors > Operators of Residential Mobile Home Sites
6517 > Finance, Insurance, And Real Estate > Real Estate > Real Estate Operators (except Developers) And Lessors > Lessors of Railroad Property
6519 > Finance, Insurance, And Real Estate > Real Estate > Real Estate Operators (except Developers) And Lessors > Lessors of Real Property, Not Elsewhere Classified
6531 > Finance, Insurance, And Real Estate > Real Estate > Real Estate Agents And Managers > Real Estate Agents and Managers
*6532 > Finance, Insurance, And Real Estate > Real Estate > Real Estate Agents And Managers > Real Estate Dealers (For Their Own Account) (SEC)*
6541 > Finance, Insurance, And Real Estate > Real Estate > Title Abstract Offices > Title Abstract Offices

6552 > Finance, Insurance, And Real Estate > Real Estate > Land Subdividers And Developers > Land Subdividers and Developers, Except Cemeteries
6553 > Finance, Insurance, And Real Estate > Real Estate > Land Subdividers And Developers > Cemetery Subdividers and Developers
6712 > Finance, Insurance, And Real Estate > Holding And Other Investment Offices > Holding Offices > Offices of Bank Holding Companies
6719 > Finance, Insurance, And Real Estate > Holding And Other Investment Offices > Holding Offices > Offices of Holding Companies, Not Elsewhere Classified
6722 > Finance, Insurance, And Real Estate > Holding And Other Investment Offices > Investment Offices > Management Investment Offices, Open-End
6726 > Finance, Insurance, And Real Estate > Holding And Other Investment Offices > Investment Offices > Unit Investment Trusts, Face-Amount Certificate Offices, and Closed-End Management Investment Offices
6732 > Finance, Insurance, And Real Estate > Holding And Other Investment Offices > Trusts > Educational, Religious, and Charitable Trusts
6733 > Finance, Insurance, And Real Estate > Holding And Other Investment Offices > Trusts > Trusts, Except Educational, Religious, and Charitable
***6770 > Finance, Insurance, And Real Estate > Holding And Other Investment Offices > Blank Checks (SEC)***
6792 > Finance, Insurance, And Real Estate > Holding And Other Investment Offices > Miscellaneous Investing > Oil Royalty Traders
6794 > Finance, Insurance, And Real Estate > Holding And Other Investment Offices > Miscellaneous Investing > Patent Owners and Lessors
***6795 > Finance, Insurance, And Real Estate > Holding And Other Investment Offices > Miscellaneous Investing > Mineral Royalty Traders (SEC)***
6798 > Finance, Insurance, And Real Estate > Holding And Other Investment Offices > Miscellaneous Investing > Real Estate Investment Trusts
6799 > Finance, Insurance, And Real Estate > Holding And Other Investment Offices > Miscellaneous Investing > Investors, Not Elsewhere Classified
7011 > Services > Hotels, Rooming Houses, Camps, And Other Lodging Places > Hotels And Motels > Hotels and Motels
7021 > Services > Hotels, Rooming Houses, Camps, And Other Lodging Places > Rooming And Boarding Houses > Rooming and Boarding Houses
7032 > Services > Hotels, Rooming Houses, Camps, And Other Lodging Places > Camps And Recreational Vehicle Parks > Sporting and Recreational Camps
7033 > Services > Hotels, Rooming Houses, Camps, And Other Lodging Places > Camps And Recreational Vehicle Parks > Recreational Vehicle Parks and Campsites
7041 > Services > Hotels, Rooming Houses, Camps, And Other Lodging Places > Organization Hotels And Lodging Houses, On > Organization Hotels and Lodging Houses, on Membership Basis
7211 > Services > Personal Services > Laundry, Cleaning, And Garment Services > Power Laundries, Family and Commercial
7212 > Services > Personal Services > Laundry, Cleaning, And Garment Services > Garment Pressing, and Agents for Laundries and Drycleaners
7213 > Services > Personal Services > Laundry, Cleaning, And Garment Services > Linen Supply
7215 > Services > Personal Services > Laundry, Cleaning, And Garment Services > Coin-Operated Laundries and Drycleaning
7216 > Services > Personal Services > Laundry, Cleaning, And Garment Services > Drycleaning Plants, Except Rug Cleaning
7217 > Services > Personal Services > Laundry, Cleaning, And Garment Services > Carpet and Upholstery Cleaning
7218 > Services > Personal Services > Laundry, Cleaning, And Garment Services > Industrial Launderers
7219 > Services > Personal Services > Laundry, Cleaning, And Garment Services > Laundry and Garment Services, Not Elsewhere Classified
7221 > Services > Personal Services > Photographic Studios, Portrait > Photographic Studios, Portrait
7231 > Services > Personal Services > Beauty Shops > Beauty Shops
7241 > Services > Personal Services > Barber Shops > Barber Shops
7251 > Services > Personal Services > Shoe Repair Shops And Shoeshine Parlors > Shoe Repair Shops and Shoeshine Parlors
7261 > Services > Personal Services > Funeral Service And Crematories > Funeral Service and Crematories
7291 > Services > Personal Services > Miscellaneous Personal Services > Tax Return Preparation Services
7299 > Services > Personal Services > Miscellaneous Personal Services > Miscellaneous Personal Services, Not Elsewhere Classified
7311 > Services > Business Services > Advertising > Advertising Agencies
7312 > Services > Business Services > Advertising > Outdoor Advertising Services

7313 > Services > Business Services > Advertising > Radio, Television, and Publishers' Advertising Representatives
7319 > Services > Business Services > Advertising > Advertising, Not Elsewhere Classified
7322 > Services > Business Services > Consumer Credit Reporting Agencies, Mercantile > Adjustment and Collection Services
7323 > Services > Business Services > Consumer Credit Reporting Agencies, Mercantile > Credit Reporting Services
7331 > Services > Business Services > Mailing, Reproduction, Commercial Art And Photography, and Stenographic Services > Direct Mail Advertising Services
7334 > Services > Business Services > Mailing, Reproduction, Commercial Art And Photography, and Stenographic Services > Photocopying and Duplicating Services
7335 > Services > Business Services > Mailing, Reproduction, Commercial Art And Photography, and Stenographic Services > Commercial Photography
7336 > Services > Business Services > Mailing, Reproduction, Commercial Art And Photography, and Stenographic Services > Commercial Art and Graphic Design
7338 > Services > Business Services > Mailing, Reproduction, Commercial Art And Photography, and Stenographic Services > Secretarial and Court Reporting Services
7342 > Services > Business Services > Services To Dwellings And Other Buildings > Disinfecting and Pest Control Services
7349 > Services > Business Services > Services To Dwellings And Other Buildings > Building Cleaning and Maintenance Services, Not Elsewhere
7352 > Services > Business Services > Miscellaneous Equipment Rental And Leasing > Medical Equipment Rental and Leasing
7353 > Services > Business Services > Miscellaneous Equipment Rental And Leasing > Heavy Construction Equipment Rental and Leasing
7359 > Services > Business Services > Miscellaneous Equipment Rental And Leasing > Equipment Rental and Leasing, Not Elsewhere Classified
7361 > Services > Business Services > Personnel Supply Services > Employment Agencies
7363 > Services > Business Services > Personnel Supply Services > Help Supply Services
7371 > Services > Business Services > Computer Programming, Data Processing, And Other Computer Related Services > Computer Programming Services
7372 > Services > Business Services > Computer Programming, Data Processing, And Other Computer Related Services > Prepackaged Software
7373 > Services > Business Services > Computer Programming, Data Processing, And Other Computer Related Services > Computer Integrated Systems Design
7374 > Services > Business Services > Computer Programming, Data Processing, And Other Computer Related Services > Computer Processing and Data Preparation and Processing Services
7375 > Services > Business Services > Computer Programming, Data Processing, And Other Computer Related Services > Information Retrieval Services
7376 > Services > Business Services > Computer Programming, Data Processing, And Other Computer Related Services > Computer Facilities Management Services
7377 > Services > Business Services > Computer Programming, Data Processing, And Other Computer Related Services > Computer Rental and Leasing
7378 > Services > Business Services > Computer Programming, Data Processing, And Other Computer Related Services > Computer Maintenance and Repair
7379 > Services > Business Services > Computer Programming, Data Processing, And Other Computer Related Services > Computer Related Services, Not Elsewhere Classified
7381 > Services > Business Services > Miscellaneous Business Services > Detective, Guard, and Armored Car Services
7382 > Services > Business Services > Miscellaneous Business Services > Security Systems Services
7383 > Services > Business Services > Miscellaneous Business Services > News Syndicates
7384 > Services > Business Services > Miscellaneous Business Services > Photofinishing Laboratories
***7385 > Services > Business Services > Miscellaneous Business Services > Telephone Interconnect Systems (SEC)***
7389 > Services > Business Services > Miscellaneous Business Services > Business Services, Not Elsewhere Classified
7513 > Services > Automotive Repair, Services, And Parking > Automotive Rental And Leasing, Without Drivers > Truck Rental and Leasing, Without Drivers
7514 > Services > Automotive Repair, Services, And Parking > Automotive Rental And Leasing, Without Drivers > Passenger Car Rental
7515 > Services > Automotive Repair, Services, And Parking > Automotive Rental And Leasing, Without Drivers > Passenger Car Leasing

```
7519 > Services > Automotive Repair, Services, And Parking > Automotive Rental And Leasing, Without Drivers > Utility Trailer and Recreational Vehicle Rental
7521 > Services > Automotive Repair, Services, And Parking > Automobile Parking > Automobile Parking
7532 > Services > Automotive Repair, Services, And Parking > Automotive Repair Shops > Top, Body, and Upholstery Repair Shops and Paint Shops
7533 > Services > Automotive Repair, Services, And Parking > Automotive Repair Shops > Automotive Exhaust System Repair Shops
7534 > Services > Automotive Repair, Services, And Parking > Automotive Repair Shops > Tire Retreading and Repair Shops
7536 > Services > Automotive Repair, Services, And Parking > Automotive Repair Shops > Automotive Glass Replacement Shops
7537 > Services > Automotive Repair, Services, And Parking > Automotive Repair Shops > Automotive Transmission Repair Shops
7538 > Services > Automotive Repair, Services, And Parking > Automotive Repair Shops > General Automotive Repair Shops
7539 > Services > Automotive Repair, Services, And Parking > Automotive Repair Shops > Automotive Repair Shops, Not Elsewhere Classified
7542 > Services > Automotive Repair, Services, And Parking > Automotive Services, Except Repair > Carwashes
7549 > Services > Automotive Repair, Services, And Parking > Automotive Services, Except Repair > Automotive Services, Except Repair and Carwashes
7622 > Services > Miscellaneous Repair Services > Electrical Repair Shops > Radio and Television Repair Shops
7623 > Services > Miscellaneous Repair Services > Electrical Repair Shops > Refrigeration and Air-Conditioning Service and Repair Shops
7629 > Services > Miscellaneous Repair Services > Electrical Repair Shops > Electrical and Electronic Repair Shops, Not Elsewhere Classified
7631 > Services > Miscellaneous Repair Services > Watch, Clock, And Jewelry Repair > Watch, Clock, and Jewelry Repair
7641 > Services > Miscellaneous Repair Services > Reupholstery And Furniture Repair > Reupholstery and Furniture Repair
7692 > Services > Miscellaneous Repair Services > Miscellaneous Repair Shops And Related Services > Welding Repair
7694 > Services > Miscellaneous Repair Services > Miscellaneous Repair Shops And Related Services > Armature Rewinding Shops
7699 > Services > Miscellaneous Repair Services > Miscellaneous Repair Shops And Related Services > Repair Shops and Related Services, Not Elsewhere Classified
7812 > Services > Motion Pictures > Motion Picture Production And Allied Services > Motion Picture and Video Tape Production
7819 > Services > Motion Pictures > Motion Picture Production And Allied Services > Services Allied to Motion Picture Production
7822 > Services > Motion Pictures > Motion Picture Distribution And Allied Services > Motion Picture and Video Tape Distribution
7829 > Services > Motion Pictures > Motion Picture Distribution And Allied Services > Services Allied to Motion Picture Distribution
7832 > Services > Motion Pictures > Motion Picture Theaters > Motion Picture Theaters, Except Drive-In
7833 > Services > Motion Pictures > Motion Picture Theaters > Drive-In Motion Picture Theaters
7841 > Services > Motion Pictures > Video Tape Rental > Video Tape Rental
7911 > Services > Amusement And Recreation Services > Dance Studios, Schools, And Halls > Dance Studios, Schools, and Halls
7922 > Services > Amusement And Recreation Services > Theatrical Producers (except Motion Picture), > Theatrical Producers (Except Motion Picture) and Miscellaneous Theatrical Services
7929 > Services > Amusement And Recreation Services > Theatrical Producers (except Motion Picture), > Bands, Orchestras, Actors, and Other Entertainers and Entertainment Groups
7933 > Services > Amusement And Recreation Services > Bowling Centers > Bowling Centers
7941 > Services > Amusement And Recreation Services > Commercial Sports > Professional Sports Clubs and Promoters
7948 > Services > Amusement And Recreation Services > Commercial Sports > Racing, Including Track Operation
7991 > Services > Amusement And Recreation Services > Miscellaneous Amusement And Recreation > Physical Fitness Facilities
7992 > Services > Amusement And Recreation Services > Miscellaneous Amusement And Recreation > Public Golf Courses
7993 > Services > Amusement And Recreation Services > Miscellaneous Amusement And Recreation > Coin-Operated Amusement Devices
7996 > Services > Amusement And Recreation Services > Miscellaneous Amusement And Recreation > Amusement Parks
7997 > Services > Amusement And Recreation Services > Miscellaneous Amusement And Recreation > Membership Sports and Recreation Clubs
7999 > Services > Amusement And Recreation Services > Miscellaneous Amusement And Recreation > Amusement and Recreation Services, Not Elsewhere Classified
8011 > Services > Health Services > Offices And Clinics Of Doctors Of Medicine > Offices and Clinics of Doctors of Medicine
8021 > Services > Health Services > Offices And Clinics Of Dentists > Offices and Clinics of Dentists
8031 > Services > Health Services > Offices And Clinics Of Doctors Of Osteopathy > Offices and Clinics of Doctors of Osteopathy
```

```
8041 > Services > Health Services > Offices And Clinics Of Other Health Practitioners > Offices and Clinics of Chiropractors
8042 > Services > Health Services > Offices And Clinics Of Other Health Practitioners > Offices and Clinics of Optometrists
8043 > Services > Health Services > Offices And Clinics Of Other Health Practitioners > Offices and Clinics of Podiatrists
8049 > Services > Health Services > Offices And Clinics Of Other Health Practitioners > Offices and Clinics of Health Practitioners, Not Elsewhere Classified
8051 > Services > Health Services > Nursing And Personal Care Facilities > Skilled Nursing Care Facilities
8052 > Services > Health Services > Nursing And Personal Care Facilities > Intermediate Care Facilities
8059 > Services > Health Services > Nursing And Personal Care Facilities > Nursing and Personal Care Facilities, Not Elsewhere Classified
8062 > Services > Health Services > Hospitals > General Medical and Surgical Hospitals
8063 > Services > Health Services > Hospitals > Psychiatric Hospitals
8069 > Services > Health Services > Hospitals > Specialty Hospitals, Except Psychiatric
8071 > Services > Health Services > Medical And Dental Laboratories > Medical Laboratories
8072 > Services > Health Services > Medical And Dental Laboratories > Dental Laboratories
8082 > Services > Health Services > Home Health Care Services > Home Health Care Services
8092 > Services > Health Services > Miscellaneous Health And Allied Services, Not > Kidney Dialysis Centers
8093 > Services > Health Services > Miscellaneous Health And Allied Services, Not > Specialty Outpatient Facilities, Not Elsewhere Classified
8099 > Services > Health Services > Miscellaneous Health And Allied Services, Not > Health and Allied Services, Not Elsewhere Classified
8111 > Services > Legal Services > Legal Services > Legal Services
8211 > Services > Educational Services > Elementary And Secondary Schools > Elementary and Secondary Schools
8221 > Services > Educational Services > Colleges, Universities, Professional Schools, And > Colleges, Universities, and Professional Schools
8222 > Services > Educational Services > Colleges, Universities, Professional Schools, And > Junior Colleges and Technical Institutes
8231 > Services > Educational Services > Libraries > Libraries
8243 > Services > Educational Services > Vocational Schools > Data Processing Schools
8244 > Services > Educational Services > Vocational Schools > Business and Secretarial Schools
8249 > Services > Educational Services > Vocational Schools > Vocational Schools, Not Elsewhere Classified
8299 > Services > Educational Services > Schools And Educational Services, Not Elsewhere > Schools and Educational Services, Not Elsewhere Classified
8322 > Services > Social Services > Individual And Family Social Services > Individual and Family Social Services
8331 > Services > Social Services > Job Training And Vocational Rehabilitation > Job Training and Vocational Rehabilitation Services
8351 > Services > Social Services > Child Day Care Services > Child Day Care Services
8361 > Services > Social Services > Residential Care > Residential Care
8399 > Services > Social Services > Social Services, Not Elsewhere Classified > Social Services, Not Elsewhere Classified
8412 > Services > Museums, Art Galleries, And Botanical And Zoological Gardens > Museums And Art Galleries > Museums and Art Galleries
8422 > Services > Museums, Art Galleries, And Botanical And Zoological Gardens > Arboreta And Botanical Or Zoological Gardens > Arboreta and Botanical or Zoological Gardens
8611 > Services > Membership Organizations > Business Associations > Business Associations
8621 > Services > Membership Organizations > Professional Membership Organizations > Professional Membership Organizations
8631 > Services > Membership Organizations > Labor Unions And Similar Labor Organizations > Labor Unions and Similar Labor Organizations
8641 > Services > Membership Organizations > Civic, Social, And Fraternal Associations > Civic, Social, and Fraternal Associations
8651 > Services > Membership Organizations > Political Organizations > Political Organizations
8661 > Services > Membership Organizations > Religious Organizations > Religious Organizations
8699 > Services > Membership Organizations > Membership Organizations, Not Elsewhere > Membership Organizations, Not Elsewhere Classified
8711 > Services > Engineering, Accounting, Research, Management, And Related Services > Engineering, Architectural, And Surveying > Engineering Services
```

8712 > Services > Engineering, Accounting, Research, Management, And Related Services > Engineering, Architectural, And Surveying > Architectural Services
8713 > Services > Engineering, Accounting, Research, Management, And Related Services > Engineering, Architectural, And Surveying > Surveying Services
8721 > Services > Engineering, Accounting, Research, Management, And Related Services > Accounting, Auditing, And Bookkeeping Services > Accounting, Auditing, and Bookkeeping Services
8731 > Services > Engineering, Accounting, Research, Management, And Related Services > Research, Development, And Testing Services > Commercial Physical and Biological Research
8732 > Services > Engineering, Accounting, Research, Management, And Related Services > Research, Development, And Testing Services > Commercial Economic, Sociological, and Educational Research
8733 > Services > Engineering, Accounting, Research, Management, And Related Services > Research, Development, And Testing Services > Noncommercial Research Organizations
8734 > Services > Engineering, Accounting, Research, Management, And Related Services > Research, Development, And Testing Services > Testing Laboratories
8741 > Services > Engineering, Accounting, Research, Management, And Related Services > Management And Public Relations Services > Management Services
8742 > Services > Engineering, Accounting, Research, Management, And Related Services > Management And Public Relations Services > Management Consulting Services
8743 > Services > Engineering, Accounting, Research, Management, And Related Services > Management And Public Relations Services > Public Relations Services
8744 > Services > Engineering, Accounting, Research, Management, And Related Services > Management And Public Relations Services > Facilities Support Management Services
8748 > Services > Engineering, Accounting, Research, Management, And Related Services > Management And Public Relations Services > Business Consulting Services, Not Elsewhere Classified
8811 > Services > Private Households > Private Households > Private Households
8999 > Services > Miscellaneous Services > Miscellaneous Services > Services, Not Elsewhere Classified
9111 > Public Administration > Executive, Legislative, And General Government, Except Finance > Executive Offices > Executive Offices
9121 > Public Administration > Executive, Legislative, And General Government, Except Finance > Legislative Bodies > Legislative Bodies
9131 > Public Administration > Executive, Legislative, And General Government, Except Finance > Executive And Legislative Offices Combined > Executive and Legislative Offices Combined
9199 > Public Administration > Executive, Legislative, And General Government, Except Finance > General Government, Not Elsewhere Classified > General Government, Not Elsewhere Classified
9211 > Public Administration > Justice, Public Order, And Safety > Courts > Courts
9221 > Public Administration > Justice, Public Order, And Safety > Public Order And Safety > Police Protection
9222 > Public Administration > Justice, Public Order, And Safety > Public Order And Safety > Legal Counsel and Prosecution
9223 > Public Administration > Justice, Public Order, And Safety > Public Order And Safety > Correctional Institutions
9224 > Public Administration > Justice, Public Order, And Safety > Public Order And Safety > Fire Protection
9229 > Public Administration > Justice, Public Order, And Safety > Public Order And Safety > Public Order and Safety, Not Elsewhere Classified
9311 > Public Administration > Public Finance, Taxation, And Monetary Policy > Public Finance, Taxation, And Monetary Policy > Public Finance, Taxation, and Monetary Policy
9411 > Public Administration > Administration Of Human Resource Programs > Administration Of Educational Programs > Administration of Educational Programs
9431 > Public Administration > Administration Of Human Resource Programs > Administration Of Public Health Programs > Administration of Public Health Programs
9441 > Public Administration > Administration Of Human Resource Programs > Administration Of Social, Human Resource And > Administration of Social, Human Resource and Income Maintenance Programs
9451 > Public Administration > Administration Of Human Resource Programs > Administration Of Veteran's Affairs, Except > Administration of Veterans' Affairs, Except Health and Insurance 9511 > Public Administration > Administration Of Environmental Quality And Housing Programs > Administration Of Environmental Quality > Air and Water Resource and Solid Waste Management
9512 > Public Administration > Administration Of Environmental Quality And Housing Programs > Administration Of Environmental Quality > Land, Mineral, Wildlife, and Forest Conservation
9531 > Public Administration > Administration Of Environmental Quality And Housing Programs > Administration Of Housing And Urban > Administration of Housing Programs
9532 > Public Administration > Administration Of Environmental Quality And Housing Programs > Administration Of Housing And Urban > Administration of Urban Planning and Community and Rural Development
9611 > Public Administration > Administration Of Economic Programs > Administration Of General Economic Programs > Administration of General Economic Programs
9621 > Public Administration > Administration Of Economic Programs > Regulation And Administration Of Transportation > Regulation and Administration of Transportation Programs
9631 > Public Administration > Administration Of Economic Programs > Regulation And Administration Of > Regulation and Administration of Communications, Electric, Gas, and Other Utilities
9641 > Public Administration > Administration Of Economic Programs > Regulation Of Agricultural Marketing And > Regulation of Agricultural Marketing and Commodities
9651 > Public Administration > Administration Of Economic Programs > Regulation, Licensing, And Inspection Of > Regulation, Licensing, and Inspection of Miscellaneous Commercial Sectors
9661 > Public Administration > Administration Of Economic Programs > Space Research And Technology > Space and Research and Technology
9711 > Public Administration > National Security And International Affairs > National Security > National Security
9721 > Public Administration > National Security And International Affairs > International Affairs > International Affairs
*9995 > Public Administration > Nonclassifiable Establishments > Nonclassifiable Establishments > Non-operating Establishments (SEC)*
9999 > Public Administration > Nonclassifiable Establishments > Nonclassifiable Establishments > Nonclassifiable Establishments
*0888 > UNKNOWN SIC – 0888 (SEC)*
*8880 > AMERICAN DEPOSITARY RECEIPTS (SEC)*
*8888 > FOREIGN GOVERNMENTS (SEC)*

# Appendix B: SEC SIC Codes and Industry Names

```
0100    AGRICULTURAL PRODUCTION-CROPS
0200    AGRICULTURAL PROD-LIVESTOCK & ANIMAL SPECIALTIES
0700    AGRICULTURAL SERVICES
0800    FORESTRY
0888    UNKNOWN SIC - 0888
0900    FISHING, HUNTING AND TRAPPING
1000    METAL MINING
1040    GOLD AND SILVER ORES
1044    SILVER ORES
1090    MISCELLANEOUS METAL ORES
1220    BITUMINOUS COAL & LIGNITE MINING
1221    BITUMINOUS COAL & LIGNITE SURFACE MINING
1311    CRUDE PETROLEUM & NATURAL GAS
1381    DRILLING OIL & GAS WELLS
1382    OIL & GAS FIELD EXPLORATION SERVICES
1389    OIL & GAS FIELD SERVICES, NEC
1400    MINING & QUARRYING OF NONMETALLIC MINERALS (NO FUELS)
1520    GENERAL BLDG CONTRACTORS - RESIDENTIAL BLDGS
1531    OPERATIVE BUILDERS
1540    GENERAL BLDG CONTRACTORS - NONRESIDENTIAL BLDGS
1600    HEAVY CONSTRUCTION OTHER THAN BLDG CONST - CONTRACTORS
1623    WATER, SEWER, PIPELINE, COMM & POWER LINE CONSTRUCTION
1700    CONSTRUCTION - SPECIAL TRADE CONTRACTORS
1731    ELECTRICAL WORK
2000    FOOD AND KINDRED PRODUCTS
2011    MEAT PACKING PLANTS
2013    SAUSAGES & OTHER PREPARED MEAT PRODUCTS
2015    POULTRY SLAUGHTERING AND PROCESSING
2020    DAIRY PRODUCTS
2024    ICE CREAM & FROZEN DESSERTS
2030    CANNED, FROZEN & PRESERVD FRUIT, VEG & FOOD SPECIALTIES
2033    CANNED, FRUITS, VEG, PRESERVES, JAMS & JELLIES
2040    GRAIN MILL PRODUCTS
2050    BAKERY PRODUCTS
2052    COOKIES & CRACKERS
2060    SUGAR & CONFECTIONERY PRODUCTS
2070    FATS & OILS
2080    BEVERAGES
2082    MALT BEVERAGES
2086    BOTTLED & CANNED SOFT DRINKS & CARBONATED WATERS
2090    MISCELLANEOUS FOOD PREPARATIONS & KINDRED PRODUCTS
2092    PREPARED FRESH OR FROZEN FISH & SEAFOODS
2100    TOBACCO PRODUCTS
2111    CIGARETTES
2200    TEXTILE MILL PRODUCTS
2211    BROADWOVEN FABRIC MILLS, COTTON
2221    BROADWOVEN FABRIC MILLS, MAN MADE FIBER & SILK
2250    KNITTING MILLS
2253    KNIT OUTERWEAR MILLS
2273    CARPETS & RUGS
2300    APPAREL & OTHER FINISHD PRODS OF FABRICS & SIMILAR MATL
2320    MEN'S & BOYS' FURNISHGS, WORK CLOTHG, & ALLIED GARMENTS
2330    WOMEN'S, MISSES', AND JUNIORS OUTERWEAR
2340    WOMEN'S, MISSES', CHILDREN'S & INFANTS' UNDERGARMENTS
2390    MISCELLANEOUS FABRICATED TEXTILE PRODUCTS
2400    LUMBER & WOOD PRODUCTS (NO FURNITURE)
2421    SAWMILLS & PLANTING MILLS, GENERAL
2430    MILLWOOD, VENEER, PLYWOOD, & STRUCTURAL WOOD MEMBERS
2451    MOBILE HOMES
2452    PREFABRICATED WOOD BLDGS & COMPONENTS
2510    HOUSEHOLD FURNITURE
2511    WOOD HOUSEHOLD FURNITURE, (NO UPHOLSTERED)
2520    OFFICE FURNITURE
2522    OFFICE FURNITURE (NO WOOD)
2531    PUBLIC BLDG & RELATED FURNITURE
2540    PARTITIONS, SHELVG, LOCKERS, & OFFICE & STORE FIXTURES
2590    MISCELLANEOUS FURNITURE & FIXTURES
2600    PAPERS & ALLIED PRODUCTS
```

```
2611    PULP MILLS
2621    PAPER MILLS
2631    PAPERBOARD MILLS
2650    PAPERBOARD CONTAINERS & BOXES
2670    CONVERTED PAPER & PAPERBOARD PRODS (NO CONTANERS/BOXES)
2673    PLASTICS, FOIL & COATED PAPER BAGS
2711    NEWSPAPERS: PUBLISHING OR PUBLISHING & PRINTING
2721    PERIODICALS: PUBLISHING OR PUBLISHING & PRINTING
2731    BOOKS: PUBLISHING OR PUBLISHING & PRINTING
2732    BOOK PRINTING
2741    MISCELLANEOUS PUBLISHING
2750    COMMERCIAL PRINTING
2761    MANIFOLD BUSINESS FORMS
2771    GREETING CARDS
2780    BLANKBOOKS, LOOSELEAF BINDERS & BOOKBINDG & RELATD WORK
2790    SERVICE INDUSTRIES FOR THE PRINTING TRADE
2800    CHEMICALS & ALLIED PRODUCTS
2810    INDUSTRIAL INORGANIC CHEMICALS
2820    PLASTIC MATERIAL, SYNTH RESIN/RUBBER, CELLULOS (NO GLASS)
2821    PLASTIC MATERIALS, SYNTH RESINS & NONVULCAN ELASTOMERS
2833    MEDICINAL CHEMICALS & BOTANICAL PRODUCTS
2834    PHARMACEUTICAL PREPARATIONS
2835    IN VITRO & IN VIVO DIAGNOSTIC SUBSTANCES
2836    BIOLOGICAL PRODUCTS, (NO DISGNOSTIC SUBSTANCES)
2840    SOAP, DETERGENTS, CLEANG PREPARATIONS, PERFUMES, COSMETICS
2842    SPECIALTY CLEANING, POLISHING AND SANITATION PREPARATIONS
2844    PERFUMES, COSMETICS & OTHER TOILET PREPARATIONS
2851    PAINTS, VARNISHES, LACQUERS, ENAMELS & ALLIED PRODS
2860    INDUSTRIAL ORGANIC CHEMICALS
2870    AGRICULTURAL CHEMICALS
2890    MISCELLANEOUS CHEMICAL PRODUCTS
2891    ADHESIVES & SEALANTS
2911    PETROLEUM REFINING
2950    ASPHALT PAVING & ROOFING MATERIALS
2990    MISCELLANEOUS PRODUCTS OF PETROLEUM & COAL
3011    TIRES & INNER TUBES
3021    RUBBER & PLASTICS FOOTWEAR
3050    GASKETS, PACKG & SEALG DEVICES & RUBBER & PLASTICS HOSE
3060    FABRICATED RUBBER PRODUCTS, NEC
3080    MISCELLANEOUS PLASTICS PRODUCTS
3081    UNSUPPORTED PLASTICS FILM & SHEET
3086    PLASTICS FOAM PRODUCTS
3089    PLASTICS PRODUCTS, NEC
3100    LEATHER & LEATHER PRODUCTS
3140    FOOTWEAR, (NO RUBBER)
3211    FLAT GLASS
3220    GLASS & GLASSWARE, PRESSED OR BLOWN
3221    GLASS CONTAINERS
3231    GLASS PRODUCTS, MADE OF PURCHASED GLASS
3241    CEMENT, HYDRAULIC
3250    STRUCTURAL CLAY PRODUCTS
3260    POTTERY & RELATED PRODUCTS
3270    CONCRETE, GYPSUM & PLASTER PRODUCTS
3272    CONCRETE PRODUCTS, EXCEPT BLOCK & BRICK
3281    CUT STONE & STONE PRODUCTS
3290    ABRASIVE, ASBESTOS & MISC NONMETALLIC MINERAL PRODS
3310    STEEL WORKS, BLAST FURNACES & ROLLING & FINISHING MILLS
3312    STEEL WORKS, BLAST FURNACES & ROLLING MILLS (COKE OVENS)
3317    STEEL PIPE & TUBES
3320    IRON & STEEL FOUNDRIES
3330    PRIMARY SMELTING & REFINING OF NONFERROUS METALS
3334    PRIMARY PRODUCTION OF ALUMINUM
3341    SECONDARY SMELTING & REFINING OF NONFERROUS METALS
3350    ROLLING DRAWING & EXTRUDING OF NONFERROUS METALS
3357    DRAWING & INSULATING OF NONFERROUS WIRE
3360    NONFERROUS FOUNDRIES (CASTINGS)
3390    MISCELLANEOUS PRIMARY METAL PRODUCTS
3411    METAL CANS
3412    METAL SHIPPING BARRELS, DRUMS, KEGS & PAILS
3420    CUTLERY, HANDTOOLS & GENERAL HARDWARE
3430    HEATING EQUIP, EXCEPT ELEC & WARM AIR; & PLUMBING FIXTURES
```

```
3433     HEATING EQUIPMENT, EXCEPT ELECTRIC & WARM AIR FURNACES
3440     FABRICATED STRUCTURAL METAL PRODUCTS
3442     METAL DOORS, SASH, FRAMES, MOLDINGS & TRIM
3443     FABRICATED PLATE WORK (BOILER SHOPS)
3444     SHEET METAL WORK
3448     PREFABRICATED METAL BUILDINGS & COMPONENTS
3451     SCREW MACHINE PRODUCTS
3452     BOLTS, NUTS, SCREWS, RIVETS & WASHERS
3460     METAL FORGINGS & STAMPINGS
3470     COATING, ENGRAVING & ALLIED SERVICES
3480     ORDNANCE & ACCESSORIES, (NO VEHICLES/GUIDED MISSILES)
3490     MISCELLANEOUS FABRICATED METAL PRODUCTS
3510     ENGINES & TURBINES
3523     FARM MACHINERY & EQUIPMENT
3524     LAWN & GARDEN TRACTORS & HOME LAWN & GARDENS EQUIP
3530     CONSTRUCTION, MINING & MATERIALS HANDLING MACHINERY & EQUIP
3531     CONSTRUCTION MACHINERY & EQUIP
3532     MINING MACHINERY & EQUIP (NO OIL & GAS FIELD MACH & EQUIP)
3533     OIL & GAS FIELD MACHINERY & EQUIPMENT
3537     INDUSTRIAL TRUCKS, TRACTORS, TRAILORS & STACKERS
3540     METALWORKG MACHINERY & EQUIPMENT
3541     MACHINE TOOLS, METAL CUTTING TYPES
3550     SPECIAL INDUSTRY MACHINERY (NO METALWORKING MACHINERY)
3555     PRINTING TRADES MACHINERY & EQUIPMENT
3559     SPECIAL INDUSTRY MACHINERY, NEC
3560     GENERAL INDUSTRIAL MACHINERY & EQUIPMENT
3561     PUMPS & PUMPING EQUIPMENT
3562     BALL & ROLLER BEARINGS
3564     INDUSTRIAL & COMMERCIAL FANS & BLOWERS & AIR PURIFING EQUIP
3567     INDUSTRIAL PROCESS FURNACES & OVENS
3569     GENERAL INDUSTRIAL MACHINERY & EQUIPMENT, NEC
3570     COMPUTER & OFFICE EQUIPMENT
3571     ELECTRONIC COMPUTERS
3572     COMPUTER STORAGE DEVICES
3575     COMPUTER TERMINALS
3576     COMPUTER COMMUNICATIONS EQUIPMENT
3577     COMPUTER PERIPHERAL EQUIPMENT, NEC
3578     CALCULATING & ACCOUNTING MACHINES (NO ELECTRONIC COMPUTERS)
3579     OFFICE MACHINES, NEC
3580     REFRIGERATION & SERVICE INDUSTRY MACHINERY
3585     AIR-COND & WARM AIR HEATG EQUIP & COMM & INDL REFRIG EQUIP
3590     MISC INDUSTRIAL & COMMERCIAL MACHINERY & EQUIPMENT
3600     ELECTRONIC & OTHER ELECTRICAL EQUIPMENT (NO COMPUTER EQUIP)
3612     POWER, DISTRIBUTION & SPECIALTY TRANSFORMERS
3613     SWITCHGEAR & SWITCHBOARD APPARATUS
3620     ELECTRICAL INDUSTRIAL APPARATUS
3621     MOTORS & GENERATORS
3630     HOUSEHOLD APPLIANCES
3634     ELECTRIC HOUSEWARES & FANS
3640     ELECTRIC LIGHTING & WIRING EQUIPMENT
3651     HOUSEHOLD AUDIO & VIDEO EQUIPMENT
3652     PHONOGRAPH RECORDS & PRERECORDED AUDIO TAPES & DISKS
3661     TELEPHONE & TELEGRAPH APPARATUS
3663     RADIO & TV BROADCASTING & COMMUNICATIONS EQUIPMENT
3669     COMMUNICATIONS EQUIPMENT, NEC
3670     ELECTRONIC COMPONENTS & ACCESSORIES
3672     PRINTED CIRCUIT BOARDS
3674     SEMICONDUCTORS & RELATED DEVICES
3677     ELECTRONIC COILS, TRANSFORMERS & OTHER INDUCTORS
3678     ELECTRONIC CONNECTORS
3679     ELECTRONIC COMPONENTS, NEC
3690     MISCELLANEOUS ELECTRICAL MACHINERY, EQUIPMENT & SUPPLIES
3695     MAGNETIC & OPTICAL RECORDING MEDIA
3711     MOTOR VEHICLES & PASSENGER CAR BODIES
3713     TRUCK & BUS BODIES
3714     MOTOR VEHICLE PARTS & ACCESSORIES
3715     TRUCK TRAILERS
3716     MOTOR HOMES
3720     AIRCRAFT & PARTS
3721     AIRCRAFT
3724     AIRCRAFT ENGINES & ENGINE PARTS
```

```
3728    AIRCRAFT PARTS & AUXILIARY EQUIPMENT, NEC
3730    SHIP & BOAT BUILDING & REPAIRING
3743    RAILROAD EQUIPMENT
3751    MOTORCYCLES, BICYCLES & PARTS
3760    GUIDED MISSILES & SPACE VEHICLES & PARTS
3790    MISCELLANEOUS TRANSPORTATION EQUIPMENT
3812    SEARCH, DETECTION, NAVAGATION, GUIDANCE, AERONAUTICAL SYS
3821    LABORATORY APPARATUS & FURNITURE
3822    AUTO CONTROLS FOR REGULATING RESIDENTIAL & COMML ENVIRONMENTS
3823    INDUSTRIAL INSTRUMENTS FOR MEASUREMENT, DISPLAY, AND CONTROL
3824    TOTALIZING FLUID METERS & COUNTING DEVICES
3825    INSTRUMENTS FOR MEAS & TESTING OF ELECTRICITY & ELEC SIGNALS
3826    LABORATORY ANALYTICAL INSTRUMENTS
3827    OPTICAL INSTRUMENTS & LENSES
3829    MEASURING & CONTROLLING DEVICES, NEC
3841    SURGICAL & MEDICAL INSTRUMENTS & APPARATUS
3842    ORTHOPEDIC, PROSTHETIC & SURGICAL APPLIANCES & SUPPLIES
3843    DENTAL EQUIPMENT & SUPPLIES
3844    X-RAY APPARATUS & TUBES & RELATED IRRADIATION APPARATUS
3845    ELECTROMEDICAL & ELECTROTHERAPEUTIC APPARATUS
3851    OPHTHALMIC GOODS
3861    PHOTOGRAPHIC EQUIPMENT & SUPPLIES
3873    WATCHES, CLOCKS, CLOCKWORK OPERATED DEVICES/PARTS
3910    JEWELRY, SILVERWARE & PLATED WARE
3911    JEWELRY, PRECIOUS METAL
3931    MUSICAL INSTRUMENTS
3942    DOLLS & STUFFED TOYS
3944    GAMES, TOYS & CHILDREN'S VEHICLES (NO DOLLS & BICYCLES)
3949    SPORTING & ATHLETIC GOODS, NEC
3950    PENS, PENCILS & OTHER ARTISTS' MATERIALS
3960    COSTUME JEWELRY & NOVELTIES
3990    MISCELLANEOUS MANUFACTURING INDUSTRIES
4011    RAILROADS, LINE-HAUL OPERATING
4013    RAILROAD SWITCHING & TERMINAL ESTABLISHMENTS
4100    LOCAL & SUBURBAN TRANSIT & INTERURBAN HWY PASSENGER TRANS
4210    TRUCKING & COURIER SERVICES (NO AIR)
4213    TRUCKING (NO LOCAL)
4220    PUBLIC WAREHOUSING & STORAGE
4231    TERMINAL MAINTENANCE FACILITIES FOR MOTOR FREIGHT TRANSPORT
4400    WATER TRANSPORTATION
4412    DEEP SEA FOREIGN TRANSPORTATION OF FREIGHT
4512    AIR TRANSPORTATION, SCHEDULED
4513    AIR COURIER SERVICES
4522    AIR TRANSPORTATION, NONSCHEDULED
4581    AIRPORTS, FLYING FIELDS & AIRPORT TERMINAL SERVICES
4610    PIPE LINES (NO NATURAL GAS)
4700    TRANSPORTATION SERVICES
4731    ARRANGEMENT OF TRANSPORTATION OF FREIGHT & CARGO
4812    RADIOTELEPHONE COMMUNICATIONS
4813    TELEPHONE COMMUNICATIONS (NO RADIOTELEPHONE)
4822    TELEGRAPH & OTHER MESSAGE COMMUNICATIONS
4832    RADIO BROADCASTING STATIONS
4833    TELEVISION BROADCASTING STATIONS
4841    CABLE & OTHER PAY TELEVISION SERVICES
4899    COMMUNICATIONS SERVICES, NEC
4900    ELECTRIC, GAS & SANITARY SERVICES
4911    ELECTRIC SERVICES
4922    NATURAL GAS TRANSMISSION
4923    NATURAL GAS TRANSMISISON & DISTRIBUTION
4924    NATURAL GAS DISTRIBUTION
4931    ELECTRIC & OTHER SERVICES COMBINED
4932    GAS & OTHER SERVICES COMBINED
4941    WATER SUPPLY
4950    SANITARY SERVICES
4953    REFUSE SYSTEMS
4955    HAZARDOUS WASTE MANAGEMENT
4961    STEAM & AIR-CONDITIONING SUPPLY
4991    COGENERATION SERVICES & SMALL POWER PRODUCERS
5000    WHOLESALE-DURABLE GOODS
5010    WHOLESALE-MOTOR VEHICLES & MOTOR VEHICLE PARTS & SUPPLIES
5013    WHOLESALE-MOTOR VEHICLE SUPPLIES & NEW PARTS
```

```
5020    WHOLESALE-FURNITURE & HOME FURNISHINGS
5030    WHOLESALE-LUMBER & OTHER CONSTRUCTION MATERIALS
5031    WHOLESALE-LUMBER, PLYWOOD, MILLWORK & WOOD PANELS
5040    WHOLESALE-PROFESSIONAL & COMMERCIAL EQUIPMENT & SUPPLIES
5045    WHOLESALE-COMPUTERS & PERIPHERAL EQUIPMENT & SOFTWARE
5047    WHOLESALE-MEDICAL, DENTAL & HOSPITAL EQUIPMENT & SUPPLIES
5050    WHOLESALE-METALS & MINERALS (NO PETROLEUM)
5051    WHOLESALE-METALS SERVICE CENTERS & OFFICES
5063    WHOLESALE-ELECTRICAL APPARATUS & EQUIPMENT, WIRING SUPPLIES
5064    WHOLESALE-ELECTRICAL APPLIANCES, TV & RADIO SETS
5065    WHOLESALE-ELECTRONIC PARTS & EQUIPMENT, NEC
5070    WHOLESALE-HARDWARE & PLUMBING & HEATING EQUIPMENT & SUPPLIES
5072    WHOLESALE-HARDWARE
5080    WHOLESALE-MACHINERY, EQUIPMENT & SUPPLIES
5082    WHOLESALE-CONSTRUCTION & MINING (NO PETRO) MACHINERY & EQUIP
5084    WHOLESALE-INDUSTRIAL MACHINERY & EQUIPMENT
5090    WHOLESALE-MISC DURABLE GOODS
5094    WHOLESALE-JEWELRY, WATCHES, PRECIOUS STONES & METALS
5099    WHOLESALE-DURABLE GOODS, NEC
5110    WHOLESALE-PAPER & PAPER PRODUCTS
5122    WHOLESALE-DRUGS, PROPRIETARIES & DRUGGISTS' SUNDRIES
5130    WHOLESALE-APPAREL, PIECE GOODS & NOTIONS
5140    WHOLESALE-GROCERIES & RELATED PRODUCTS
5141    WHOLESALE-GROCERIES, GENERAL LINE
5150    WHOLESALE-FARM PRODUCT RAW MATERIALS
5160    WHOLESALE-CHEMICALS & ALLIED PRODUCTS
5171    WHOLESALE-PETROLEUM BULK STATIONS & TERMINALS
5172    WHOLESALE-PETROLEUM & PETROLEUM PRODUCTS (NO BULK STATIONS)
5180    WHOLESALE-BEER, WINE & DISTILLED ALCOHOLIC BEVERAGES
5190    WHOLESALE-MISCELLANEOUS NONDURABLE GOODS
5200    RETAIL-BUILDING MATERIALS, HARDWARE, GARDEN SUPPLY
5211    RETAIL-LUMBER & OTHER BUILDING MATERIALS DEALERS
5271    RETAIL-MOBILE HOME DEALERS
5311    RETAIL-DEPARTMENT STORES
5331    RETAIL-VARIETY STORES
5399    RETAIL-MISC GENERAL MERCHANDISE STORES
5400    RETAIL-FOOD STORES
5411    RETAIL-GROCERY STORES
5412    RETAIL-CONVENIENCE STORES
5500    RETAIL-AUTO DEALERS & GASOLINE STATIONS
5531    RETAIL-AUTO & HOME SUPPLY STORES
5600    RETAIL-APPAREL & ACCESSORY STORES
5621    RETAIL-WOMEN'S CLOTHING STORES
5651    RETAIL-FAMILY CLOTHING STORES
5661    RETAIL-SHOE STORES
5700    RETAIL-HOME FURNITURE, FURNISHINGS & EQUIPMENT STORES
5712    RETAIL-FURNITURE STORES
5731    RETAIL-RADIO, TV & CONSUMER ELECTRONICS STORES
5734    RETAIL-COMPUTER & COMPUTER SOFTWARE STORES
5735    RETAIL-RECORD & PRERECORDED TAPE STORES
5810    RETAIL-EATING & DRINKING PLACES
5812    RETAIL-EATING  PLACES
5900    RETAIL-MISCELLANEOUS RETAIL
5912    RETAIL-DRUG STORES AND PROPRIETARY STORES
5940    RETAIL-MISCELLANEOUS SHOPPING GOODS STORES
5944    RETAIL-JEWELRY STORES
5945    RETAIL-HOBBY, TOY & GAME SHOPS
5960    RETAIL-NONSTORE RETAILERS
5961    RETAIL-CATALOG & MAIL-ORDER HOUSES
5990    RETAIL-RETAIL STORES, NEC
6021    NATIONAL COMMERCIAL BANKS
6022    STATE COMMERCIAL BANKS
6025    NATIONAL BANKS-FEDERAL RESERVE SYSTEM
6029    COMMERCIAL BANKS, NEC
6035    SAVINGS INSTITUTION, FEDERALLY CHARTERED
6036    SAVINGS INSTITUTIONS, NOT FEDERALLY CHARTERED
6099    FUNCTIONS RELATED TO DEPOSITORY BANKING, NEC
6111    FEDERAL & FEDERALLY-SPONSORED CREDIT AGENCIES
6120    SAVINGS & LOAN ASSOCIATIONS
6141    PERSONAL CREDIT INSTITUTIONS
6153    SHORT-TERM BUSINESS CREDIT INSTITUTIONS
```

```
6159    MISCELLANEOUS BUSINESS CREDIT INSTITUTION
6162    MORTGAGE BANKERS & LOAN CORRESPONDENTS
6163    LOAN BROKERS
6172    FINANCE LESSORS
6189    ASSET-BACKED SECURITIES
6199    FINANCE SERVICES
6200    SECURITY & COMMODITY BROKERS, DEALERS, EXCHANGES & SERVICES
6211    SECURITY BROKERS, DEALERS & FLOTATION COMPANIES
6221    COMMODITY CONTRACTS BROKERS & DEALERS
6282    INVESTMENT ADVICE
6311    LIFE INSURANCE
6321    ACCIDENT & HEALTH INSURANCE
6324    HOSPITAL & MEDICAL SERVICE PLANS
6331    FIRE, MARINE & CASUALTY INSURANCE
6351    SURETY INSURANCE
6361    TITLE INSURANCE
6399    INSURANCE CARRIERS, NEC
6411    INSURANCE AGENTS, BROKERS & SERVICE
6500    REAL ESTATE
6510    REAL ESTATE OPERATORS (NO DEVELOPERS) & LESSORS
6512    OPERATORS OF NONRESIDENTIAL BUILDINGS
6513    OPERATORS OF APARTMENT BUILDINGS
6519    LESSORS OF REAL PROPERTY, NEC
6531    REAL ESTATE AGENTS & MANAGERS (FOR OTHERS)
6532    REAL ESTATE DEALERS (FOR THEIR OWN ACCOUNT)
6552    LAND SUBDIVIDERS & DEVELOPERS (NO CEMETERIES)
6770    BLANK CHECKS
6792    OIL ROYALTY TRADERS
6794    PATENT OWNERS & LESSORS
6795    MINERAL ROYALTY TRADERS
6798    REAL ESTATE INVESTMENT TRUSTS
6799    INVESTORS, NEC
7000    HOTELS, ROOMING HOUSES, CAMPS & OTHER LODGING PLACES
7011    HOTELS & MOTELS
7200    SERVICES-PERSONAL SERVICES
7310    SERVICES-ADVERTISING
7311    SERVICES-ADVERTISING AGENCIES
7320    SERVICES-CONSUMER CREDIT REPORTING, COLLECTION AGENCIES
7330    SERVICES-MAILING, REPRODUCTION, COMMERCIAL ART & PHOTOGRAPHY
7331    SERVICES-DIRECT MAIL ADVERTISING SERVICES
7340    SERVICES-TO DWELLINGS & OTHER BUILDINGS
7350    SERVICES-MISCELLANEOUS EQUIPMENT RENTAL & LEASING
7359    SERVICES-EQUIPMENT RENTAL & LEASING, NEC
7361    SERVICES-EMPLOYMENT AGENCIES
7363    SERVICES-HELP SUPPLY SERVICES
7370    SERVICES-COMPUTER PROGRAMMING, DATA PROCESSING, ETC.
7371    SERVICES-COMPUTER PROGRAMMING SERVICES
7372    SERVICES-PREPACKAGED SOFTWARE
7373    SERVICES-COMPUTER INTEGRATED SYSTEMS DESIGN
7374    SERVICES-COMPUTER PROCESSING & DATA PREPARATION
7377    SERVICES-COMPUTER RENTAL & LEASING
7380    SERVICES-MISCELLANEOUS BUSINESS SERVICES
7381    SERVICES-DETECTIVE, GUARD & ARMORED CAR SERVICES
7384    SERVICES-PHOTOFINISHING LABORATORIES
7385    SERVICES-TELEPHONE INTERCONNECT SYSTEMS
7389    SERVICES-BUSINESS SERVICES, NEC
7500    SERVICES-AUTOMOTIVE REPAIR, SERVICES & PARKING
7510    SERVICES-AUTO RENTAL & LEASING (NO DRIVERS)
7600    SERVICES-MISCELLANEOUS REPAIR SERVICES
7812    SERVICES-MOTION PICTURE & VIDEO TAPE PRODUCTION
7819    SERVICES-ALLIED TO MOTION PICTURE PRODUCTION
7822    SERVICES-MOTION PICTURE & VIDEO TAPE DISTRIBUTION
7829    SERVICES-ALLIED TO MOTION PICTURE DISTRIBUTION
7830    SERVICES-MOTION PICTURE THEATERS
7841    SERVICES-VIDEO TAPE RENTAL
7900    SERVICES-AMUSEMENT & RECREATION SERVICES
7948    SERVICES-RACING, INCLUDING TRACK OPERATION
7990    SERVICES-MISCELLANEOUS AMUSEMENT & RECREATION
7997    SERVICES-MEMBERSHIP SPORTS & RECREATION CLUBS
8000    SERVICES-HEALTH SERVICES
8011    SERVICES-OFFICES & CLINICS OF DOCTORS OF MEDICINE
```

```
8050    SERVICES-NURSING & PERSONAL CARE FACILITIES
8051    SERVICES-SKILLED NURSING CARE FACILITIES
8060    SERVICES-HOSPITALS
8062    SERVICES-GENERAL MEDICAL & SURGICAL HOSPITALS, NEC
8071    SERVICES-MEDICAL LABORATORIES
8082    SERVICES-HOME HEALTH CARE SERVICES
8090    SERVICES-MISC HEALTH & ALLIED SERVICES, NEC
8093    SERVICES-SPECIALTY OUTPATIENT FACILITIES, NEC
8111    SERVICES-LEGAL SERVICES
8200    SERVICES-EDUCATIONAL SERVICES
8300    SERVICES-SOCIAL SERVICES
8351    SERVICES-CHILD DAY CARE SERVICES
8600    SERVICES-MEMBERSHIP ORGANIZATIONS
8700    SERVICES-ENGINEERING, ACCOUNTING, RESEARCH, MANAGEMENT
8711    SERVICES-ENGINEERING SERVICES
8731    SERVICES-COMMERCIAL PHYSICAL & BIOLOGICAL RESEARCH
8734    SERVICES-TESTING LABORATORIES
8741    SERVICES-MANAGEMENT SERVICES
8742    SERVICES-MANAGEMENT CONSULTING SERVICES
8744    SERVICES-FACILITIES SUPPORT MANAGEMENT SERVICES
8880    AMERICAN DEPOSITARY RECEIPTS
8888    FOREIGN GOVERNMENTS
8900    SERVICES-SERVICES, NEC
9721    INTERNATIONAL AFFAIRS
9995    NON-OPERATING ESTABLISHMENTS
```

Note: NEC = Not Elsewhere Classified

## Appendix C: R Source Code

```
sec.all.sic <- function (run.all.sic = F, incl.otc = F)
{
    require(XML)
    require(httr)
    strip.white <- function(x)
            gsub("^ *|(?<= ) | *$", "", x, perl = TRUE)

    curr.time <- Sys.time()
    curr.time <- gsub(" ", "\\.", curr.time)
    curr.time <- gsub(":", "\\.", curr.time)
    log.file <- paste("log.", curr.time, ".txt", sep = "")
    write(as.character(Sys.time()), log.file, append = F)
    if(run.all.sic | !file.exists("SIC.Codes.txt"))
            try.sic <- 100:9999
    else
    {
            try.sic <- read.delim("SIC.Codes.txt", header = F, as.is = T)
            try.sic <- as.matrix(try.sic)
            try.sic <- as.numeric(try.sic[, 1])
    }
    take <- try.sic < 1000
    try.sic[take] <- paste(0, try.sic[take], sep = "")
    mat <- c("CIK", "Name", "SIC", "Industry", "Location")
    for(ix in try.sic)
    {
            get.pg <- T
            print(paste("Processing ", ix, sep = ""))
            s <- 0
            while(get.pg)
            {
                    url <- paste(
```

```
"https://www.sec.gov/cgi-bin/browse-edgar?action=getcompany&SIC=",
ix, "&owner=include&match=&start=",
s, "&count=100", "&hidefilings=0", sep = "")

            x <- sec.get.webpage(url)
            y <- sec.parse.html(x, keyword = "div")
            y1 <- grep(paste("SIC ", ix, sep = ""), y)
            y2 <- grep(paste("SIC: ", ix, sep = ""), y)
            if(length(y1) > 0)
            {
                y <- y[y1]
                y <- unlist(strsplit(y, paste("SIC ", ix, " - ",
                    sep = "")))[2]
                ind <- unlist(strsplit(y, " Click on CIK"))[1]
                x <- sec.parse.html(x, keyword = "table")
                x <- x[-(1:3)]
                cik <- !is.na(as.numeric(x))
                z <- matrix("", sum(cik), 5)
                j <- 0
                for(i in 1:length(x))
                {
                    if(cik[i])
                        z[j <- j + 1, 1] <- x[k <- i]
                    else if(i == k + 1)
                        z[j, 2] <- strip.white(x[i])
                    else if(i == k + 2)
                        z[j, 5] <- x[i]
                }
                z[, 3] <- ix
                z[, 4] <- ind
                mat <- rbind(mat, z)
                get.pg <- nrow(z) == 100
            }
            else if(length(y2) > 0)
            {
                z <- rep("", 5)
                y1 <- grep("CIK#: ", y)
                y1 <- y[y1[1]]
                y1 <- unlist(strsplit(y1, " CIK#: "))
                z[2] <- strip.white(y1[1])
                y1 <- y1[2]
                z[1] <- unlist(strsplit(y1, " "))[1]
                y2 <- grep("SIC: ", y)
                y2 <- y[y2[1]]
                y2 <- unlist(strsplit(y2, paste("SIC: ",
                    ix, " - ", sep = "")))[2]
                y2 <- unlist(strsplit(y2, "State location: "))
                z[4] <- y2[1]
                y2 <- unlist(strsplit(y2[2], " "))
                z[5] <- y2[1]
                z[3] <- ix
                mat <- rbind(mat, z)
                get.pg <- F
            }
            else
                get.pg <- F
```

```r
                        s <- s + 100
                }
        }
        sec.write.table(mat, "SIC.Download.txt", T)
        write(as.character(Sys.time()), log.file, append = T)
        sec.sic(incl.otc = incl.otc)
}

sec.dump <- function ()
{
        x <- sapply(ls(pos = 1), function(x) storage.mode(get(x)))
        y <- names(x)[x == "function"]
        z <- grep("sec\\.", y)
        y <- y[z]
        save(list = y, file = "SEC.RData")
        dump(y, "sec.code.txt")
}

sec.fix.names <- function(x, drop.suffix, drop.last = T)
{
        x <- toupper(x)
        x <- gsub("D/B/A", "", x)
        x <- gsub("\\.COM", " COM", x)
        x <- gsub(",", "", x)
        x <- gsub("\'", "", x)
        x <- gsub("\\.", "", x)
        x <- gsub(";", "", x)
        x <- gsub("\\(", "", x)
        x <- gsub("\\)", "", x)
        x <- gsub("/", " ", x)
        x <- gsub("!", " ", x)
        x <- gsub("&", "AND", x)
        x <- gsub("&", "AND", x)
        x <- gsub("-", " ", x)
        x <- gsub("  ", " ", x)
        x <- gsub("CORPORATION", "CORP", x)
        x <- gsub("^ *|(?<= ) | *$", "", x, perl = TRUE)
        if(!drop.last)
                return(x)
        for(i in 1:length(x))
        {
                while(drop.last){
                        y <- unlist(strsplit(x[i], " "))
                        if(y[1] == "THE")
                                y <- y[-1]
                        n <- length(y)
                        z <- y[n]
                        drop.last <- any(z == drop.suffix) & n > 1
                        if(drop.last)
                        {
                                x[i] <- y[1]
                                if(n > 2)
                                        for(j in 2:(n - 1))
                                                x[i] <- paste(x[i], y[j], sep = " ")
                        }
                }
                drop.last <- T
```

```r
        }
        return(x)
}

sec.get.webpage <- function (url)
{
        response <- try(GET(url))
        if(mode(response) == "list")
        {
                x <- unlist(response)
                x <- grep("status_code", x)
                if(length(x) == 1)
                        if(response$status_code != 200)
                                print(c(url, ",
                                        http status = ", response$status_code))
        }
        u <- htmlParse(response, useInternalNodes = T, asText = T)
        webpage <- as(u, "character")
        webpage <- readLines(tc <- textConnection(webpage))
        close(tc)
        return(webpage)
}

sec.match <- function (a, b)
{
        x <- match(a, b)
        x <- x[!is.na(x)]
        return(x)
}

sec.nasdaq.ticker <- function (x)
{
        if(vec.x <- is.vector(x))
                x <- as.matrix(x, length(x), 1)
        y <- x[, 1]
        take <- nchar(y) == 5
        if(sum(take) == 0)
        {
                if(vec.x)
                        x <- as.vector(x)
                return(x)
        }
        z <- y[take]
        for(i in 1:length(z))
        {
                v <- unlist(strsplit(z[i], ""))
                if(any(v[5] == c("G","H","I","R","T","V","W","X")))
                        z[i] <- "99999"
        }
        y[take] <- z
        x[, 1] <- y
        bad <- y == "99999"
        x <- x[!bad, ]
        if(vec.x)
                x <- as.vector(x)
        return(x)
}
```

```r
sec.nq.amex <- function (db)
{
    y <- sec.read.nq(db$file.nq.amex)
    y <- sec.nyse.ticker(y)
    db$tickers.nq.amex <- as.character(y[, 1])
    db$mkt.cap.nq.amex <- as.character(y[, 2])
    db$last.nq.amex <- as.numeric(y[, 3])
    db$exch.nq.amex <- rep("A", nrow(y))
    db$name.nq.amex <- as.character(y[, 4])
}

sec.nq.nasdaq <- function (db)
{
    y <- sec.read.nq(db$file.nq.nasdaq)
    y <- sec.nasdaq.ticker(y)
    db$tickers.nq.nasdaq <- as.character(y[, 1])
    db$mkt.cap.nq.nasdaq <- as.character(y[, 2])
    db$last.nq.nasdaq <- as.numeric(y[, 3])
    db$exch.nq.nasdaq <- rep("Q", nrow(y))
    db$name.nq.nasdaq <- as.character(y[, 4])
}

sec.nq.nyse <- function (db)
{
    y <- sec.read.nq(db$file.nq.nyse)
    y <- sec.nyse.ticker(y)
    db$tickers.nq.nyse <- as.character(y[, 1])
    db$mkt.cap.nq.nyse <- as.character(y[, 2])
    db$last.nq.nyse <- as.numeric(y[, 3])
    db$exch.nq.nyse <- rep("N", nrow(y))
    db$name.nq.nyse <- as.character(y[, 4])
}

sec.nt.nasdaq <- function (db)
{
    x <- sec.read.nt(db$file.nt.nasdaq)
    y <- x[, c("Symbol", "Test.Issue")]
    y <- sec.nasdaq.ticker(y)
    test <- y[, 2] == "Y"
    db$tickers.nt.nasdaq <- y[!test, 1]
    db$exch.nt.nasdaq <- rep("Q", length(db$tickers.nt.nasdaq))
    db$test.tickers.nt.nasdaq <- y[test, 1]
}

sec.nt.other <- function (db)
{
    x <- sec.read.nt(db$file.nt.other)
    exch <- x[, "Exchange"]
    exch[exch == "P"] <- "N"
    x[, "Exchange"] <- exch
    y <- x[, c("CQS.Symbol", "Exchange", "Test.Issue")]
    y <- sec.nyse.ticker(y)
    test <- y[, 3] == "Y"
    db$tickers.nt.other <- y[!test, 1]
    db$exch.nt.other <- y[!test, 2]
    db$test.tickers.nt.other <- y[test, 1]
}
```

```r
sec.nyse.ticker <- function (x)
{
	if(vec.x <- is.vector(x))
		x <- as.matrix(x, length(x), 1)
	y <- x[, 1]
	z <- gsub("[A-Z]", "", y)
	take <- z == "" | z == "."
	y[!take] <- "99999"

	for(i in 1:length(y))
	{
		if(y[i] == "99999")
			next
		z <- unlist(strsplit(y[i], "\\."))
		if(length(z) == 2)
			if(!any(c("A", "B", "C") == z[2]))
				y[i] <- "99999"
	}

	x[, 1] <- y
	bad <- y == "99999"
	x <- x[!bad, ]
	if(vec.x)
		x <- as.vector(x)
	return(x)
}

sec.osha <- function ()
{
	require(XML)
	require(httr)
	get.ind <- function(id, code)
	{
		url <- paste(
	"https://www.osha.gov/pls/imis/sic_manual.display?id=",
	id, "&tab=group", sep = "")

		y <- sec.get.webpage(url)
		x <- sec.parse.html(y, keyword = "div")
		y1 <- grep(paste("Industry Group ", code, "*.:", sep = ""), x)[1]
		y2 <- grep("SIC Search Division Structure", x)[1]
		x <- x[y1:(y2-1)]
		z <- c("Code", "Industry Group", "Industry")
		for(i in 1:length(x))
		{
			if(length(grep("Industry Group", x[i])) > 0)
			{
				u <- unlist(strsplit(x[i], " "))
				u <- u[-(1:3)]
				for(j in 1:length(u))
					if(j == 1)
						ind.grp <- u[1]
					else
						ind.grp <- paste(ind.grp, u[j],
							sep = " ")
			}
			else
```

```r
            {
                v <- unlist(strsplit(x[i], " "))
                code <- v[1]
                ind <- gsub(paste(code, " ", sep = ""), "", x[i])
                z <- rbind(z, c(code, ind.grp, ind))
            }
        }
        return(z)
    }
    url <- "https://www.osha.gov/pls/imis/sic_manual.html"
    y <- sec.get.webpage(url)
    x <- sec.parse.html(y, keyword = "div")
    y1 <- grep("SIC Division Structure", x)[1]
    y2 <- grep("Major Group 99", x)[1]
    x <- x[(y1+1):y2]
    z <- c("Code", "Division", "Major Group", "Industry Group", "Industry")
    for(i in 1:length(x))
    {
        u <- unlist(strsplit(x[i], ": "))
        if(length(grep("Division", u[1])) > 0)
            div <- u[2]
        else
        {
            maj <- u[2]
            code <- unlist(strsplit(u[1], " "))[3]
            v <- grep(paste("Major Group ", code, sep = ""), y)
            v <- y[v]
            v <- unlist(strsplit(v, "id="))[2]
            id <- unlist(strsplit(v, "&"))[1]
            q <- get.ind(id, code)
            q <- cbind(q[, 1], rep(div, nrow(q)), rep(maj, nrow(q)),
                q[, 2:3])
            q <- q[-1, ]
            z <- rbind(z, q)
        }
    }
    sec.write.table(z, "SIC.table.txt", T)
}

sec.parse.html <- function (webpage, keyword)
{
    pagetree <- htmlTreeParse(webpage, error=function(...){},
        useInternalNodes = TRUE)
    # parse the tree
    x <- xpathSApply(pagetree, paste("//*/", keyword, sep = ""), xmlValue)
    if(!is.character(x))
        return(keyword)
    # do some clean up with regular expressions
    x <- unlist(strsplit(x, "\n"))
    x <- gsub("\t","",x)
    x <- sub("^[[:space:]]*(.*?)[[:space:]]*$", "\\1", x, perl=TRUE)
    x <- x[!(x %in% c("", "|"))]
    return(x)
}

sec.read.nq <- function (file)
{
```

```r
    x <- read.csv(file = file, header = T, strip.white = T, sep = ",")
    x <- as.matrix(x)
    mode(x) <- "character"
    x1 <- sec.strip.white(x)
    y <- x1[, c("Symbol", "MarketCap", "LastSale")]
    y <- cbind(y, x[, "Name"])
    return(y)
}

sec.read.nt <- function (file)
{
    x <- readLines(file)
    x <- gsub("\'", "", x)
    file <- paste(file, ".cleaned.txt", sep = "")
    write(x, file = file)
    x <- read.table(file = file, sep = "|", header = T, comment.char = "@",
        fill = T, strip.white = T)
    x <- as.matrix(x)
    mode(x) <- "character"
    test <- x[, "Test.Issue"]
    bad <- test == ""
    x <- x[!bad, ]
    return(x)
}

sec.sic <- function (incl.otc = F)
{
    fund.etc <- function(x)
    {
        if(length(grep("FUND", x)) > 0)
            return(T)
        if(length(grep("TRUST", x)) > 0)
            return(T)
        if(length(grep("PORTFOLIO", x)) > 0)
            return(T)
        if(length(grep("ETF", x)) > 0)
            return(T)
        if(length(grep("INCOME", x)) > 0)
            return(T)
        if(length(grep("DIVIDEND", x)) > 0)
            return(T)
        if(length(grep("SHARES", x)) > 0)
            return(T)
        if(length(grep("DIVIDEND", x)) > 0)
            return(T)
        if(length(grep(" BOND", x)) > 0)
            return(T)
        if(length(grep(" RETURN", x)) > 0)
            return(T)
        if(length(grep(" SECURITIES", x)) > 0)
            return(T)
        if(length(grep(" INVESTMENT", x)) > 0)
            return(T)
        if(length(grep(" INVESTOR", x)) > 0)
            return(T)
        if(length(grep(" MUNICIPAL", x)) > 0)
            return(T)
```

```
                if(length(grep(" GROWTH", x)) > 0)
                        return(T)
                if(length(grep(" INFLATION", x)) > 0)
                        return(T)
                if(length(grep(" VOLATILITY", x)) > 0)
                        return(T)
                if(length(grep("DOW 30", x)) > 0)
                        return(T)
                if(length(grep("TREASURY", x)) > 0)
                        return(T)
                if(length(grep("CONTINGENT", x)) > 0)
                        return(T)
                if(length(grep("FLOATING", x)) > 0)
                        return(T)
                if(length(grep(" RATE", x)) > 0)
                        return(T)
                if(length(grep("INDEX", x)) > 0)
                        return(T)
                return(F)
        }
        drop.suffix <- c("INC", "INCORPORATED", "LP", "CORPORATION",
                "CORP", "PLC", "LTD", "LIMITED", "COMPANY", "AG", "SA",
                "LLC", "PLLC", "DBA", "THE", "NEW", "NV", "HOLDING",
                "HOLDINGS", "CO", "HLDGS", "HLDG", "PARTNERSHIP")

        drop.suffix <- c(drop.suffix, state.abb[])
        u <- read.delim("SIC.Codes.txt", header = F)
        u <- as.matrix(u)
        sic.names <- u[, 2]
        names(sic.names) <- sic.codes <- as.character(u[, 1])
        w <- read.delim("SIC.Download.txt", header = T)
        w <- as.matrix(w)
        mode(w) <- "character"
        sec.names <- sec.fix.names(w[, 2], "", drop.last = F)
        sec.sic.codes <- w[, 3]
        db <- new.env()
        db$file.nq.amex = "NQ_AMEX.csv"
        db$file.nq.nyse = "NQ_NYSE.csv"
        db$file.nq.nasdaq = "NQ_NASDAQ.csv"
        db$file.nt.other = "NT_otherlisted.txt"
        db$file.nt.nasdaq = "NT_nasdaqlisted.txt"
        sec.nq.amex(db)
        sec.nq.nyse(db)
        sec.nq.nasdaq(db)
        sec.nt.other(db)
        sec.nt.nasdaq(db)
        sec.tickers(db)
        tickers <- db$tickers
        co.names <- sec.fix.names(db$name, "", drop.last = F)
        exch <- db$exch.nq
        cap <- db$mkt.cap.nq
        if(incl.otc)
        {
                x <- read.csv("otctickers.csv", header = T)
                x <- as.matrix(x)
                tickers <- c(tickers, x[, 1])
                co.names <- c(co.names,
```

```r
                sec.fix.names(x[, 10], "", drop.last = F))
        exch <- c(exch, x[, 4])
        cap <- c(cap, rep(NA, nrow(x)))
    }
    n <- length(tickers)
    sic <- c("TICKER", "EXCH", "SIC", "SIC.NAME", "MKT.CAP")
    no.sic <- c("TICKER", "EXCH", "MKT.CAP", "FUND.ETC", "NO.MATCH")
    for(i in 1:n)
    {
        take <- co.names[i] == sec.names
        x <- unique(sec.sic.codes[take])
        if(length(x) == 1)
        {
            sic <- rbind(sic,
                 c(tickers[i], exch[i], x, sic.names[x], cap[i]))
            next
        }
        co.name <- sec.fix.names(co.names[i], drop.suffix)
        y <- grep(co.name, sec.names)
        if(length(y) == 0)
        {
            no.sic <- rbind(no.sic,
                 c(tickers[i], exch[i], cap[i], fund.etc(co.name),
                     "TRUE"))
            next
        }
        sec.name <- sec.fix.names(sec.names[y], drop.suffix)
        sec.sic.code <- sec.sic.codes[y]
        take <- co.name == sec.name
        if(sum(take) == 0)
        {
            no.sic <- rbind(no.sic,
                 c(tickers[i], exch[i], cap[i], fund.etc(co.name),
                     "TRUE"))
            next
        }
        x <- unique(sec.sic.code[take])
        if(length(x) > 1)
        {
            no.sic <- rbind(no.sic,
                 c(tickers[i], exch[i], cap[i], fund.etc(co.name),
                     "FALSE"))
            next
        }
        sic <- rbind(sic,
             c(tickers[i], exch[i], x, sic.names[x], cap[i]))
    }
    y <- unique(sic[-1, 3])
    ind <- matrix(0, k <- nrow(sic) - 1, length(y))
    for(i in 1:k)
        ind[i, ] <- as.numeric(sic[i + 1, 3] == y)
    dimnames(ind)[[1]] <- sic[-1, 1]
    dimnames(ind)[[2]] <- y
    sec.write.table(sic, file = "TICKER.SIC.txt", last.return = T)
    sec.write.table(no.sic, file = "NO.SIC.txt", last.return = T)
    write.table(ind, file = "SIC.IND.CLASS.txt", quote = F, sep = "\t")
}
```

```r
sec.strip.white <- function (x)
{
	x <- gsub("\t", "", x)
	x <- gsub(" ", "", x)
	return(x)
}

sec.tickers <- function (db)
{
	db$tickers.nq <- c(db$tickers.nq.amex, db$tickers.nq.nyse,
		db$tickers.nq.nasdaq)
	db$exch.nq <- c(db$exch.nq.amex, db$exch.nq.nyse, db$exch.nq.nasdaq)
	db$mkt.cap.nq <- c(db$mkt.cap.nq.amex, db$mkt.cap.nq.nyse,
		db$mkt.cap.nq.nasdaq)
	db$last.nq <- c(db$last.nq.amex, db$last.nq.nyse, db$last.nq.nasdaq)
	db$name.nq <- c(db$name.nq.amex, db$name.nq.nyse, db$name.nq.nasdaq)
	db$tickers.nt <- c(db$tickers.nt.other, db$tickers.nt.nasdaq)
	db$exch.nt <- c(db$exch.nt.other, db$exch.nt.nasdaq)
	db$test.tickers.nt <- c(db$test.tickers.nt.other,
		db$test.tickers.nt.nasdaq)
	x <- sec.match(db$test.tickers.nt, db$tickers.nq)
	db$tickers.nq <- db$tickers.nq[-x]
	db$exch.nq <- db$exch.nq[-x]
	db$mkt.cap.nq <- db$mkt.cap.nq[-x]
	db$last.nq <- db$last.nq[-x]
	db$name.nq <- db$name.nq[-x]
	db$tickers <- db$tickers.nq
	db$name <- db$name.nq
}

sec.write.table <- function (x, file, last.return = F)
{
	if(last.return)
	{
		write.table(x, file = file, quote = F, row.names = F,
			col.names = F, sep = "\t")
		return(1)
	}
	single.line <- F
	if(is.matrix(x))
		if(nrow(x) == 1)
			single.line <- T
	if(is.vector(x))
		single.line <- T
	if(single.line)
	{
		write.table(x, file = file, quote = F, row.names = F,
			col.names = F, sep = "\t", eol = "")
		return(1)
	}
	y <- x[nrow(x), ]
	x <- x[-nrow(x), ]
	z <- y[1]
	if(length(y) > 1)
		for(i in 2:length(y))
			z <- paste(z, "\t", y[i], sep = "")
```

```
        write.table(x, file = file, quote = F, row.names = F,
            col.names = F, sep = "\t")
        write.table(z, file = file, quote = F, row.names = F,
            col.names = F, sep = "\t", eol = "", append = T)
}
```

**Appendix C.1: Ticker Symbol Downloads**

● `wget.exe "http://www.nasdaq.com/screening/companies-by-name.aspx?exchange=amex&render=download" -O NQ_AMEX.csv /y`
● `wget.exe "http://www.nasdaq.com/screening/companies-by-name.aspx?exchange=nyse&render=download" -O NQ_NYSE.csv /y`
● `wget.exe "http://www.nasdaq.com/screening/companies-by-name.aspx?exchange=nasdaq&render=download" -O NQ_NASDAQ.csv /y`
● `wget.exe "http://www.nasdaqtrader.com/dynamic/SymDir/nasdaqlisted.txt" -O NT_nasdaqlisted.txt /y`
● `wget.exe "http://www.nasdaqtrader.com/dynamic/SymDir/otherlisted.txt" -O NT_otherlisted.txt /y`

**Appendix D: DISCLAIMERS**

Wherever the context so requires, the masculine gender includes the feminine and/or neuter, and the singular form includes the plural and vice-versa. The author of this paper ("Author") and his affiliates including without limitation Quantigic® Solutions LLC ("Author's Affiliates" or "his Affiliates") make no implied or express warranties or any other representations whatsoever, including without limitation implied warranties of merchantability and fitness for a particular purpose, in connection with or with regard to the content of this paper including without limitation any code or algorithms contained herein ("Content").

The reader may use the Content solely at his/her/its own risk and the reader shall have no claims whatsoever against the Author or his Affiliates and the Author and his Affiliates shall have no liability whatsoever to the reader or any third party whatsoever for any loss, expense, opportunity cost, damages or any other adverse effects whatsoever relating to or arising from the use of the Content by the reader including without any limitation whatsoever: any direct, indirect, incidental, special, consequential or any other damages incurred by the reader, however caused and under any theory of liability; any loss of profit (whether incurred directly or indirectly), any loss of goodwill or reputation, any loss of data suffered, cost of procurement of substitute goods or services, or any other tangible or intangible loss; any reliance placed by the reader on the completeness, accuracy or existence of the Content or any other effect of using the Content; and any and all other adversities or negative effects the reader might encounter in using the Content irrespective of whether the Author or his Affiliates is or are or should have been aware of such adversities or negative effects.

The R code included in Appendix C hereof is part of the copyrighted R code of Quantigic® Solutions LLC and is provided herein with the express permission of Quantigic® Solutions LLC. The copyright owner retains all rights, title and interest in and to its copyrighted source code included in Appendix C hereof and any and all copyrights therfor.

Katzen, S. (1995) Economic classification policy committee: Standard industrial classification replacement–The North American industry classification system proposed industry classification structure. *Federal Register* **60**(143): 38436-38452.

Kile, C.O.; Phillips, M.E. (2009) Using Industry Classification Codes to Sample High-Technology Firms: Analysis and Recommendations. *Journal of Accounting, Auditing & Finance* **24**(1): 35-58.

King, B.F. (1966) Market and Industry Factors in Stock Price Behavior. *Journal of Business* **39**(1): 139-190.

Kort, J.R. (2001) The North American industry classification system in BEA's economic accounts. *Survey of Current Business* **81**(5): 7-13.

Krishnan, J.; Press, E. (2002) The North American Industry Classification System and Its Implications for Accounting Research. Temple University. Working paper.

Laestadius, S. (2005) The classification and taxonomy of industries – measuring the right thing. In: Hirsch-Kreinsen, H.; Jacobson, D.; Laestadius, S. (eds.) *Low-tech innovation in the knowledge economy*. New York, NY: Peter Lang Publishing, pp. 63-84.

Lamponi, D. (2014) Is industry classification useful to predict U.S. stock price co-movements? *Journal of Wealth Management* **17**(1): 71-77.

Lee, C.M.C.; Ma, P.; Wang, C.C.Y. (2015) Search-based peer firms: Aggregating investor perceptions through internet co-searches. *Journal of Financial Economics* **116**(2): 410-431.

Mantegna, R.N. (1999) Hierarchical structure in financial markets. *The European Physical Journal B: Condensed Matter and Complex Systems* **11**(1): 193-197.

Miccichè, S.; Lillo, F.; Mantegna, R.N. (2005) Correlation based hierarchical clustering in financial time series. In: Beck, C.; Benedek, G.; Rapisarda, A.; Tsallis, C. (eds.) Complexity, Metastability and Nonextensivity: Proceedings of the 31st Workshop of the International School of Solid State Physics, Erice, Sicily, Italy, 20-26 July 2004. Singapore: World Scientific, pp. 327-335.

Mross, E.L.; McGuigan, G.S. (2016) The Necessity of NAICS: Industrial Classification as an Indexing Search Tool. *Academic BRASS* **11**(2): Fall, 2016.

Murphy, J.B. (1998) Introducing the North American industry classification system. *Monthly Labor Review* **121**(7): 43-47.

Nadig, D.; Crigger, L. (2011) Signal from noise. *Journal of Indexes* **14**(2): 40-43.
65

Yaros, J.R.; Imieliński, T. (2015) Data-driven methods for equity similarity prediction. *Quantitative Finance* **15**(10): 1657-1681.

## Tables

| Run/Ticker # | Total | w/ SIC | w/o SIC | No match | Multiple matches | Funds, etc. |
|---|---|---|---|---|---|---|
| Listed only | 6040 | 4759 | 1281 | 1209 | 72 | 889 |
| Listed + OTC | 15640 | 9592 | 6048 | 5927 | 121 | 1067 |

**Table 1.** Ticker matching statistics for runs on 04/12/2017 (see Subsection 2.3 for details).

| Download | Start Time | End Time | Download Time (min:sec) |
|---|---|---|---|
| All SIC | 2017-04-08 17:08:50 | 2017-04-08 17:38:34 | 29:44 |
| All SIC | 2017-04-08 23:58:40 | 2017-04-09 00:27:46 | 29:06 |
| All SIC | 2017-04-09 23:09:37 | 2017-04-09 23:38:45 | 29:12 |
| SEC SIC | 2017-04-08 23:50:45 | 2017-04-08 23:56:37 | 5:52 |
| SEC SIC | 2017-04-09 15:55:30 | 2017-04-09 15:59:46 | 4:16 |
| SEC SIC | 2017-04-09 22:57:33 | 2017-04-09 23:01:54 | 4:21 |

**Table 2.** Download times for all 0100-9999 ("All SIC") and `SIC.Codes.txt` ("SEC SIC") codes.

| Classification | ROC | SR | CPS |
|---|---|---|---|
| BICS, sub-industries | 46.96% | 17.61 | 2.077 |
| BICS, industries | 46.05% | 16.62 | 2.018 |
| BICS, sectors | 42.85% | 14.55 | 1.868 |
| GICS, sub-industries | 47.43% | 17.96 | 2.096 |
| SIC, industries | 45.05% | 16.85 | 1.987 |
| FF48 | 44.19% | 15.94 | 1.929 |
| FF49 | 44.33% | 15.98 | 1.936 |

**Table 3.** Annualized Return on Capital (ROC), Annualized Sharpe Ratio (SR) and Cents-Per-Share (CPS) for backtests in Section 3 for heterotic risk models using the classifications in column 1. GICS = Global Industry Classification Standard; BICS = Bloomberg Industry Classification System; SIC = Standard Industrial Classification; FF48 (FF49) = Fama-French [1997] industry classification with 48 (49) SIC-based "industries". The mean-reversion alphas discussed in Subsection 3.4 are optimized using heterotic risk models, which in turn are built based on these industry classifications listed in column 1 of this table.

**Figures**

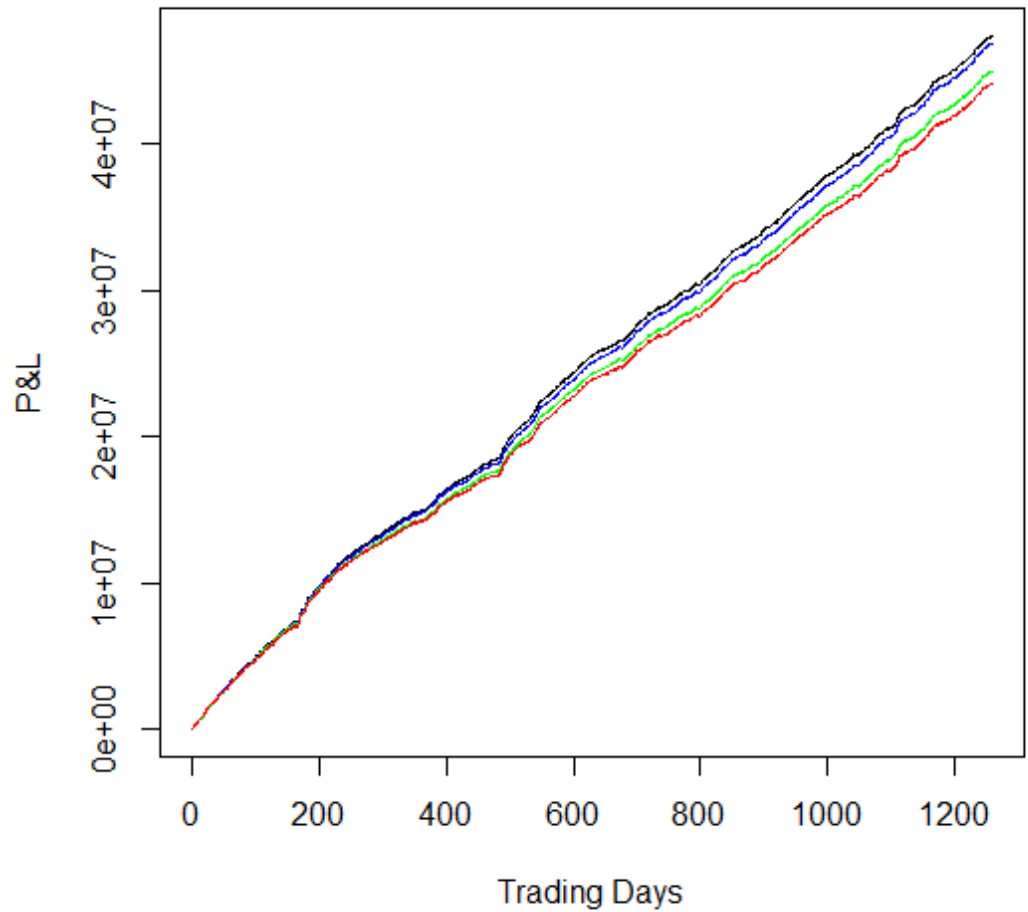

**Figure 1.** P&L graphs for the intraday alphas summarized in Table 3. Bottom-to-top-performing: i) FF48, ii) SIC industries, iii) BICS sub-industries, iv) GICS sub-industries. The investment level is $10M long plus $10M short. See Section 3 for details.